\documentclass[twocolumn,aps,prb,superscriptaddress]{revtex4-2}
\makeindex
\usepackage{amsmath}    
\usepackage{graphicx}   
\usepackage{verbatim}   
\usepackage{color}      
\usepackage{subfigure}  
\usepackage{hyperref}   
\usepackage{gensymb}    
\usepackage{xcolor}     
\usepackage{nicefrac,xfrac} 
\usepackage{soul}       
\byrevtex

\usepackage{chemformula}     
\usepackage{afterpage}       
\usepackage{upgreek}         
\usepackage{array}           

\usepackage{amssymb}         
\usepackage{mathtools}       
\usepackage{xspace}          
\usepackage{colortbl}        
\usepackage{ragged2e}        
\usepackage{rotating}        
\usepackage{textcomp}        
\usepackage{revsymb4-2}      
\usepackage{mathrsfs}        
\usepackage{natbib}          
\usepackage{tabulary}        
\usepackage{siunitx}         
\usepackage[version=4]{mhchem} 
\usepackage{float}           
\usepackage[shortlabels]{enumitem} 

\begin{document}
\title{Soft phonon and the central peak at the cubic-to-tetragonal phase transition in SrTiO$_3$}

\author{Avishek Maity}
\email{Contact author: maitya@ornl.gov}
\email{Present address: Neutron Scattering Division, Oak Ridge National Laboratory, Oak Ridge, Tennessee 37831, USA}
\affiliation{Heinz Maier-Leibnitz Zentrum (MLZ), Technische Universität München, D-85747 Garching, Germany}

\author{Klaus Habicht}
\affiliation{Department Dynamics and Transport in Quantum Materials, Helmholtz-Zentrum Berlin f\"{u}r Materialien und Energie, Hahn-Meitner-Platz 1, D-14109 Berlin, Germany}
\affiliation{Institut f\"{u}r Physik und Astronomie, Universit\"{a}t Potsdam, Karl-Liebknecht-Stra{\ss}e 24-25, D-14476 Potsdam, Germany}

\author{Michael Merz}
\affiliation{Institute for Quantum Materials and Technologies, Karlsruhe Institute of Technology, Kaiserstr. 12, D-76131 Karlsruhe, Germany}
\affiliation{Karlsruhe Nano Micro Facility (KNMFi), Karlsruhe Institute of Technology, Kaiserstr. 12, D-76131 Karlsruhe, Germany}

\author{Ayman H. Said}
\affiliation{Advanced Photon Source, Argonne National Laboratory, Lemont, Illinois 60439, USA}

\author{Christo Guguschev}
\affiliation{Leibniz-Institut für Kristallzüchtung, Max-Born-Straße 2, D-12489 Berlin, Germany}

\author{Danny Kojda}
\affiliation{Department Dynamics and Transport in Quantum Materials, Helmholtz-Zentrum Berlin f\"{u}r Materialien und Energie, Hahn-Meitner-Platz 1, D-14109 Berlin, Germany}

\author{Britta Ryll}
\affiliation{Department Dynamics and Transport in Quantum Materials, Helmholtz-Zentrum Berlin f\"{u}r Materialien und Energie, Hahn-Meitner-Platz 1, D-14109 Berlin, Germany}

\author{Jan-Ekkehard Hoffmann}
\affiliation{Department Dynamics and Transport in Quantum Materials, Helmholtz-Zentrum Berlin f\"{u}r Materialien und Energie, Hahn-Meitner-Platz 1, D-14109 Berlin, Germany}

\author{Andrea Dittmar}
\affiliation{Leibniz-Institut für Kristallzüchtung, Max-Born-Straße 2, D-12489 Berlin, Germany}

\author{Thomas Keller}
\affiliation{Max-Planck-Institut für Festkörperforschung, Heisenbergstraße 1, D-70569 Stuttgart, Germany}
\affiliation{Max Planck Society Outstation at the Heinz Maier-Leibnitz Zentrum (MLZ), D-85748 Garching, Germany}

\author{Frank Weber}
\email{Contact author: frank.weber@kit.edu}
\affiliation{Institute for Quantum Materials and Technologies, Karlsruhe Institute of Technology, Kaiserstr. 12, D-76131 Karlsruhe, Germany}

\begin{abstract}
The continuous displacive phase transition in SrTiO$_3$ near $T_c \approx 105$ K features a central elastic peak in neutron scattering investigations at temperatures above $T_c$, \textit{i.e.}, before the corresponding soft phonon mode is overdamped upon cooling. The origin of this central peak is still not understood. Here, we report an inelastic x-ray scattering investigation of the cubic-to-tetragonal phase transition in SrTiO$_3$. We compare quantitatively measurements of the soft phonon mode on two differently grown samples and discuss the findings regarding results from thermodynamic and transport probes such as specific heat and thermal conductivity. Furthermore, we use inelastic x-ray scattering to perform elastic scans with both high momentum- and milli-electronvolt energy-resolution and, thus, be able to separate elastic intensities of the central peak from low-energy quasielastic phonon scattering. Our results indicate that the evolution of the soft mode is similar in both samples though the intensities of the central peak differ by a factor of four. Measurements revealing anisotropic correlation lengths on cooling towards $T_c$, indicate that local properties of the crystals to which collective lattice excitations are insensitive are likely at the origin of the central elastic line in SrTiO$_3$.

\end{abstract}

\maketitle

\section{INTRODUCTION}

Among the class of transition metal oxides, SrTiO$_3$ has not only served as a model material to demonstrate a variety of basic physical phenomena, but is also known for its remarkable physical properties, most notably polar and structural characteristics, which allow for the realization of exceptional functional properties. Besides static incidences of emergent states due to collective order such as dilute superconductivity \citep{Schooley_prl_1964,Fauqu_prr_2023,Collignon_arcmp_2019,Gastiasoro_aop_2020} and emergent phenomena which originate at interfaces in artificial heterostructures based on SrTiO$_3$ \citep{Ohtomo_nature_2004,Bi_ncom_2014,Li_nphys_2011,Zubko_arcmp_2011}, recent research has demonstrated that the interplay of quantum fluctuations and antiferrodistortive (AFD) structural instabilities in SrTiO$_3$, which normally leads to its quantum paraelectric state, can be dynamically manipulated resulting in light-induced or strain-induced ferroelectricity \citep{Fechner_nmat_2024,Haeni_nature_2004,Xu_ncom_2020} or even electric-field driven dynamical multiferroicity in SrTiO$_3$ \citep{Basini_nature_2024}. Moreover, a plethora of unusual thermal transport properties derive from the interplay of ferroelectricity, phonon softening, quantum fluctuations and topological properties \citep{Enderlein_ncom_2020,Marel_prr_2019,Ahadi_sciadv_2019,Edge_prl_2015,Tomioka_ncom_2019,Coak_pnas_2020,He_prl_2020,Peng_sciadv_2020}, including Poiseuille flow of phonons \citep{Martelli_prl_2018} and the elusive coupling of phonons in SrTiO$_3$ to magnetic fields \citep{Li_prl_2020,Sim_prl_2021}.

The model perovskite SrTiO$_3$ is well-known for its strongly anharmonic phonon properties underlying the intriguing physics of soft phonon modes for phase transition \citep{Cochran_prl_1959,Cowley_aip_1980}. The cubic-to-tetragonal structural phase transition at $T_C=105$ K \citep{Unoki_jsps_1967,Fleury_prl_1968,Cowley_ssc_1969} has been studied in detail by neutron  \citep{Shirane_physrev_1969,Yamada_jpsj_1969,Otnes_ssc_1971,Riste_ssc_1971,Shapiro_ssc_1972,Stirling_jpcssp_1972,Topler_jpcssp_1977,Hastings_prl_1978,Shirane_prb_1993}
and x-ray scattering \citep{Andrews_jpcssp_1986,McMorrow_ssc_1990,Hirota_prb_1995,Hunnefeld_prb_2002,Holt_prl_2007,Ravy_prl_2007,Hong_prb_2008}
 since the late sixties, and is considered a text-book example of a phonon driven second-order structural phase transition \citep{Cowley_aip_1980}. At temperatures above $T_C$ (cubic phase), the soft mode corresponds to a rotation of the TiO$_6$ octahedra close to the R-point of the Brillouin zone. Approaching the phase transition from higher temperatures, this phonon mode softens and finally locks into a static rotation angle in the tetragonal phase.

In addition to the soft phonon modes, early neutron scattering experiments have detected quasi-elastic scattering centered around zero energy transfer, known as central peak (CP), revealing the presence of a second time scale apart from that of the phonon softening \citep{Riste_ssc_1971,Shapiro_ssc_1972,Shirane_prb_1993}. Subsequent x-ray experiments reported a peak much narrower in momentum space [also known as narrow component] than that observed in neutron scattering and, thus, revealed a second length scale  in momentum space \citep{Andrews_jpcssp_1986,McMorrow_ssc_1990,Hirota_prb_1995}.

Defects or dislocations have been proposed as the origin of the CP enigma \citep{Hirota_prb_1995,Hunnefeld_prb_2002}. Indeed, introduction of point defects changes the transition temperature $T_C$ \citep{Hunnefeld_prb_2002,McCalla_cm_2016}. Yet, neutron scattering found only a doubling of the CP intensity upon increasing the oxygen defect concentration by two orders of magnitude \citep{Hastings_prl_1978}. Thus, the precise mechanism for the emergence of the CP remains a matter of ongoing debate.

In our work, we carefully characterize differently grown samples of SrTiO$_3$ and employ inelastic x-ray scattering (IXS) to study soft-mode as well as CP properties on cooling towards the cubic-to-tetragonal phase transition. We compare scattering intensities quantitatively and find that the soft-mode behavior does not depend on defect concentrations. On the other hand, the CP intensity increases four-fold, surprisingly in the sample with less defects.

\section{EXPERIMENTAL DETAILS}
Single crystals of SrTiO$_3$ used for our studies were grown using the edge-defined film-fed growth (EFG) \citep{Guguschev_cec_2015} and Verneuil (VER) techniques. Specific heat and thermal conductivity measurements were performed using a commercial physical property measurement system (PPMS) from Quantum Design, Inc. For more details see supplemental information (SI) \citep{Supple_STO_2024} (see also references \citep{1s16_Gane_acsb_2015,2s6_Choi_adm_2013,3s19_Li_ceri_2019,4s5_Lashley_cryo_2003,5s3_Rekveldt_eurol_2001,6s15_Kubacki_jcp_2018,7s9_Duraen_jpcm_2008,8s21_Keller_nnews_2007,9s7_Ahrens_jpbcm_2007,10s34_Currat_prb_1978,11s13_Buban_prb_2004,12s17_Spinelli_prb_2010,13s8_McCalla_prm_2019,14s12_Ricca_prr_2020,15s25_Brockhouse_physrev_1962,16s31_Silberglitt_ssc_1972,17s36_Lu_ssc_2015,18s11_Baeuerle_ssc_1978,19s18_sole_sapbas_2007,s23_Habicht_unipotsdam_2016} therein). 

IXS experiments were performed at the HERIX (high energy resolution inelastic x-ray scattering) spectrometer located at beam line 30 ID of the Advanced Photon Source (APS) at the Argonne National Laboratory (ANL) \citep{Said_jsr_2011}. Both samples had rectangular platelet shape with dimensions of $[2\times1.2\times0.06]$ mm$^3$ and $[3.2\times2.1\times0.06]$ mm$^3$ respectively, and were mounted on the same Be plate [see inset in Fig. \ref{fig2}(b)]. The samples were placed in a closed cycle refrigerator to investigate in the temperature range $100<T<200$ K and the measurements were carried out in the transmission mode with x-ray energies of 23.78 keV. Different circular analyzer diameters ($15-95$ mm) were used providing different momentum resolutions. The components of the scattering vectors are expressed in reciprocal lattice units ($r.l.u$) as $(Q_h,Q_k,Q_l)=2\pi\times(H/a,K/b,L/c)$ where $a,b,c$ denote the cubic lattice constants. Phonon excitations measured from constant$-$\textbf{\textit{Q}} scans were approximated using damped harmonic oscillator (DHO) functions convoluted with the experimental resolution. From the fits, we have obtained phonon energy $\omega_q=\sqrt{\tilde{\omega_q}^2 - \Gamma^2}$, where $\tilde{\omega_q}$ represents the unrenormalized phonon energy and $\Gamma$ the damping of the DHO \citep{Fak_pbcm_1997}.

\section{RESULTS}

\begin{figure}
\includegraphics[width=0.475\textwidth]{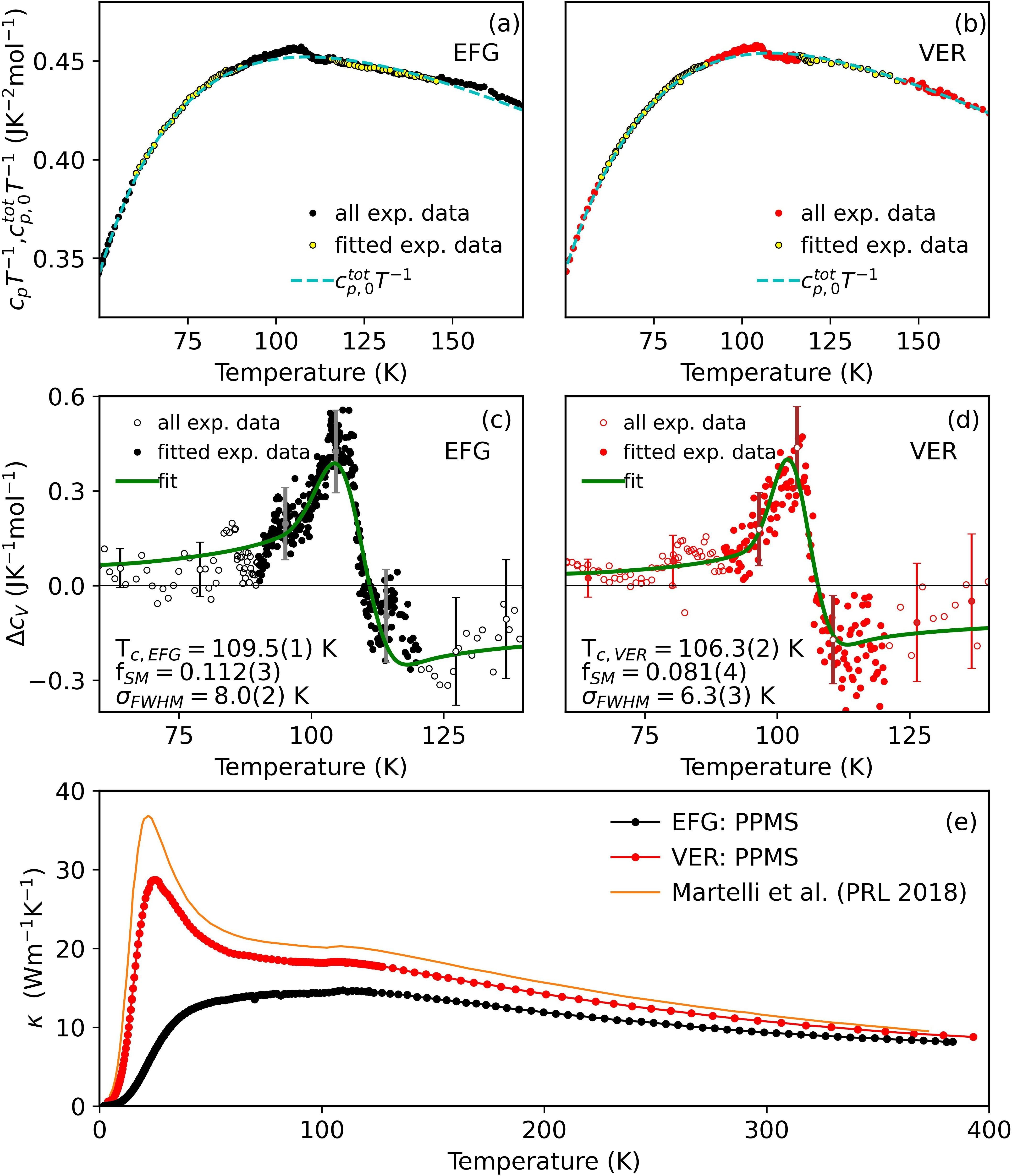}
\caption{\label{fig1}(a,b) Specific heat ($C_p$) data for the EFG and VER samples, respectively. (c,d) Difference molar specific heat ($\Delta C_v$) for the (c) EFG and (d) VER samples, respectively. Here, the estimated contributions to the specific heat from phonons other than the soft mode [dashed lines in panels (a,b)] was subtracted. The fit parameters including $T_c$ are indicated in the corresponding text inserts. For more detailed explanations see SI \citep{Supple_STO_2024}. (e) Thermal conductivity ($\kappa$) data, obtained from PPMS (Physical Property Measurement System), are presented indicating the suppression of the low temperature peak with the increasing defect concentration and compared to published data from pristine SrTiO$_3$ \citep{Martelli_prl_2018}. }
\end{figure}

Specific heat ($C_p$) measurements for the EFG and VER grown SrTiO$_3$ samples are presented in Figs. \ref{fig1}(a,b). Based on a model of phonons and their contribution to the specific heat in SrTiO$_3$ (see SI \citep{Supple_STO_2024}), we can separate the phase transition anomaly at $T_c \sim 105$ K related to the strongly temperature dependent soft phonon mode from the smoothly varying contributions of the other phonons. Subtracting the latter normal phononic background [dashed lines in Figs. \ref{fig1}(a,b)], we describe the remaining difference molar specific heat ($\Delta C_v$) data [Figs. \ref{fig1}(c,d)] by the contribution from the soft phonon mode of the cubic to tetragonal phase transition [solid lines in Figs. \ref{fig1}(c,d)]. Thus, we determine the transition temperatures $T_c$ for the antiferrodistortive structural transition $T_{c,EFG}=109.5(1)$ K and $T_{c,VER}=106.3(2)$ K in reasonable agreement with corresponding values reported in the literature \citep{Hunnefeld_prb_2002}. We note that the precise transition temperature of SrTiO$_3$ strongly depends on the defect concentration \citep{Ranjan_prl_2000}.

Our thermal conductivity ($\kappa$) measurements [Fig. \ref{fig1}(e)] show a clear suppression of the low temperature peak compared to pristine SrTiO$_3$ [orange line in Fig. \ref{fig1}(e)]. Previous work identified the number of point defects as origin for this suppression \citep{Martelli_prl_2018,Jiang_pnas_2022}. Consequently, we conducted a comprehensive characterization of the samples to identify of possible candidate impurities by x-ray fluorescence ($\mu$XRF) and to determine the defect concentrations on a quantitative level by inductively coupled plasma optical emission spectrometry (ICP-OES). In addition, we performed Larmor diffraction (LD) measurements to assess the sample mosaicity (see detailed report in the SI \citep{Supple_STO_2024}). Indeed, we detect significant concentrations of Fe, Ca and Ni defects by ICP-OES in our samples. While several types of point defects can be detected, the main difference between the VER and EFG samples originates from a larger concentration of Fe defects in the EFG sample. The presence of more defects in the EFG sample can also be identified by its greater opacity compared to the nearly transparent VER samples [see Fig. S1 in SI \citep{Supple_STO_2024}] \citep{Guguschev_cec_2015,Kok_cec_2016}. In summary, our characterization reveals that the EFG sample shows an impurity level about 1.5 times higher than the VER sample, despite its superior mosaicity and low dislocation density.

In the following, we present quantitatively comparable IXS results for both the EFG and VER samples. We study the behavior of the soft phonon mode as well as that of the CP measured in the same experimental setup and in nearly identically shaped/prepared sample. Samples were mounted on one sample holder and measurements were taken in one experimental run. Thus, we can discuss the relation between our results deduced from IXS and compare them to our results on the phase transition and defect concentration described above. The data sets for the inelastic and elastic x-ray scattering results presented in Figures 2-4 are available at the open data repository KITopen \citep{KITopen_STO_2025}.

\begin{figure}
\includegraphics[width=0.475\textwidth]{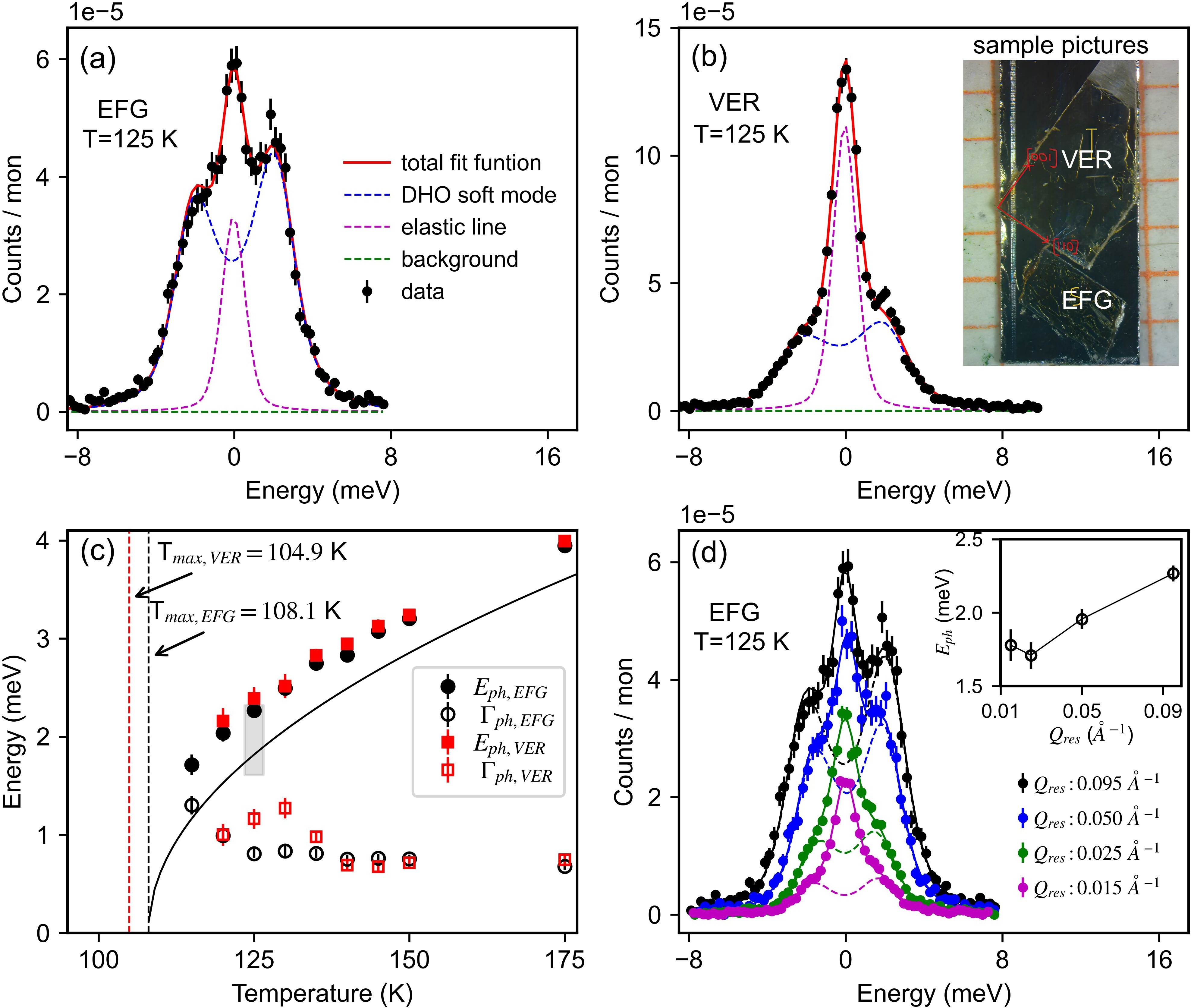}
\caption{\label{fig2}(a,b) Energy scans performed at constant momentum \textbf{\textit{Q$_R$}} $ =(3/2,3/2,5/2)$ at $T=125$ K for the EFG and VER grown samples, respectively. Solid (red) lines are fits consisting of a DHO function convoluted with the experimental resolution (blue dashed line), estimated background (straight green dashed line) and a resolution limited pseudo-Voigt function for the elastic line (purple dashed line). Error bars represent standard deviations. (Inset in b) Photo of the VER and EFG samples mounted on the Be plate. Red arrows indicate directions with respect to the cubic unit cell. Orientation was the same for both samples. Measurements were done in transmission.  (c) $T-$dependence of the phonon energies (filled symbols) and linewidths (open symbols) of the soft mode extracted from the fits of the energy scans at \textbf{\textit{Q$_R$}} for EFG (squares) and VER (dots/circles) samples, respectively. The grey-shaded rectangle at $T=125$ K shows the range of deduced phonon energies for varying momentum-resolution performed on the EFG sample [see (d) and text]. Solid line  represents a fit to phonon energies deduced from published neutron scattering data \citep{Cowley_ssc_1969,Shirane_physrev_1969,Otnes_ssc_1971,Stirling_jpcssp_1972}. Dashed vertical lines present the $T_c$ extracted from the $Q-$integrated intensity measurements at \textbf{\textit{Q$_R$}} [see Fig. \ref{fig3}(c) and text]. (d) Energy scans for the EFG sample performed at $T=125$ K with varying momentum resolutions (see text). (Inset) Extracted phonon energies  as a function of different momentum resolutions. The data sets for the inelastic x-ray scattering are available at the open data repository KITopen \citep{KITopen_STO_2025}.}
\end{figure}

To this end, we investigated the behavior of the soft phonon mode for both samples via a series of constant$-$\textbf{\textit{Q}} scans performed at the R-point of the cubic Brillouin zone, \textbf{\textit{Q$_R$}} $=(3/2,3/2,5/2)$, in the temperature range 175 K down to 100 K. The spectra shown in Figs. \ref{fig2}(a,b)] present the characteristic features observed at T = 125 K: (1) a soft phonon mode at finite energy and (2) a CP at zero energy transfer. We find that the CP intensity is generally four times larger for the VER sample compared to the EFG one (see below for a more detailed discussion), whereas the phonon intensities of the soft phonon mode  are practically the same.

Energy scans at \textbf{\textit{Q$_R$}} have been fitted with a DHO function for the soft mode convoluted with the experimental resolution [blue dashed lines in Figs. 2(a,b)] while the CP was approximated by a resolution-limited pseudo-voigt function [purple dashed lines]. The deduced energy of the DHO function ($E_{ph}$) and its line width ($\Gamma_{ph}$) are presented in Fig. \ref{fig2}(c). The CP dominates the energy spectrum close to $T_c$ and we cannot determine the soft mode properties anymore below $T_{min,VER} = 120$ K and $T_{min,EFG} = 115$ K. Within these limits, we find that there is no significant difference between the two samples with respect to the phonon softening on cooling towards the transition temperature. However, we find that the deduced energies are generally larger by an offset of $\sim 0.5$ meV than those reported previously [solid line in Fig. \ref{fig2}(c)] \citep{Cowley_ssc_1969,Otnes_ssc_1971,Stirling_jpcssp_1972,Shirane_physrev_1969}.

\begin{figure}
\includegraphics[width=0.475\textwidth]{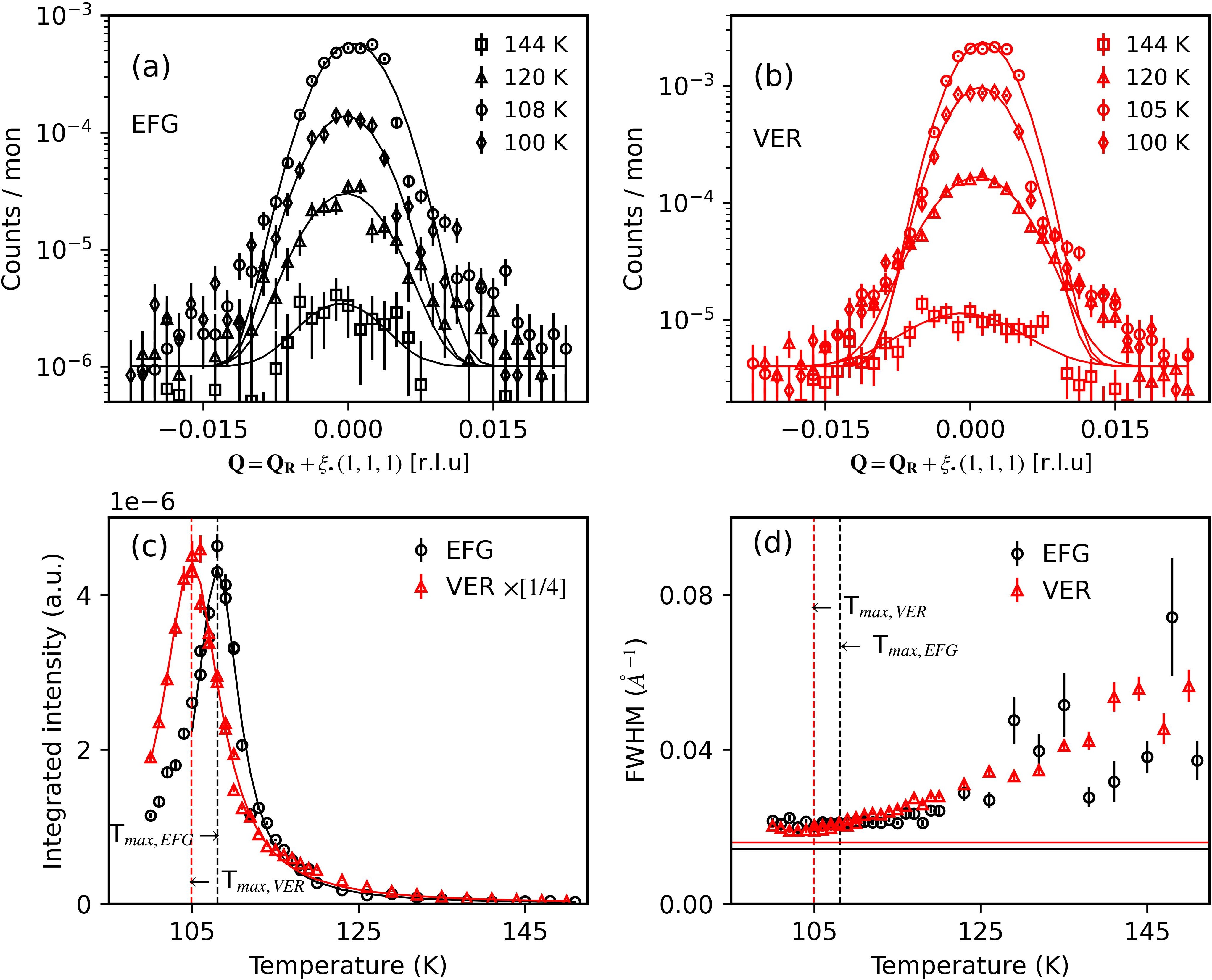}
\caption{\label{fig3}(a,b) Elastic scans through \textbf{\textit{Q$_R$}} $=(3/2,3/2,5/2)$ along $[111]$ for (a) EFG and (b) VER samples, respectively, using high-momentum resolution (see text). A phonon-background was subtracted from the data (see text and SI \citep{Supple_STO_2024}). The solid lines represent Gaussian fits. Temperature dependent (c) integrated intensities and (d) full width at half maximum (FWHM)   of the approximated Gaussians. The vertical dashed lines indicate the temperatures featuring maximum intensities, $T_{max,EFG}$ and $T_{max,VER}$, extracted from a peak fit of the integrated intensities for both samples [solid lines in panel (c)]. Color-coded horizontal solid lines in (d) denote the momentum resolutions along $[111]$,  0.01425 \AA$^{-1}$ for EFG sample and 0.01592 \AA$^{-1}$ for VER sample, extracted from the high resolution scans (15 mm analyzer opening) of the Bragg reflection (222) (see SI \citep{Supple_STO_2024}). The data sets for the elastic x-ray scattering are available at the open data repository KITopen \citep{KITopen_STO_2025}.}
\end{figure}

As we will show in the following this is due to the momentum resolution of our IXS experiment combined with the steep dispersion of the soft mode away from \textbf{\textit{Q$_R$}} \citep{Stirling_jpcssp_1972}. Though IXS is known to have a good momentum resolution, the typical size of the analyzer (in our case: circular shape with an opening of 95 mm at a distance of 9 m from the sample) is a compromise between scattering intensity and resolution and leads to a momentum resolution of FWHM $= 0.095$ \AA$^{-1}$ along $[111]$ direction. Measurements at $T = 125$ K in the EFG sample with continuously reduced analyzer openings from 95 mm to 15 mm [Fig. \ref{fig2}(d)] yield a reduction of the deduced phonon energy of about 0.5 meV [inset in Fig. \ref{fig2}(d)]. The result for the best momentum resolution is in good agreement with results obtained using low-energy neutrons \citep{Otnes_ssc_1971,Stirling_jpcssp_1972,Shirane_physrev_1969}. On the other hand, scattering intensities are dramatically reduced and, thus, the highest-resolution setup can only be employed for few selected phonon scans.

In contrast, elastic intensities at \textbf{\textit{Q$_R$}} close to $T_c$ are large and we have performed high momentum resolution scans (15 mm analyzer opening) along $[111]$ direction for both EFG [Fig. \ref{fig3}(a)] and VER [Fig. \ref{fig3}(b)] samples. Here, we have subtracted a phonon-background estimated from (1) the ratio between the DHO intensity and  the total observed one, both at zero energy transfer, and (2) the momentum dependence of the high-temperature scattering where the CP contribution is still small. The data were approximated by a Gaussian [lines in Figs. \ref{fig3}(a,b)].

We find that the elastic intensities increases on cooling well above $T_c$ for both samples and that the CP intensity is about four times larger in the VER sample, \textit{i.e.}, the sample having less defects [Fig. \ref{fig3}(c)]. The temperatures where the scattering intensities display a maximum, $T_{max,EFG} = 108.1$ K and $T_{max,VER} = 104.9$ K, are slightly below the transition temperatures deduced from specific heat $T_{c,EGF} = 109.5(1)$ K and  $T_{c,VER} = 106.3(2)$ K which is likely an effect of the finite momentum resolution. The reduction of the peak intensity originates in the splitting of the $(3/2,3/2,5/2)$ superlattice reflection below $T_c$ in different domains. However the splitting only affects the observed intensities once it is larger than the experimental momentum resolution. We note that the qualitative temperature dependence of the integrated intensities are the same except for a shift because of the $\Delta T_c$ already detected in our specific heat measurements.

The CP linewidths in both samples monotonically decrease on cooling but level off at a value of 0.02 \AA$^{-1}$ around $T = 115$ K, \textit{i.e.}, above the corresponding resolution limits along $[111]$ direction [color-coded solid horizontal lines in Fig. \ref{fig3}(d)] and the respective transition temperatures [color-coded dashed vertical lines in Fig. \ref{fig3}(d)] and the respective values of $T_{max}$ [color-coded dashed vertical lines in Fig. \ref{fig3}(d)]. Differences between the samples close to the phase transition are small, though it seems that results for the EFG sample level off at slightly higher temperatures and at a slightly higher value of the FWHM. Overall, the evolution of the width of the CP is similar for both samples along the $[111]$ direction. Thus, the most striking difference between the two samples is the four times higher CP intensity in the VER sample [Fig. \ref{fig3}(c)]. We emphasize the relevance of this discrepancy in comparison to the observed phonon intensities which are the same for both samples within the experimental uncertainty [see Figs. \ref{fig2}(a,b) and SI Fig. S12].

\begin{figure}
\includegraphics[width=0.79\textwidth]{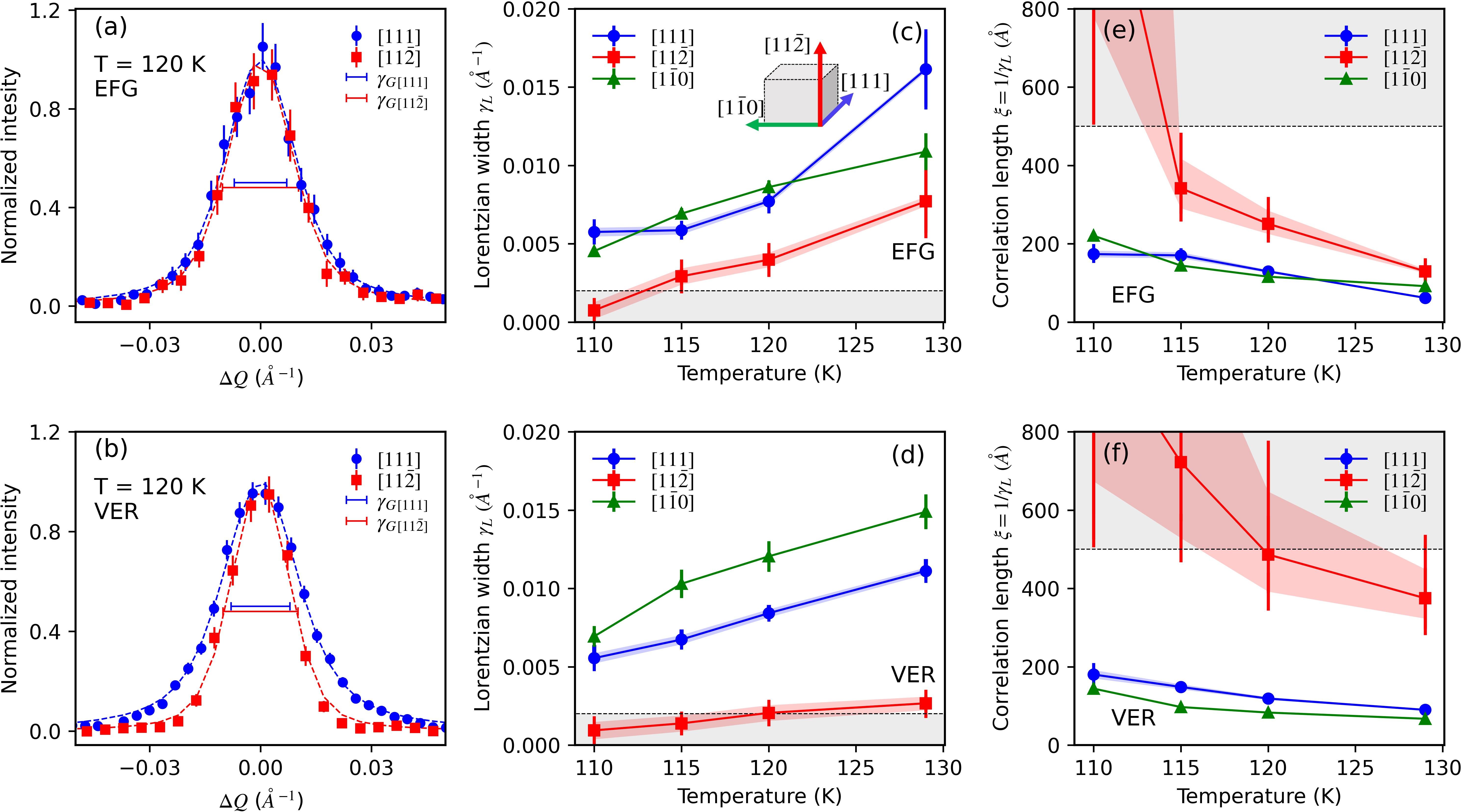}
\caption{\label{fig4}(a,b)
Momentum scans at zero energy transfer through \textbf{\textit{Q$_R$}} $=(3/2,3/2,5/2)$ along two orthogonal directions $[111]$ (blue dots) and $[11\bar{2}]$ (red squares) in the (a) EFG and (b) VER samples, respectively, using high-momentum resolution (see text). The data have been normalized to a peak intensity of one for a better comparability of the peak shape observed in different momentum directions. The normalized data are fitted using a Voigt model (dashed lines) with fixed Gaussian width ($\gamma_G$, horizontal bar). (c,d) Temperature dependent Lorentzian width ($\gamma_L$) deduced from scans along the three orthogonal directions $[111]$ (blue dots), $[11\bar{2}]$ (red squares) and $[1\bar{1}0]$ (green triangles) are presented for the EFG and VER samples, respectively. (e,f)   Calculated correlation length   $\xi=(1/\gamma_L)$  along the three orthogonal directions are presented for the EFG and VER samples, respectively. The shaded areas around the data points in (c-f) show effects of an estimated uncertainty ($\pm5\%$) of the experimental resolution. Error bars represent statistical deviations. Shaded area in grey in (c,d,e,f) above and below the dashed black lines indicate the resolution limitation for the measurements. The data sets for the elastic x-ray scattering are available at the open data repository KITopen \citep{KITopen_STO_2025}.}
\end{figure}

We performed additional scans through \textbf{\textit{Q$_R$}} at selected temperatures, 110 K $\leq T \leq$ 129 K, along two directions, $[11\bar{2}]$ and $[1\bar{1}0]$ orthogonal to $[111]$ to probe a possible anisotropies of the correlation lengths of the AFD structural distortion in the cubic crystal [Figs. \ref{fig4}(a,b)]. The scans were approximated by Voigt functions [dashed lines in Figs. \ref{fig4}(a,b)], where the Gaussian width $\gamma_G$ was fixed to the experimental resolution along the scan directions. Data taken at $T=120$ K [Figs. \ref{fig4}(a,b)] reveal that scans along $[111]$ [blue dots] features a more pronounced broadening (beyond the experimental resolution $\gamma_G$) than those scans performed along $[11\bar{2}]$ [red squares]. This difference is larger in the VER sample [Fig. \ref{fig4}(b)] in comparison to EFG sample [Fig. \ref{fig4}(a)]. The Lorentzian linewidth $\gamma_L$ [HWHM, Figs. \ref{fig4}(c,d)] of the approximated Voigt function describes this broadening beyond the experimental resolution $\gamma_G$ and is related to the inverse of the correlation length $\xi=1/\gamma_L$ represented by the CP. Thus, we find a smaller correlation along the $[111]$ direction than along $[11\bar{2}]$ [Fig. \ref{fig4}(e,f)].

\onecolumngrid
\onecolumngrid

\newpage
\twocolumngrid

 As expected, $\gamma_L$ decreases ($\xi$ increases) on cooling towards $T_c$ in both samples [Figs. \ref{fig4}(c,f)]. Unfortunately our scans along $[11\bar{2}]$ feature only resolution limited peaks in both EFG and VER samples below 115 K and 129 K respectively, \textit{i.e.}, $\xi$ becomes too large ($\geq 500$ \AA) to be determined in our experimental setup. Our analysis shows that correlation lengths along $[111]$ and $[1\bar{1}0]$ have similar values, $100$ \AA $-$ $200$ \AA, for both samples in the investigated temperature range. However, $\xi$ deduced from scans along $[11\bar{2}]$ is significantly larger for the VER sample [Fig. \ref{fig4}(f)] than for the EFG sample [Fig. \ref{fig4}(e)]. It will be the subject of future investigations to find the origin of these anisotropies and to clarify the impact on the CP intensity.

\section{DISCUSSION}

The soft phonon at the AFD transition in SrTiO$_3$ has been previously investigated by inelastic neutron scattering (INS) \citep{Cowley_ssc_1969,Shirane_physrev_1969,Stirling_jpcssp_1972,Fleury_prl_1968} and inelastic x-ray scattering (IXS) \citep{Hong_prb_2008}. Our results for the phonon softening are in good agreement with the previous reports if one takes into account different experimental resolutions and sample properties, \textit{i.e.}, different transition temperatures. Importantly, our measurements in two samples show very similar phonon intensities and, thus, verify that we can quantitatively compare the scattering intensities in our samples. 

An intensely discussed subject is the presence or absence of the narrow component of the CP in neutron and x-ray scattering measurements. Our results [\textit{e.g.} see Figs. 3(a,b)] show a similar momentum width of the CP [see Fig. 3(d)] to that reported by INS \citep{Shirane_prb_1993}, where only the broad component was observed. The observed absolute value for the linewidth, 0.02 \AA$^{-1}$ along $[111]$ direction, is still above our experimental resolution of $\sim 0.015$ \AA$^{-1}$. This is consistent with an x-ray diffraction (XRD) study \citep{Hunnefeld_prb_2002} which reported that the narrow component was present only in measurements of a crystallographically rather perfect sample and not in a less perfect Verneuil-grown sample, the latter having a transition temperature of 105.8 K, close to our value of $T_{c,VER} = 106.3$ K. The presence of the narrow component was associated with the much larger amount of oxygen vacancies in the low$-T_c$ sample – in particular close to the surface of the single crystal. Yet other XRD studies \citep{Holt_prl_2007} report a narrow component in a Verneuil grown sample with a transition temperature of about 104 K but do not provide details about the defect concentrations. Nevertheless, the reported linewidth of the narrow component of the order of $10^{-3}$ \AA$^{-1}$ would have yielded a resolution-limited peak in our measurements. 

The results of the latter XRD study \citep{Holt_prl_2007} were compared to data taken by IXS \citep{Hong_prb_2008} at the HERIX spectrometer previously located at sector 3 at the APS, ANL, in a very similar setup to ours. Whereas results for the phonon softening are compatible, Hong \textit{et al.} report the observation of a narrow component visible in both thermal diffuse scattering \citep{Holt_prl_2007} and IXS \citep{Hong_prb_2008}. The sample was the same specimen grown by the Verneuil method discussed above. We note that the momentum resolution in the data shown in Figure 5 of \citep{Hong_prb_2008} is supposed to be about an order of magnitude smaller than what we could achieve at a very similar instrument at sector 30. Unfortunately, no details are given on how such a superb resolution could be achieved. 

XRD investigations typically assign the broad component to effects of the soft phonon mode whereas the narrow component reflects Bragg scattering from small volumes of the sample \citep{Hong_prb_2008}. This is true in the sense that XRD integrates over the whole phonon spectrum and, thus, observes increased thermal diffuse scattering due to phonon softening at the R-point in SrTiO$_3$ close to $T_c$. However, neutron scattering, similar to our own IXS results, shows a CP clearly separate from the soft phonon mode but having a much larger width in momentum space than the narrow component. Thus, XRD averages over both the soft phonon mode and the CP observed in neutron scattering and IXS. The commonly used stereotype that the momentum resolution of INS is too coarse to distinguish between narrow component and broad component (\textit{e.g.}, see \citep{Hong_prb_2008}) is not true for the case of SrTiO$_3$. For instance, Shirane and co-workers achieved a very good resolution of 0.0045 \AA$^{-1}$ but only observed linewidths for the CP about twice that value \citep{Shirane_prb_1993}.

In contrast to earlier x-ray-scattering experiments \citep{Hunnefeld_prb_2002} which identify the narrow component to arise from the surface of the crystal, the CP intensities we have measured by IXS are a volume response rather than a surface response. Both samples used for IXS had a thickness of 60 microns, and our x-ray measurements were carried out in transmission mode with an incident energy of 23.78 keV ($\sim 0.5213$ \AA). We note that the surfaces of the two samples have been treated identically in that the thickness was reduced to 60 microns by polishing both surfaces, front and back, using a grinding tool with 0.1 $\mu$m grit specification. Laterally, the sample sizes by far exceed the beam size, so that the same volume is illuminated for both samples during IXS measurements. Hence, we do neither expect any variation of the CP intensity due to geometrical effects, a significant surface contribution nor effects related to surface roughness that contribute to the CP intensity of the two samples.

We summarize from the above given discussion that we observe a broad component similar to neutron scattering in our IXS data. The new feature in our study is that the intensity of this CP is sample dependent in the way that the sample with less point defects features a higher CP intensity. This is in contrast to the point of view on the role of defects in SrTiO$_3$ presented in previous reports \citep{Hastings_prl_1978, Cowley_jpsj_2006}. Here, defects are identified as nucleation centers around which the AFD transition occurs at a higher temperature than the overall one observed, \textit{e.g.}, in specific heat measurements.

Different to our study, defects were introduced into samples investigated in Ref. \citep{Hastings_prl_1978} intentionally by hydrogen treatment. Indeed, the defect concentrations of different samples were not measured quantitatively, but the measured charge carrier concentrations (increasing with defect concentration) were taken as a relative scale. The argument that defects increase the CP intensity was taken up in recent publications, \textit{e.g.}, on the incomplete phonon softening and rise of a CP in underdoped YBCO  \citep{Tacon_nphys_2013} or the presence of a CP above the charge-density-wave transition in ZrTe$_3$ \citep{Hoesch_prl_2009}. In our study, the VER sample featuring a more intense CP has less defects. Moreover, the observed low-temperature specific heat (see Fig. S6) shows lower values for the VER sample than the EFG one and, thus, corroborates lower charge carrier and defect concentrations in the VER sample.

Research on the structural phase transition in SrTiO$_3$ encompasses more than 50 years (see reviews \citep{Cowley_jpsj_2006} and \citep{Cowley_IntFerro_2012} for a detailed account) and, originally, two different approaches, one based on dynamic processes, \textit{e.g.}, anharmonic phonon processes \citep{Shapiro_ssc_1972}, and another based on defects \citep{Halperin_prb_1976}, have been put forward to explain the origin of the observed CP. The former necessitates a finite energy linewidth of the CP. This has not been observed until now, and an upper limit of the CP’s energy linewidth is 0.08 $\mu$eV \citep{Darlington_physletta_1975}. On the other hand, various reports and our own work on the sample dependence of the CP properties in SrTiO$_3$ point towards a prominent role of defects. It is evident that defects modify the CP intensity, in line with the pioneering experiments from the 1970s \citep{Topler_jpcssp_1977,Hastings_prl_1978}. The surprising insight from our work is that the nature of defects matters to that extent that different kinds of defects have either a detrimental or a beneficial impact on the CP intensity. This is in contrast to the paradigm that the central peak intensity generally increases with increasing point defect density. Moreover, the direction the CP intensity changes is not correlated to the direction the transition temperature $T_c$ changes. A larger oxygen vacancy concentration decreases $T_c$, in contrast, increasing Nb substitutional doping increases $T_c$. However, both defect types are known to increase the CP intensity. Our work suggests that the CP intensity decreases with increasing Fe concentration while $T_c$ is slightly increased. We are only aware of systematic investigations of the CP in $n$-type doped SrTiO$_3$ that quantitatively relate the CP intensity to the defect concentration \citep{Hastings_prl_1978,Hunnefeld_prb_2002}. Given that our study identifies Fe as the main difference in the impurity spectrum and Fe$^{3+}$ is known to act as $p$-type dopant opens up the possibility that electronic properties, altered by adding defects, play an important role for the CP phenomenon. The role of defects in this scenario could then be the modification of the electronic structure and subsequently electron-phonon coupling properties. Thus, defects acting either as $n$- or $p$-type dopants could have opposite impacts in this scenario. Speculating that electron-phonon interaction becomes significant to change the electronic eigenstates, the cations in the regular cubic crystal structure located at high symmetry positions experience modified potentials and lock into symmetry-breaking positions much alike the frozen impurities as discussed by Halperin and Varma \citep{Halperin_prb_1976}, thus giving rise to diffuse but static scattering. An obvious route for future experiments to verify these speculations is a systematic study of the evolution of the CP intensity with controlled concentrations of charge-neutral defects, such as Ca, and $p$-type dopants, such as Fe, complementing earlier investigations with varied concentrations of $n$-type dopants or oxygen vacancies \citep{Topler_jpcssp_1977,Hastings_prl_1978}.

\section{CONCLUSION}  
In summary, our IXS measurements reveal that the evolution of the soft mode is similar in both EFG and VER samples irrespective of the different defect concentrations. Regarding the CP issue, we do only observe a broad component consistent with inelastic data from neutron scattering \citep{Shirane_prb_1993} and in contrast to previous IXS measurements \citep{Hong_prb_2008}. The observed CP intensities differ by a factor of four in the two samples investigated and are larger in the sample with less point defects. On the other hand, the VER sample featuring the more intense CP shows a larger anisotropy in the CP’s 3D correlation lengths on cooling towards the structural phase transition. It needs more dedicated experiments to answer the question whether this anisotropy can be linked to the CP strength.

\section{ACKNOWLEDGEMENTS}

This research used resources of the Advanced Photon Source, a U.S. Department of Energy (DOE) Office of Science User Facility operated for the DOE Office of Science by Argonne National Laboratory under Contract No. DE-AC02-06CH11357. The SrTiO$_3$ crystal growth activities at the Leibniz-Institut f\"{u}r Kristallz\"{u}chtung (IKZ) using the EFG technique were supported by a project of the Leibniz Association under reference number SAW-2013-IKZ-2. Larmor diffraction measurements have been performed at the TRISP triple axis spectrometer with neutron resonance spin echo located at the Heinz Maier-Leibnitz Zentrum (MLZ), Garching, Germany and operated by Max-Planck-Institut f\"{u}r Festk\"{o}rperforschung. We thank the MLZ for the allocation of beamtime. We thank Ralph Feyerherm for assistance with the thermal conductivity and heat capacity measurements which were performed in the CoreLab Quantum Materials, HZB, Berlin, Germany. We thank Martin Rusu for assistance with the $\mu$XRF measurements which were performed in the X-Ray CoreLab, HZB, Berlin, Germany.

%

\newpage

\setcounter{section}{0}      
\setcounter{subsection}{0}   
\setcounter{equation}{0}     
\setcounter{figure}{0}       
\setcounter{table}{0}        
\setcounter{page}{1}         

\renewcommand{\thesection}{\arabic{section}}              
\renewcommand{\thesubsection}{\thesection.\arabic{subsection}} 
\renewcommand{\thesubsubsection}{\thesubsection.\arabic{subsubsection}} 

\makeatletter
\renewcommand{\p@subsection}{}    
\renewcommand{\p@subsubsection}{} 
\makeatother

\renewcommand{\theequation}{S\arabic{equation}} 
\renewcommand{\thefigure}{S\arabic{figure}}     
\renewcommand{\thetable}{S\arabic{table}}       
\renewcommand{\bibnumfmt}[1]{[S#1]}              
\renewcommand{\citenumfont}[1]{S#1}               

\newcolumntype{C}[1]{>{\centering\arraybackslash}m{#1}}

\onecolumngrid
\begin{center}
{\Large \textbf{Supplementary Information:}}\\[2em]
\large \textbf{Soft phonon and the central peak at the cubic-to-tetragonal phase transition in SrTiO$_3$}\\[0.75em]
\normalsize{Avishek Maity},\textsuperscript{1,$\ast$} \normalsize{Klaus Habicht},\textsuperscript{2,3} \normalsize{Michael Merz},\textsuperscript{4,5} \normalsize{Ayman H. Said},\textsuperscript{6} \normalsize{Christo Guguschev},\textsuperscript{7} \normalsize{Danny\\Kojda},\textsuperscript{2} \normalsize{Britta Ryll},\textsuperscript{2} \normalsize{Jan-Ekkehard Hoffmann},\textsuperscript{2} \normalsize{Andrea Dittmar},\textsuperscript{7} \normalsize{Thomas Keller},\textsuperscript{8,9} and \normalsize{Frank Weber}\textsuperscript{4,$\dagger$}\\[0.5em]
\small\textsuperscript{1}\textit{Heinz Maier-Leibnitz Zentrum (MLZ), Technische Universität München, D-85747 Garching, Germany}\\
\textsuperscript{2}\textit{Department Dynamics and Transport in Quantum Materials,\\Helmholtz-Zentrum Berlin f\"{u}r Materialien und Energie,\\Hahn-Meitner-Platz 1, D-14109 Berlin, Germany}\\
\textsuperscript{3}\textit{Institut f\"{u}r Physik und Astronomie, Universit\"{a}t Potsdam,\\Karl-Liebknecht-Stra{\ss}e 24-25, D-14476 Potsdam, Germany}\\    
\textsuperscript{4}\textit{Institute for Quantum Materials and Technologies,\\Karlsruhe Institute of Technology, Kaiserstr. 12, D-76131 Karlsruhe, Germany}\\
\textsuperscript{5}\textit{Karlsruhe Nano Micro Facility (KNMFi), Karlsruhe Institute of Technology, Kaiserstr. 12, D-76131 Karlsruhe, Germany}\\
\textsuperscript{6}\textit{Advanced Photon Source, Argonne National Laboratory, Lemont, Illinois 60439, USA}\\    
\textsuperscript{7}\textit{Leibniz-Institut für Kristallzüchtung, Max-Born-Straße 2, D-12489 Berlin, Germany}\\
\textsuperscript{8}\textit{Max-Planck-Institut für Festkörperforschung, Heisenbergstraße 1, D-70569 Stuttgart, Germany}\\
\textsuperscript{9}\textit{Max Planck Society Outstation at the Heinz Maier-Leibnitz Zentrum (MLZ), D-85748 Garching, Germany}
\end{center}

\bigskip
\bigskip
\bigskip
\thispagestyle{empty}

\section{Defect characterization}\label{SI_A}

In order to assess the defects present in the single-crystalline \ce{SrTiO_{3}} samples investigated by inelastic x-ray scattering (IXS), we have performed a series of characterization measurements on specimens which were synthesized by the same crystal growth methods, i.e.~by the standard, commercially used Verneuil (VER) method and by the more advanced edge-defined film fed growth (EFG) method \cite{Guguschev2015}. Whenever possible, specimens used to characterize defects were taken from the same batch from which the IXS specimens were taken. 

We use low temperature thermal conductivity data to demonstrate that the EFG crystals have a larger number of defects which give rise to phonon scattering than the VER crystals have. Caloric measurements probe the specific heat anomaly which occurs at the transition temperature $T_{C}$ for the cubic-to-tetragonal structural phase transition, for pure samples at $T_{C}=\SI{105}{\kelvin}$. Substitutional defects and oxygen vacancies are known to alter $T_{C}$, and for a large variety of substituents, the differential rate of change of the transition temperature with substitution $dT_{C}/dx$ has been experimentally determined. Moreover, $dT_{C}/dx$ follows a universal power law dependent on the ionic valence mismatch for isovalent and aliovalent substitution \cite{McCalla2016}. Here, we apply a new method to accurately determine the phase transition temperature from specific heat data; the method is outlined in section \ref{SI_B} of the Supplementary Information (SI) and will be detailed elsewhere. The observed differences in $T_{C}$ for the VER and EFG samples support the hypothesis that the EFG sample has a larger number of point defects, i.e.~substitutional defects or oxygen vacancies. A Sommerfeld analysis of specific heat data in the low temperature limit confirms the absence of charge carriers in the VER samples and indicates the presence of minute amounts of aliovalent impurities giving rise to detectable charge carrier concentrations in the EFG samples. Assuming that these charge carriers exclusively arise from oxygen vacancies and neglecting electron localization allows to conclude that the  concentration of charge-uncompensated oxygen vacancies is negligible in both, VER and EFG samples. Measurements of the electrical conductivity $\sigma$ and Hall effect measurements to characterize charge carrier concentration $n$ and Hall mobility $\mu$ of the nominally undoped EFG sample were performed. While thermal conductivity $\kappa$, specific heat $c_{p}$ and electrical conductivity are all sensitive to defects, a quantitative determination of the defect concentration is impeded when multiple defects are present simultaneously. To address this challenge, we first use micro X-ray fluorescence spectrometry ($\mu$XRF) to identify the chemical elements of the major impurities, and then use inductively coupled plasma optical emission spectrometry (ICP-OES) for elemental analysis on a quantitative level. In addition to rocking scans performed during the course of the IXS measurements, we use Larmor diffraction (LD) with neutrons \cite{Rekveldt2001} to assess the crystal quality by determining the mosaicity of VER and EFG samples, thus probing planar defects. These measurements confirm the superior crystal quality of EFG samples consistent with earlier measurements of the etch-pit density \cite{Guguschev2015}.   

Tab.~\ref{Tab_sample_overview} shows which samples from which charges were used for the IXS experiments and the various defect characterization measurements. Verneuil-grown specimens used for the defect characterization measurements were all purchased from the same commercial supplier, Crystal GmbH, Berlin, Germany. EFG grown samples were grown at Leibniz-Institut für Kristallz\"uchtung (IKZ) \cite{Guguschev2015}. We anticipate that marked differences in the sample quality, i.e.~impurity levels, oxygen vacancy concentrations and crystallite orientations, are strongly dominated by the degree of purity of the starting materials used in the different synthesis routes. Except for the LD experiments on EFG and VER samples, which were performed on large single crystal specimens, and the $\mu$XRF spectroscopy on the VER samples, the specimens used to characterize defects were taken from the same charge from which the IXS specimens were taken. 

\begin{table*}[b]

\centering
\begin{ruledtabular}
\begin{tabular}{C{2cm}|C{2cm}|C{2cm}|C{2cm}|C{2cm}|C{2cm}|C{2cm}|C{2cm}}
IXS       		& $\kappa$ 	& $c_{p}$ 		& $\sigma$ & $n,\mu$ & $\upmu$XRF 	& ICP-OES  & LD          \\ 
\hline
\hline
           		&          		&         & &		&            		&         		   			&     			\\[-2mm]
 \textbf{EFG1} 	         	&   \textbf{EFG1}       	&   \textbf{EFG1}       &   EFG1   &	EFG1   &   EFG1          	&EFG1         	     		& 	  		\\
(\#25\_8)  	&   (\#25\_7)  	&   (\#25\_7) 	& (\#25\_6) & (\#25\_6) &  (\#25\_10)	&(\#25\_8)      		&     			\\
            		&          		&  & &       		&            		&         		     			& EFG2		\\
            		&          		&  & &       		&            		&         		     			& (\#37\_1) 		\\
            		&          		&  & &        		&            		&         		     			&     			\\[-2mm]
\hline
           		&          		&        		&        & &    		&         		     			&     			\\[-2mm]
\textbf{VER1}			&   \textbf{VER1}       	&  \textbf{VER1}       	&    & &        		&VER1       		      		& 			\\
 (\#30452\_27)	&   (\#30452\_26)	&  (\#30452\_21b)&          & &  		& (\#30452\_25)  &      			\\
          		&          		&        		&            & &		&         		 			&     			\\[-3mm]
           		&   VER2    		&         		&        & &    		&VER2  		      		&    			\\
           		&   (\#d2750)    	&         		&        & &    		&(\#d2750)   	     	&   			\\
           		&          		&    VER3   	& & 	&  VER3           	&        		        		&     			\\
           		&          		&    (\#d2229) 	& & &   (\#d2229) 	&        		 		    	&     			\\
           		&          		&         		&            	& &	&        		     			& VER4    		\\
          		&          		&         		&            	& &	&        		    			&  (\#21337\_1)	\\[-4mm]
          		&          		&         		&            	& &	&        		     			&     			
\end{tabular}
\end{ruledtabular}
\caption{Samples used for IXS and defect characterization measurements. The labelling scheme indicates the growth method as EFG or VER. Numbers refer to different batches. In parentheses, the \# symbol is followed by the manufacturer's charge number, and the underscore is followed by a number that identifies an individual part of that charge. \underline{\textit{\textcolor{blue}{Results from the samples printed in \textbf{BOLD font} are discussed in the main text.}}}}
\label{Tab_sample_overview}
\end{table*}

\subsection{Thermal transport probed with a physical properties measurement system (PPMS)}\label{ThermalTransportSection}

The thermal conductivity $\kappa$ of Verneuil-grown and EFG-grown specimens has been measured in the temperature range from 2 K to 380 K by using the thermal transport option (TTO) in a commercial physical property measurement system (PPMS) from Quantum Design, Inc..
Below, we describe the relevant experimental parameters of the thermal conductivity measurements (section \ref{ThermalConductivityExperimentalDetailsSection}) along with a detailed error analysis (section \ref{ThermalConductivityErrorAnalysisSection}). Results are discussed in section \ref{ThermalConductivityResultsSection}. 

\subsubsection{Experimental details of thermal conductivity measurements}\label{ThermalConductivityExperimentalDetailsSection}

\begin{figure}[b]
\centering
\includegraphics[width=0.75\linewidth]{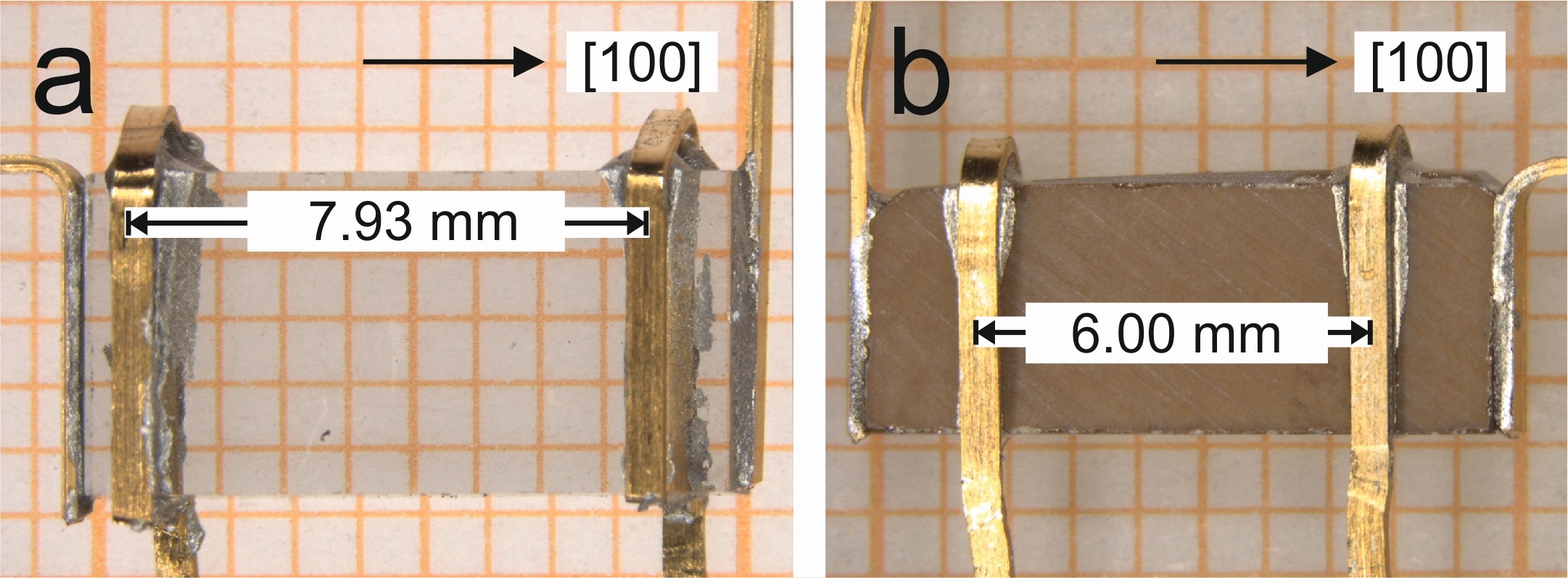}
\caption{Samples VER2 (a) and EFG1 (b) prepared for thermal conductivity measurements along the [100] direction with PPMS-TTO. Gold-plated copper leads are attached with silver-filled epoxy to provide four-probe lead configurations. Lengths indicated in each panel are mid-to-mid distances between contact leads.}
\label{fig:figS01_Photo_VER2_EFG1_Samples_PPMS}
\end{figure}

\begin{figure}
\centering
\includegraphics[width=0.6\linewidth]{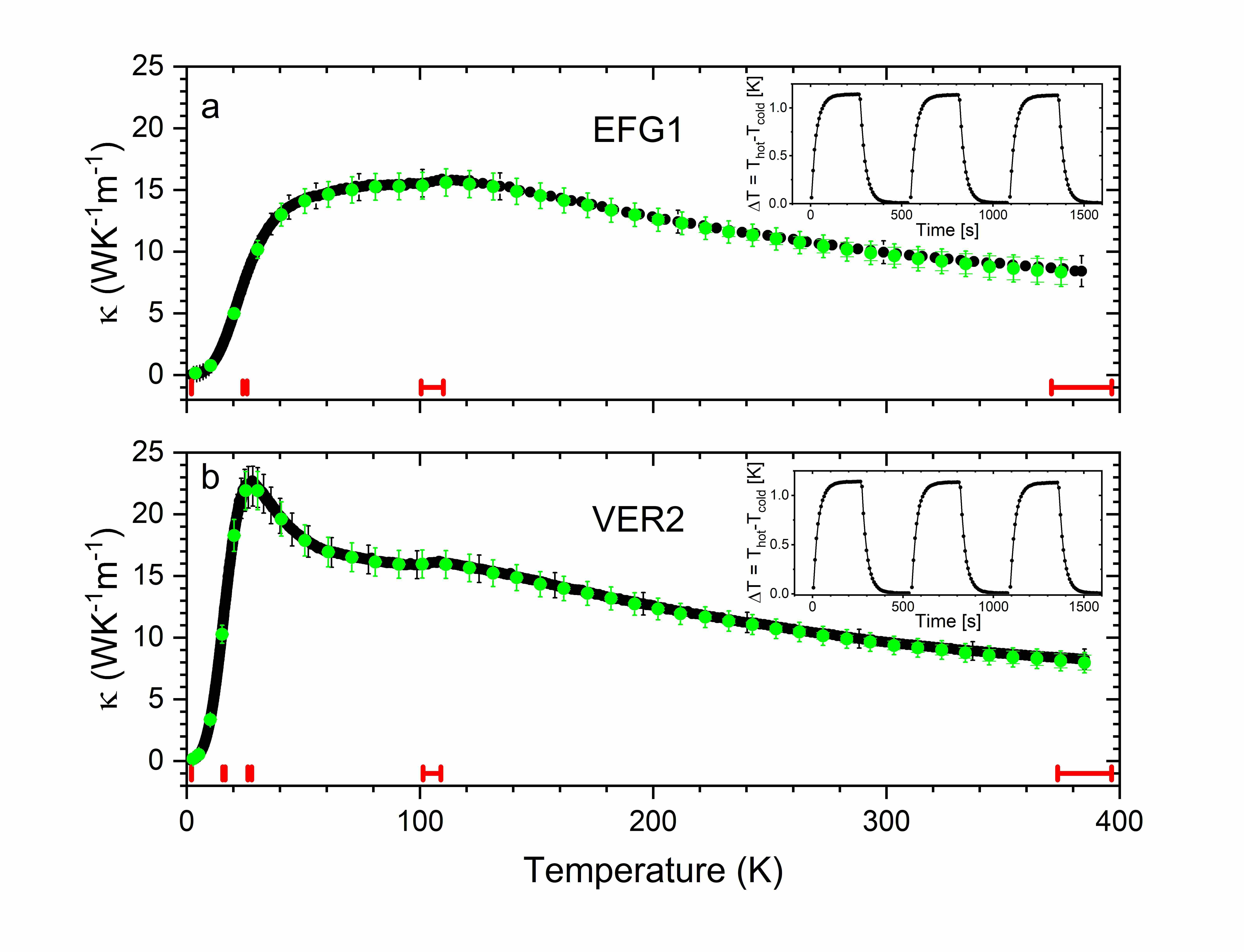}
\caption{Temperature-dependent thermal conductivity $\kappa$ of samples EFG1 (panel a) and VER2 (panel b): Data measured in continuous (transient) operation mode (black lines), and data measured in steady-state (single) operation mode (green circles) show excellent agreement as expected. For clarity, error bars for continuous-mode data are only shown for a reduced number of data points. The horizontal red bars at the bottom of each panel indicate the temperature interval over which individual transients average $\kappa$ for selected data points. Insets in both panels show raw data of the transient difference signal $\Delta T(t)= T_{\text{hot}}(t) - T_{\text{cold}}(t)$ of hot and cold thermometers (data points) and fits of the transients (lines) for three subsequent heat pulses corresponding to $T \simeq 105$ K.}
\label{fig:figS02_PPPMS_kappa_data-EFG1-VER2}
\end{figure}

Measurements of $\kappa$ along the crystallographic [011] (VER1) and [100] (EFG1, VER2) directions with the TTO installed in the PPMS were performed in four-probe lead configuration. Gold-plated copper leads were attached with an electrically conducting, silver-filled epoxy to samples with dimensions  $1.2\times3.8\times9.8$ mm$^3$ (EFG1), $1.9\times2.1\times11.3$ mm$^3$ (VER1) and $2.0\times5.0\times10.0$ mm$^3$ (VER2) (see Fig.~\ref{fig:figS01_Photo_VER2_EFG1_Samples_PPMS}). The bulk of the data was collected in continuous (dynamic) measurement mode, i.e.~continuously decreasing the sample temperature $T_{S}$ from maximum to base temperature while applying a series of heat pulses. Thermal conductivity data of crystal EFG1 were collected at ramp rates  -0.3 K/min  for $380 \ge T \ge 120$ K; -0.2 K/min  for $120 > T \ge  20$ K and -0.1 K/min  for $T < 20$ K. Thermal conductivity data of crystal VER2 were collected at a constant ramp rate -0.1 K/min.  For samples EFG1 and VER2, additional measurements with a reduced number of data points were made in static measurement mode, i.e.~establishing thermal steady-state conditions prior to the determination of temperature differences $\Delta T$ at an adjusted, fixed heater power $P_{h}$. Within error, identical results for $\kappa$ were obtained in the two different operation modes (see Fig.~\ref{fig:figS02_PPPMS_kappa_data-EFG1-VER2}), confirming the reliablity of data collected in continuous operation mode. Thermal conductivity data of crystal VER1 were collected in static measurement mode.

Applying a heat pulse to one of the outer leads generates an increase of the temperature at the cold and hot thermometers connected to the inner leads. In the continuous operation mode, the transient difference signal $\Delta T(t)= T_{\text{hot}}(t) - T_{\text{cold}}(t)$ of hot and cold thermometers is fitted for heating and cooling cycles with the routines provided by the MultiVu\textsuperscript{TM} software provided by Quantum Design, Inc.. Fits of the transients to a phenomenological model 
\begin{equation}
\Delta T_{\text{heating,cooling}}(t)=A_{\text{heating,cooling}}\pm \Delta
T_{\infty }\left( 1-\frac{\tau _{1}e^{-\frac{t}{\tau _{1}}}-\tau _{2}e^{-%
\frac{t}{\tau _{2}}}}{\tau _{1}-\tau _{2}}\right) 
\label{PPMSTransientFunction}
\end{equation}
with two time constants $\tau_{1,2}$ and the constant fit parameter $A_{\text{heating}}$ ($A_{\text{cooling}}=0$) yield the extrapolated asymptotic temperature difference $\Delta T_{\infty }$. The insets to Fig.~\ref{fig:figS02_PPPMS_kappa_data-EFG1-VER2} show exemplary fits to selected transient raw data performed with the MultiVu\textsuperscript{TM} software. The extrapolated asymptotic temperature difference $\Delta T_{\infty }$ enters the thermal conductance which is calculated from the heater power $P_{h}=I^{2}R$ (for current $I$ applied to a resistor with resistance $R$) corrected for radiative and conductive heat losses:
\begin{equation}
K=\frac{P_{h}-P_{\text{rad}}}{\Delta T_{\infty }}-K_{l}.
\label{PPMSConductance}
\end{equation}
Radiative heat losses $P_{\text{rad}}$ are calculated from Stefan-Boltzmann's law assuming that only half of the sample surface with surface area $A_{S}$ radiates: 
\begin{equation}
P_{\text{rad}}=\sigma \varepsilon \frac{A_{S}}{2}\left( T_{\text{hot}%
}^{4}-T_{\text{cold}}^{4}\right),
\label{PPMSRadLoss}
\end{equation}
where $\varepsilon$ is the emissivity of the sample, and $\sigma$ is the Stefan-Boltzmann constant. The thermal conductance $K_{l}$ in Eq.~(\ref{PPMSConductance}) takes into account conductive heat losses from the leads, approximated by a third-order polynomial 
\begin{equation}
K_{l}=aT+bT^{2}+cT^{3}  \label{PPMSLeadLoss}
\end{equation}
with fixed constants $a,b,c$ obtained from instrument calibration performed prior to our sample measurements. Finally, the thermal conductivity is calculated from 
\begin{equation}
\kappa =K\frac{A_{X,S}}{l_{S}}  \label{PPMthermalConductivity}
\end{equation}
with $A_{X,S}$ being the cross-sectional area of the sample perpendicular to the direction of heat transport and sample length $l_{S}$. 

The applied heater power (varied between $1.5$ $\mu $W $ \leq P_{h}^{\text{EFG1}} \leq 52$ mW and  $3.6$ $\mu $W $ \leq P_{h}^{\text{VER2}} \leq 72$ mW) typically resulted in an extrapolated asymptotic temperature difference $\Delta T_{\infty}$ across the heated sample which was less than 2.5\% of the sample temperature $T_{S}$  for  $T_{S}\leq25$ K and less than 1.5\% for  $T_{S}\geq25$ K. This is a reasonable compromise between maximizing signal level and minimizing the total temperature interval $\Delta T = T^{\text{max}}_{\text{hot}} - T^{\text{min}}_{\text{cold}}$ covered by each transient, thus reducing the interval over which the sample temperature and $\kappa$ are averaged and simultaneously minimizing thermal heat losses through the thermal contact leads. For measurements on sample VER1 the applied heater power varied between $240$ $\mu $W $ \leq P_{h}^{\text{EFG1}} \leq 34$ mW, corresponding to  less than 1.5\% for  $T_{S}\geq10$ K and up to 45\% of the sample temperature $T_{S}$  for  $T_{S}\leq10$ K. Radiation losses, which are a significant source of error ($>$2\% above 140 K), have been corrected for by adjusting the infrared emissivity parameter $\varepsilon$ such that the corrected data shows best agreement with thermal conductivity data obtained from laser flash analysis and differential scanning calorimetry (LFA/DSC) and thermal conductivity data obtained by the 3$\omega$ method (not shown). This is justified since the LFA/DSC technique is less sensitive to systematic errors introduced by radiative heat losses. Best agreement in the temperature range where PPMS-TTO and LFA/DSC data overlap (175 K - 385 K) is obtained with emissivity parameters $\varepsilon_{\text{VER1,2}} = \varepsilon_{\text{EFG1}} = 0.45$. 

\subsubsection{Error analysis of thermal conductivity data}\label{ThermalConductivityErrorAnalysisSection}

In contrast to the standard procedure employed by the PPMS analysis routines, we calculate the errors from

\begin{equation}
\sigma _{\kappa }=\sqrt{\left( \frac{R_{\Delta T}}{\Delta T_{\infty }}%
\right) ^{2}+\left( \frac{2IR\Delta I}{P_{h}}\right) ^{2}+\left( \frac{%
0.1\, \Delta T_{\infty }K_{l}}{P_{h}}\right) ^{2}+\left( \frac{0.1 \, P_{%
\text{rad}}}{P_{h}}\right) ^{2}+\left( \frac{\Delta l_{S}}{l_{S}}\right) ^{2} + \left( \frac{\Delta A_{X,S}}{A_{X,S}}  \right) 
^{2}}\kappa 
\label{eq:PPMSStdDeviation}
\end{equation}

where $R_{\Delta T}$ is the residual from the fit of the data to Eq.~(\ref{PPMSTransientFunction}), $\Delta I$ is the error in the current applied to the heater. The error due to conductive heat losses (third term in the sum of the squared terms Eq.~(\ref{eq:PPMSStdDeviation})) is estimated to be 10\%. In view of the fact that we have an excellent match with both, the LFA/DSC and 3$\omega$ data we have lowered the radiative loss contribution to the total error from 20\% as assumed by the PPMS software to only 10\% in our case. Also, we explicitly include the error from geometrical uncertainties which is estimated to be 5\% for both, the mid-to-mid distance of the inner leads $l_{S}$ and the cross-sectional area $A_{X,S}$ perpendicular to the direction of heat transport. 

In fact, the dominant error in the PPMS-TTO data for $\kappa$ arises from variations in the distance between the thermal contact regions $l_{\text{EFG1}}=6.00$ mm ($l_{\text{VER1}}=8.22$ mm, $l_{\text{VER2}}=7.93$ mm) and from variations in the cross-sectional area $A_{\text{X,EFG1}}=1.21\times3.83$ mm$^2$ ($A_{\text{X,VER1}}=1.88\times2.10$ mm$^2$, $A_{\text{X,VER2}}=2.0\times5.0$ mm$^2$). The error bars shown in Fig. \ref{fig:figS02_PPPMS_kappa_data-EFG1-VER2} have been calculated according to Eq.~(\ref{eq:PPMSStdDeviation}). As a result, the total relative error of the thermal conductivity varies between 15 \% (10 \%) at $T=385$ K and 7\% ( 7\%) at $T=2$ K in our PPMS-TTO measurements of samples EFG1 and VER2, respectively.   

The total temperature interval $\Delta T = T^{max}_{hot} - T^{min}_{cold}$ covered by each transient depends on the mean sample temperature $T_{S}$. For the bulk of the $\kappa$ data the relative temperature interval is $\Delta T /T_{S}$ is below 3.5\%. At $T_{S}\sim 100$ K the relative temperature interval $\Delta T /T_{S} \simeq 3.8\% \,(4.8\%)$ for measurements of VER1,2 (EFG1) samples. The absolute temperature intervals covered by the transients at the phase transition temperature $T_{C}=105$ K are $\Delta T=3.8$ K ($4.8$ K), indicated as horizontal red error bars in Fig.~\ref{fig:figS02_PPPMS_kappa_data-EFG1-VER2}. As a consequence, any sharp features in the thermal conductivity data will be broadened by this instrumental resolution. The results for $\kappa$ obtained from the analysis of transient data measured in continuous and static modes are shown in Fig.~\ref{fig:figS02_PPPMS_kappa_data-EFG1-VER2} along with typical errors.  

\subsubsection{Results and discussion of thermal conductivity }\label{ThermalConductivityResultsSection}

\begin{figure}
\centering
\includegraphics[width=0.6\linewidth]{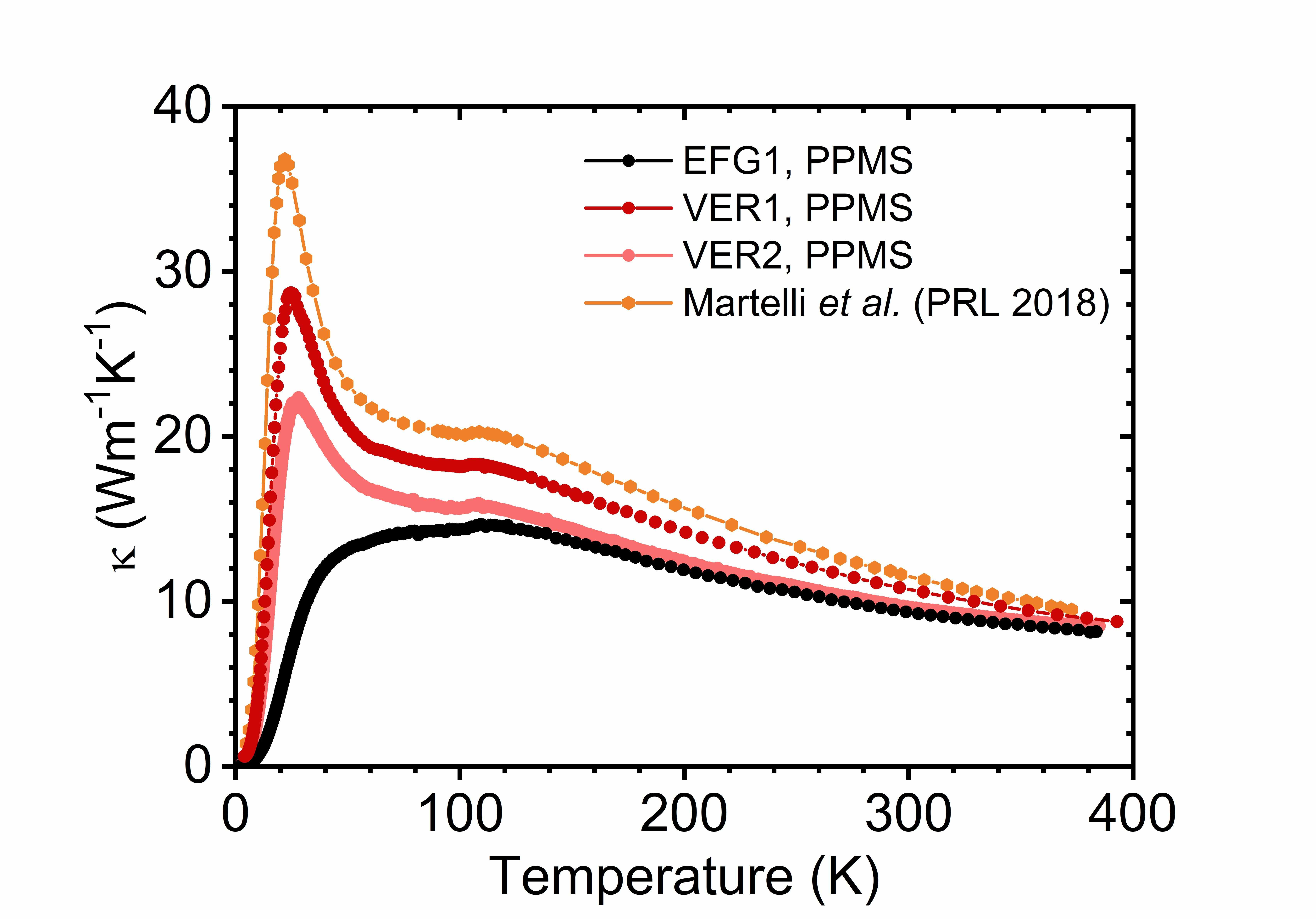}
\caption{Thermal conductivity data obtained from PPMS measurements of Verneuil grown samples from two different batches (VER1, VER2) and of an edge-defined film fed grown sample EFG1 compared to published data from undoped \ce{SrTiO_{3}} \cite{Martelli2018}. }
\label{fig:fig_S03_Thermal_Conductivity}
\end{figure}

Fig.~\ref{fig:fig_S03_Thermal_Conductivity} shows thermal conductivity data for samples EFG1 and VER1,2 compared to published data from nominally undoped, high purity \ce{SrTiO_{3}} \cite{Martelli2018}. There are marked differences in $\kappa$, most prominent in the temperature range below \SI{50}{\kelvin} where $\kappa$ shows a peak for VER1 and VER2 samples as expected for undoped \ce{SrTiO_{3}}. Notably, the peak at $T\simeq\SI{25}{\kelvin}$ is largely suppressed for the EFG1 sample, indicating the presence of a significant number of defects leading to the reduction of $\kappa$ via increased phonon-defect scattering. We also observe a significant batch-to-batch variation of the thermal conductivity in nominally undoped \ce{SrTiO_{3}} samples VER1 and VER2. 

Martelli \emph{et al.}~have performed a systematic study of $\kappa$ of \ce{SrTi_{1-x}Nb_{x}O_{3}} for increasing Nb doping \cite{Martelli2018}. Their study shows that Nb doping on the \SI{100}{atppm} level already suppresses the peak significantly while 1at\% Nb doping is sufficient for a complete suppression of the peak. The suppression of the low temperature peak in $\kappa$ in our data suggests that the defect concentration in sample EFG1 is significantly larger than the defect concentration in sample VER1. No attempt is made to estimate the defect concentration from the thermal conductivity data on a quantitative level. Even assuming that only substitutional point defects are present, the energy-dependent phonon scattering rates require knowledge of the elemental type of the defects in order to obtain quantitative information about the defect concentration.       

\subsection{Specific heat}

The specific heat at constant pressure $c_{p}$ of specimens EFG1, VER1 and VER3 was measured in the temperature range from 2 K to 200 K employing short-pulse relaxation calorimetry using the heat capacity option (HCO) in a commercial physical property measurement system (PPMS) from Quantum Design, Inc.. 

\subsubsection{Experimental details of short-pulse relaxation calorimetry}\label{SubSectionShortPulseRelaxationCalorimetry}
Specimens with masses  $m_{\text{EFG1}}=5.62(1)$ mg, $m_{\text{VER1}}=45.1(5)$ mg and $m_{\text{VER3}}=5.99(1)$ mg, experimentally determined  with a commercial analytical balance (model XS205DU by Mettler-Toledo GmbH), were used to determine their heat capacities $C_{x}$ ($x=$ EFG1, VER1, VER3). This choice of sample mass resulted in sizeable contributions of the sample heat capacity $C_{x}$ to the total heat capacity $C_{tot}=C_{x} + C_{a}$. The total heat capacity $C_{tot}$ includes the heat capacity $C_{a}$ of the calorimeter chip with the thermal contact grease addenda (also referred to as the platform). Following standard procedures, $C_{a}$ was determined in a first run without sample, here during cooling, while $C_{tot}$ was determined in a second run with the sample attached to the calorimeter chip, there during heating. Since Apiezon N has been used as addendum, we have limited the maximum temperature to $\sim$200 K. After having reached a stable (1\% settling accuracy) platform temperature $T_{p}(t=0)=T_{0}$ in thermal equilibrium with the bath temperature $T_{0}$, a constant heater power $P_{h}$ is applied for a time $\Delta t$. This leads to an increase of the platform temperature $T_{p}(t)$. For each set temperature $T_{p}(t=0)=T_{0}$, the heater power and $\Delta t$ were chosen to result in a temperature increase $\simeq 1$ \% of the set temperature. We assume that the sample and the chip are in good thermal contact with each other. Hence the sample temperature $T_{s}$ and the platform temperature  $T_{p}$  are assumed to be identical during the measurement. For $\Delta t \rightarrow \infty $, the platform temperature  $T_{p}$ would asymptotically approach the temperature difference  $\Delta T_{\infty} = P_{h} / K_{1}$, where $K_{1}$ is the thermal conductance between the chip and the thermal bath. After the heater is switched off, the chip temperature decreases exponentially towards $T_{0}$. Neglecting the finite thermal conductance between sample and platform $K_{2}$, i.e.~assuming infinite $K_{2}$, the time evolution of the sample temperature during the heating cycle $T_{s,h}$ is governed by the differential equation \cite{Lashley2003}: 
\begin{equation}
\left( C_{x}+C_{a}\right) \frac{dT_{s,h}}{dt}=P+K_{1}\left(T_{s,h}-T_{0}\right),
\label{eq:LashleyCpEquation}
\end{equation}
which has the solution 
\begin{equation}
T_{s,h}(t)=\frac{P}{K_{1}}\left( 1-e^{-\frac{t}{\tau }}\right) +T_{0}.
\label{eq:LashleyCpDGLSolutionHeatingEquation}
\end{equation}
During the cooling cycle the time dependence of the sample temperature $T_{s,c}$ is given by:
\begin{equation}
T_{s,c}(t)=\Delta Te^{-\frac{t}{\tau }}+T_{0},
\label{eq:LashleyCpDGLSolutionCoolingEquation}
\end{equation}
where $\Delta T=T_{s,h}(\Delta t)-T_{0}$ is given by the difference between the maximum sample temperature $T_{s,h}(\Delta t)$ reached upon heating the sample for a time $\Delta t$ and the bath temperature $T_{0}$. For our measurements we have chosen $\Delta t$ to be equal to one time constant $\tau$ (determined from the preceding heater-on-heater-off cycle). For each sample temperature setpoint $T_{0}$ and its corresponding heater-on-heater-off cycle, the time constant $\tau$ and the thermal conductance $K_{1}$ are obtained from a fit of the data to a piecewise defined function modeling the time evolution of the sample temperature according to Eqns.~(\ref{eq:LashleyCpDGLSolutionHeatingEquation}) and (\ref{eq:LashleyCpDGLSolutionCoolingEquation}). The total heat capacity $C_{tot}$ is then obtained from $C_{tot}=\tau_{x} K_{1}$. Similarly, the heat capacity of the platform is determined from fits applied to the data of the addendum measurement $C_{a}=\tau_{a} K_{1}$. 
Finally, the specific heat of the samples is calculated from 
\begin{equation}
c_{x}=\frac{C_{tot}-C_{a}}{m_{x}}.
\label{eq:SpecificHeatHCOEquation}
\end{equation}
Especially at low temperatures below 10 K poor heat capacity ratios  $C_{x} / C_{a}$ can introduce considerable systematic errors \cite{Lashley2003}. For our measurements the mean $C_{x} / C_{a}$ ratio averaged over the temperature interval from 2.0 K to 10 K was 0.21(2) (0.13(2)) for the EFG1 sample (VER3) sample. Following Lashley {\textit{et al.}}~\cite{Lashley2003} the corresponding systematic error is $\lesssim5$ \% ($\lesssim10$ \%).

In order to remove ambiguities arising from a small, but significant (0.65(1) \% error at $T \simeq 100$ K) dependence of the $C_{x}$ values as extracted by the PPMS software on differently chosen ramp rates $dT_{0} / dt$ for the set sample temperature $T_{0}$ ($dT_{0} / dt$ ranging between $0.1$ K$\,$min$^{-1}$ and $0.5$ K$\,$min$^{-1}$), we have post-processed the raw data using our own fitting routines to determine the time constants $\tau_{x,a}$ and the thermal conductance $K_{1}$. 

\begin{figure}
\centering
\includegraphics[width=0.75\linewidth]{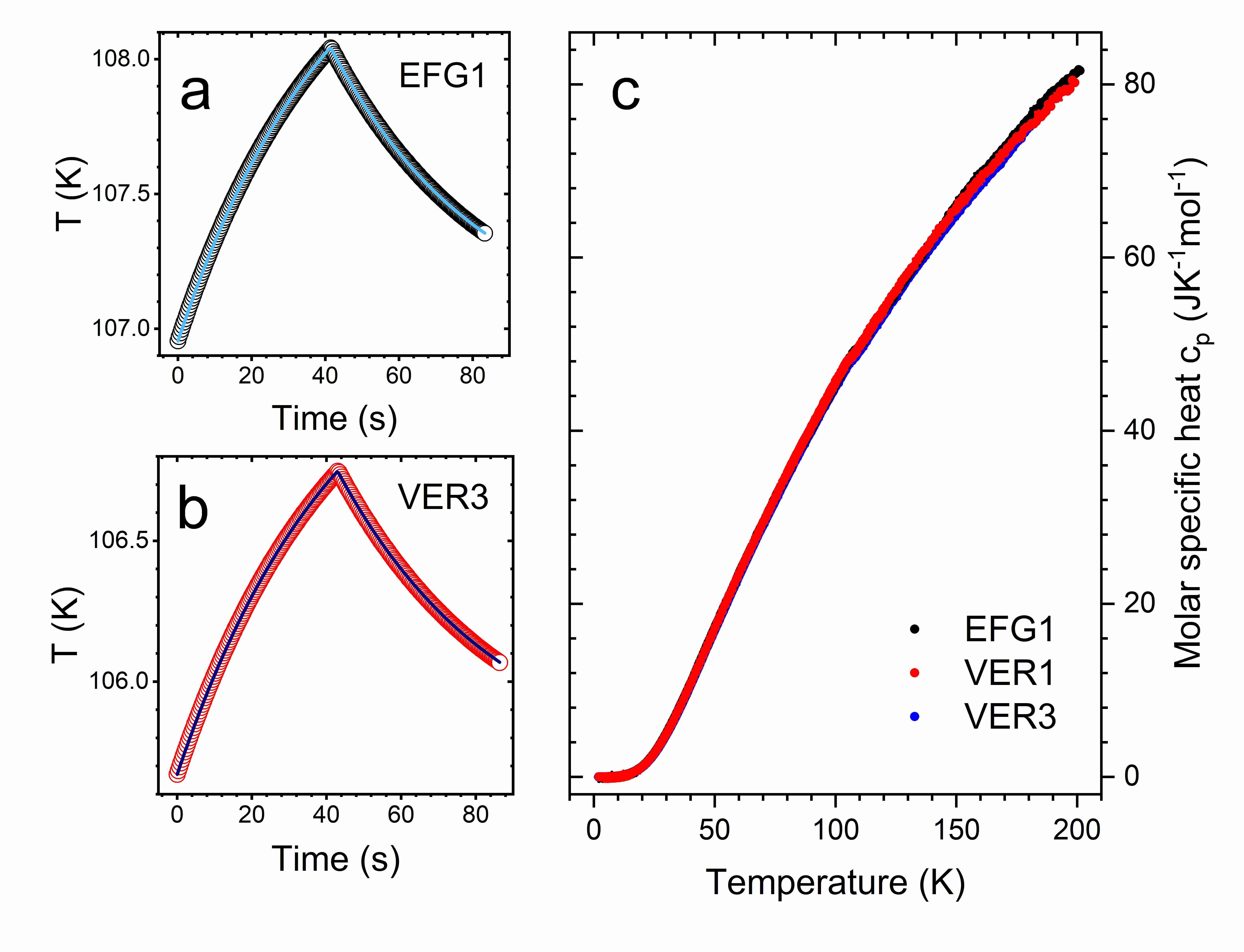}
\caption{Panels (a) and (b): Increase and relaxation of the PPMS sample platform temperature upon applying a heater on - heater off sequence. Panel (a) shows data (black circles) for sample EFG1 close to $T_{C} = 107.4 $ K, i.e. near the critical temperature of the structural phase transition, along with the corresponding fit (light blue line).  Panel (b) shows corresponding data (red circles) for the VER3 crystal close to $T_{C}= 106.1$ K along with the corresponding fit (black line). Note that the slope of the transient data in both cases changes smoothly - except upon switching the heater off - as expected for a second-order structural-phase transition. Panel (c): Specific heat $c_{p}$ for EFG1 (black symbols) and VER1,3 (red and blue symbols) crystals. Error bars are smaller than the symbol size.}
\label{fig:figS04_PPMS_HCO_Specific_Heat_EFG_VER}
\end{figure}

These fit routines are based on a Matlab\textsuperscript{\textregistered} version of fminuit. For these fits we have assumed a relative error of $2 \times 10^{-5}$ for the sample temperature measured during the heater-on-heater-off cycle. Panels a,b of Fig.~\ref{fig:figS04_PPMS_HCO_Specific_Heat_EFG_VER} show exemplary PPMS-HCO raw data for the temperature response upon applying a heater-on-heater-off sequence along with fits to Eqns. (\ref{eq:LashleyCpDGLSolutionHeatingEquation}) and (\ref{eq:LashleyCpDGLSolutionCoolingEquation}). The excellent agreement between data and fits demonstrates that $\tau$ and $K_{1}$ are reliably extracted. On purpose, we have chosen to show the transient signals close to $T_{C}\simeq 105$ K, i.e.~near the critical temperature of the second-order structural phase transition. In the absence of latent heat, a discontinuity in the specific heat $c_{p}$ is expected. Note that the slope of the transient data changes smoothly when crossing $T_{C}$ in both, heating and cooling segments, justifying our analysis also for temperatures close to the phase transition temperature. 

To be able to apply Eq.~(\ref{eq:SpecificHeatHCOEquation}) we have used a linear interpolation of the $C_{a}$ data since the addendum measurement has a lower sampling point density in $T$, which is justified by the smooth temperature dependence of $C_{a}$. We note that we have determined the temperature $T$ which is assigned to a particular $c_{p}$ value by averaging over the total temperature interval covered by each transient signal. The corresponding standard deviation is 0.25 \% of the sample temperature, this error being neglected in the further data analysis. The error of the specific heat is calculated from Gaussian error propagation including the errors $\Delta \tau_{x,a}$ and $\Delta K_{1}$ from the fit and the error of the sample mass $\Delta m_{x}$. The heater power was assumed to have negligible error.  
This fitting procedure successfully removed any significant dependence of $C_{x}$  on the temperature ramp rate $dT_{0}/dt$. 

The specific heat data of samples EFG1 and VER1,3 obtained in this way are shown in Fig.~\ref{fig:figS04_PPMS_HCO_Specific_Heat_EFG_VER} panel (c). For convenience we show the \emph{molar} heat capacity. The gross temperature dependence of the specific heat for both crystals appears to be identical. On this scale, the specific heat $c_{p}$ for EFG1 and VER1,3 crystals shows nearly identical temperature dependence

\subsubsection{Standard analysis of specific heat data}\label{SectionStandardAnalysisOfSpecificHeatData}

We first apply a standard analysis to the specific heat data for specimens EFG1 and VER1,3. We focus on the low-temperature PPMS-HCO $c_{p}$ data by (1) briefly discussing the difference $\Delta c_{p}=c_{p,\text{EFG1}}-c_{p,\text{VER1,3}}$, (2) by analyzing normalized $c_{p}T^{-3}$ data as a function of $T$, and (3) by analyzing normalized $c_{p}T^{-1}$ data as a function of $T^{2}$. The analysis of the specific-heat anomaly near the structural phase transition at $T \simeq 105$ K is analyzed by a model described in section \ref{SI_B}. 

\subsubsection{Sample-related differences}\label{SubSectionSampleRelatedDifferencesInSpecificHeat}

\begin{figure}
\centering
\includegraphics[width=0.75\linewidth]{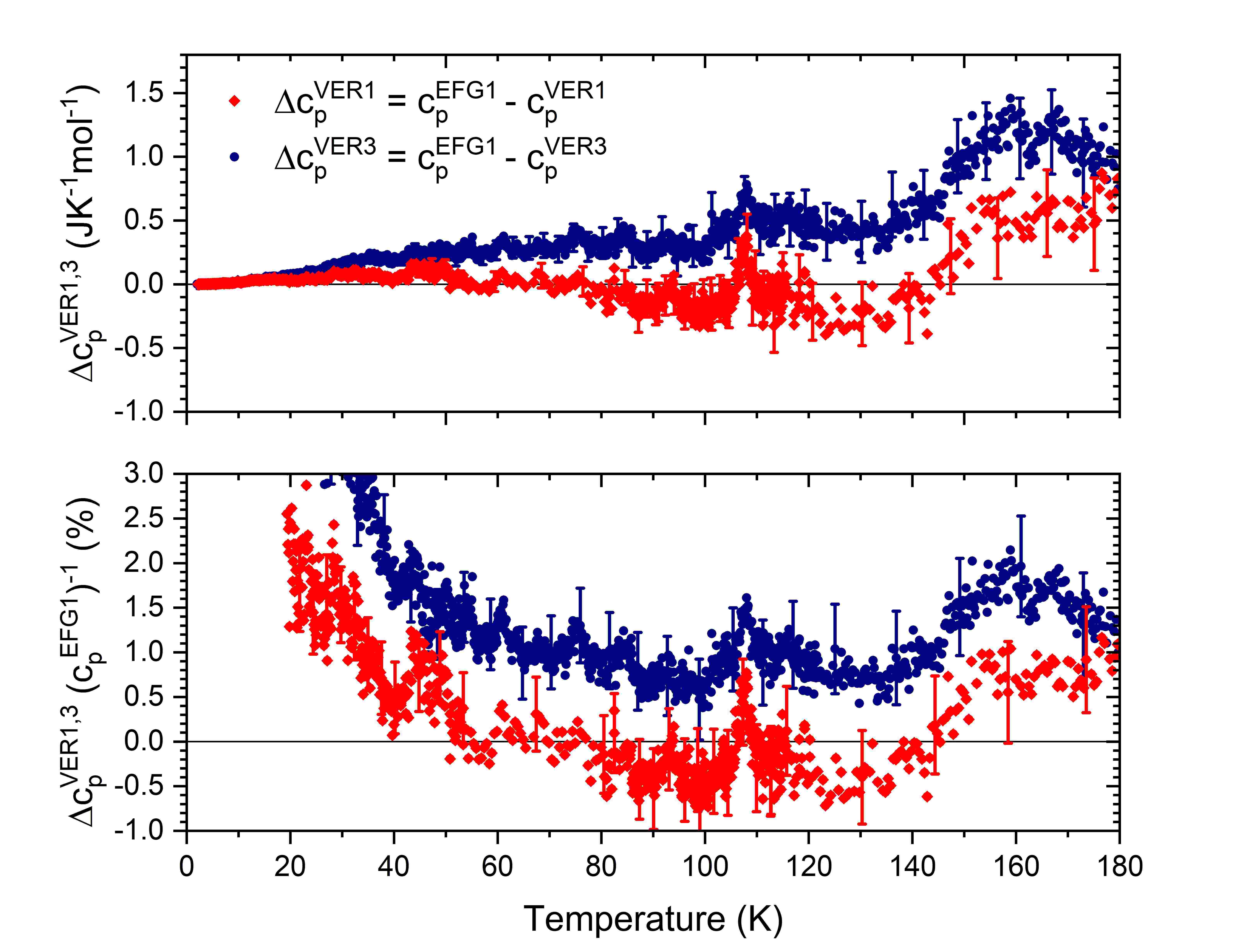}
\caption{Top: Temperature dependence of the absolute difference in specific heat $\Delta c_{p}=c_{p}^{\text{EFG1}}-c_{p}^{\text{VER1,3}}$. Bottom: Relative difference $\Delta c_{p}^{\text{VER1,3}}/c_{p}^{\text{EFG1}}$ as a function of temperature. For clarity, error bars are only shown for a reduced number of data points. }
\label{fig:figS05_PPMS_HCO_Difference_Specific_Heat_Verneuil_EFG}
\end{figure}

Fig.~\ref{fig:figS05_PPMS_HCO_Difference_Specific_Heat_Verneuil_EFG} shows the difference in specific heat $\Delta c_{p}^{\text{VER1,3}}=c_{p}^{\text{EFG1}}-c_{p}^{\text{VER1,3}}$ for the two different samples. The absolute difference $\Delta c_{p}^{\text{VER1,3}}$ (top panel of Fig.~\ref{fig:figS05_PPMS_HCO_Difference_Specific_Heat_Verneuil_EFG}) remains small over the whole temperature range investigated. In fact, up to $\sim100$ K the absolute value of the difference is less than 0.4 J$\,$K$^{-1}\,$mol$^{-1}$. Two regions with excess specific heat can be identified in the data, a peak close to $T \sim 105$ K and a broader range with larger $\Delta c_{p}$ at around $\sim 160$ K. The peak can be attributed to a slightly larger transition temperature $T_{C}$ for the EFG1 sample, accompanied by a shift of excess specific heat related to the structural phase transition. The origin of the region with increased $\Delta c_{p}$, which corresponds to an energy scale of $\sim 14$ meV, is presently unknown, excess specific heat due to oxygen-octahedron rotation induced by oxygen vacancies V$_{\text{O}}$ being a candidate. In support of this hypothesis we note that such antiferrodistortive-like oxygen-octahedron rotations in the paraelectric cubic phase (!) have been theoretically predicted by density functional studies \cite{Choi2013}. These studies conclude that combinations of different oxygen vacancies V$_{\text{O}}$ with different symmetries have binding energies below 100 meV. Given the error in $\Delta c_{p}^{\text{VER1,3}}$, which suffers from a significant contribution from the mass error $\Delta m$, and thus essentially renders the baseline for the difference specific heat unreliable, attempts to quantitatively analyze the difference data further have not been made. 

The two regions with excess specific heat for EFG1 are also visible in the relative difference $\Delta c_{p}^{\text{VER1,3}}/c_{p}^{\text{EFG1}}$ shown in the bottom panel of Fig.~\ref{fig:figS05_PPMS_HCO_Difference_Specific_Heat_Verneuil_EFG}. Above 40 K, the specific heat of the EFG1 crystal $c_{p}^{\text{EFG1}}$ exceeds the specific heat of the Verneuil-grown crystal by no more than 2 \%. In addition to the excess specific heat at higher temperatures, this representation of the data clearly shows that below 40 K $c_{p}^{\text{EFG1}}$ becomes significantly larger than $c_{p}^{\text{VER1,3}}$. At these temperatures the electronic contribution is a significant, at lowest temperature expected to be the dominant, contribution to the total specific heat indicating a slightly different charge carrier concentration in sample EFG1. On the other hand, there is a sizeable contribution of the specific heat arising from the R-corner soft phonons (see section \ref{SoftModePhononContribution}).

\subsubsection{Debye temperature}\label{SubSectionDebyeTemperature}

\begin{figure}
\centering
\includegraphics[width=0.75\linewidth]{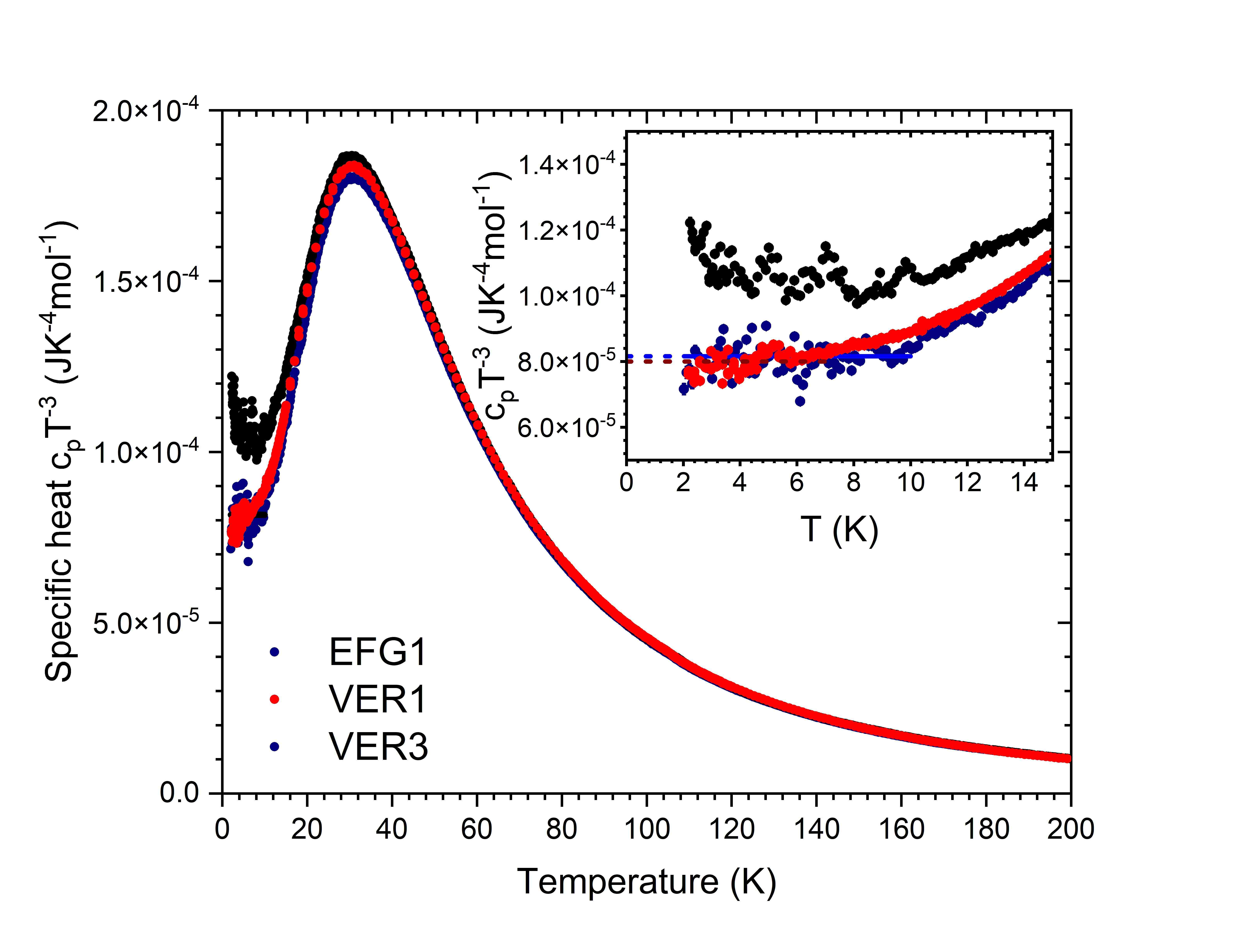}
\caption{Temperature dependence of the specific heat $c_{p}$ normalized by $T^{3}$ for the EFG1 (black circles) and VER1,3 crystals (red and blue circles). The data show excess specific heat peaking at $\sim30$ K indicating phonon contributions beyond Debye-like behavior. The inset shows a close-up of the low-temperature specific heat extrapolated towards $T=0$ for EFG1 (black circles) and VER1,3 crystals (red and blue circles). The lowest values are obtained for the Verneuil-grown samples which allows to extract the characteristic temperature by a Debye-model description of the data.}
\label{fig:figS06_PPMS_HCO_Specific_Heat_by_Tcubed_Verneuil_EFG}
\end{figure}

Fig.~\ref{fig:figS06_PPMS_HCO_Specific_Heat_by_Tcubed_Verneuil_EFG} shows the specific heat $c_{p}$ divided by $T^{3}$. In a standard Debye model and in the absence of any electronic contribution to the specific heat we expect $ c_{p}/T^{3}$ to be constant. Although the data is dominated by a prominent Einstein-like contribution peaking at $T\simeq 30$ K, the data approaches a constant value below 10 K. Explicitly, the Debye model gives in the low-temperature limit
\begin{equation}
\frac{c_{p}}{T^{3}}=\frac{12}{5}\pi ^{4}rN_{A}k_{B}\frac{1}{\Theta_{D}^{3}},
\label{eq:DebyeLowTLimit}
\end{equation}
where $c_{p}$ is the molar heat capacity, $\Theta_{D}$ is the Debye temperature,  $r=5$ is the number of atoms in the cubic unit cell, and $N_{A}$ and $k_{B}$ are Avogadro's number and the Boltzmann constant, respectively. From a fit  to the $c_{p}^{\text{VER1}}/T^{3}$ data below 7.0 K assuming 10 \% error (error bars not shown in Fig.~\ref{fig:figS06_PPMS_HCO_Specific_Heat_by_Tcubed_Verneuil_EFG}), we extract $\alpha^{\text{VER1}} \equiv \lim_{T\longrightarrow 0}\left( c_{p}^{\text{VER1}}/T^{3}\right) = 8.00(3) \times 10^{-5}$ J$\,$K$^{-4}\,$mol$^{-1}$ and a Debye temperature  $\Theta_{D}= 495 \pm 3$ K. We note that other systematic errors introduced by a low ratio $C_{x} / C_{a}$ for the heat capacities of sample $C_{x}$ and addendum $C_{a}$ ratio are not included in our analysis. With this kind of analysis we do not attempt to extract a $\Theta_{D}$ value for the EFG1 crystal, since a possible electronic contribution to $c_{p}$ would lead to an incorrect value for $\Theta_{D}$.   

Our value for $\Theta_{D}$ for the VER1 crystal is in reasonable agreement with the Debye temperature obtained by Ahrens {\textit{et al.}}~\cite{Ahrens2007} who obtain $\Theta_{D}= 513 \pm 3$ K and recent work by McCalla \textit{et al.}~\cite{McCalla2019} who obtain $\Theta_{D}= 515\pm20 $ K. We note, however, that the experimental values of the Debye temperature $\Theta_{D}$ for SrTiO\textsubscript{3} available in literature are substantially scattered, spanning a range from 304 K to 515 K for samples with various dopants and different doping or oxygen vacancy concentrations. Even if one constrains the selection to undoped, single-crystalline SrTiO\textsubscript{3} samples, the $\Theta_{D}$-values still vary between 324 K and 515 K (see Supplemental Material to reference \cite{McCalla2019}). Other interesting approaches have been pursued introducing a temperature-dependent Debye temperature $\Theta_{D}(T)$ in order to reconcile Debye fits to $c_{p}$ data at low and high temperature \cite{Duran2008}. 

\subsubsection{Sommerfeld coefficient}\label{SubSectionSommerfeldCoefficient}

\begin{figure}
\centering
\includegraphics[width=0.75\linewidth]{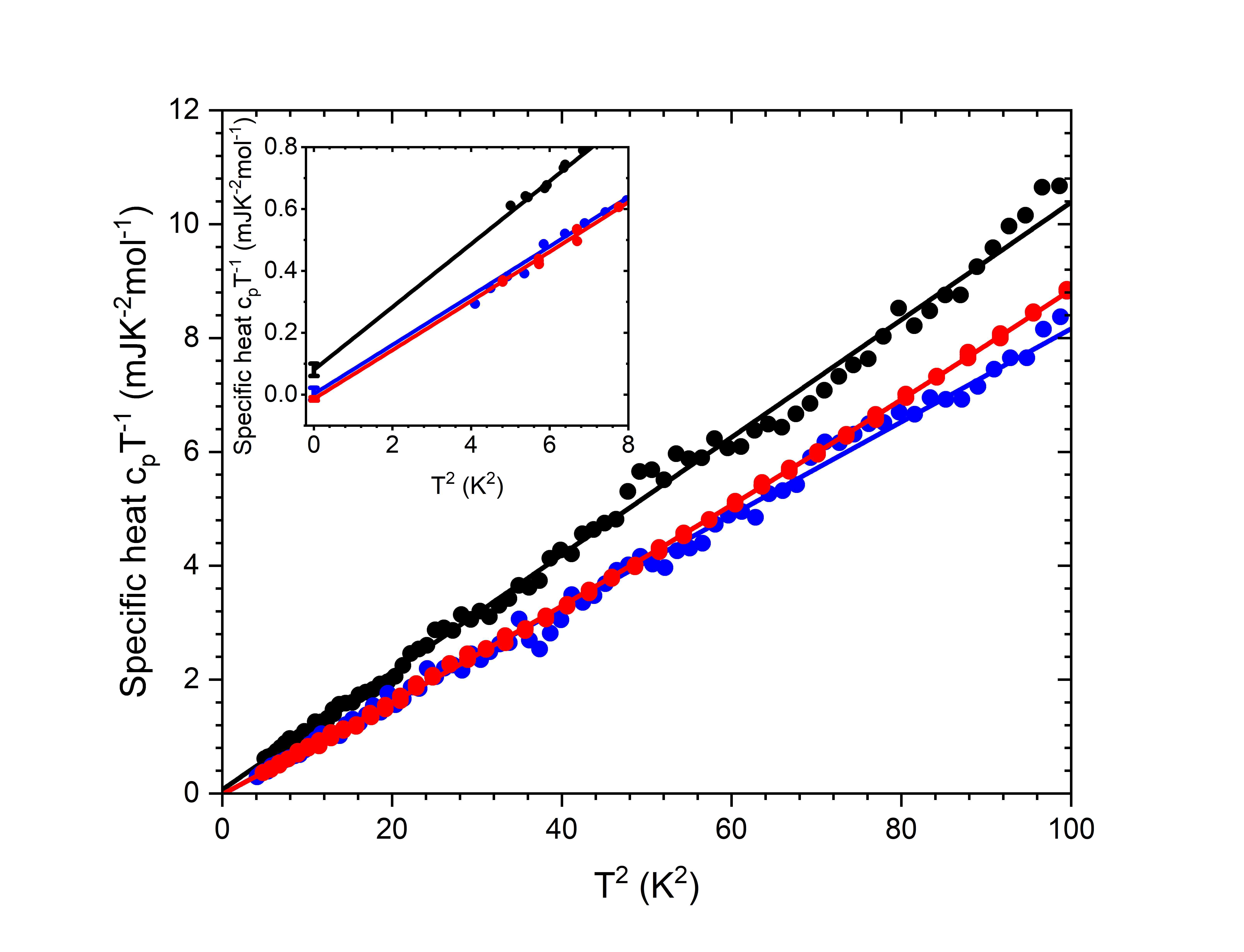}
\caption{Specific heat $c_{p}$ normalized by $T$ as a function of $T^{2}$ for the Verneuil-grown (black circles) and EFG crystals (red circles) demonstrating linear behavior as expected from a Debye model. Lines are fits described in the text. While the zero intercept of the fit with the y-axis for the Verneuil-grown sample shows a vanishing electronic contribution to the specific heat, the EFG sample displays a tiny but significant electronic contribution corresponding to a Sommerfeld coefficient $\gamma_{\text{EFG}}=8(2) \times 10^{-5}$ J$\,$K$^{-2}\,$mol$^{-1}$ . Note the different slopes indicating different Debye temperatures $\Theta^{D}$. Up to $T^{2}\sim 100$ K$^{2}$ deviations from linear behavior remain negligible.}
\label{fig:figS07_PPMS_HCO_Specific_Heat_by_T_Tsquared_Verneuil_EFG}
\end{figure}

An alternative representation of the $c_{p}$ data is shown in Fig.~\ref{fig:figS07_PPMS_HCO_Specific_Heat_by_T_Tsquared_Verneuil_EFG}. Here, $c_{p}/T$ is plotted vs. $T^{2}$. For the fit of the data over the range $4$ K$^{2}\leq T^{2} \leq 100$ K$^{2}$ we follow the approach by McCalla {\textit{et al.}$\,$}\cite{McCalla2019} and allow for a next-higher order Debye term, i.e.~modelling the data by
\begin{equation}  
c_{p}(T)=\gamma T + \alpha T^{3}  + \beta T^{5}. 
\label{eq:cp_Tpolynomial}
\end{equation}
Here, $\gamma$ is the Sommerfeld coefficient given by $\gamma =\left( \pi /3\right) ^{2/3}\left( k_{B}/\hbar \right)
^{2}m_{\text{DOS}}\,n^{1/3}$, with charge carrier concentration $n$ and density-of-states effective mass $m_{{\text{DOS}}}$, $\alpha$ is the Debye-term coefficient, and $\beta$ is the coefficient for the next-higher order Debye term. In fact, the coefficients $\beta_{{\text{VER3}}} = 3(2)\times 10^{-8}$ J$\,$K$^{-6}\,$mol$^{-1}$, $\beta_{{\text{EFG1}}} = 2(2)\times 10^{-8}$ J$\,$K$^{-6}\,$mol$^{-1}$ are close to zero in agreement with reference \cite{McCalla2019} for undoped SrTiO\textsubscript{3}. A finite value $\beta_{\text{VER1}} = 9.9(8)\times 10^{-8}$ J$\,$K$^{-6}\,$mol$^{-1}$ is obtained for specimen VER1, so that the data deviates from linear behavior for $T^{2} > 70$ K$^{2}$.

From the fit we extract $ \alpha_{\text{VER1}}=7.86(6) \times 10^{-5}$ J$\,$K$^{-4}\,$mol$^{-1}$  and $ \alpha_{\text{VER3}}=7.9(2) \times 10^{-5}$ J$\,$K$^{-4}\,$mol$^{-1}$ corresponding to Debye temperatures $\Theta_{D,\text{VER1}}= 498 \pm 1$ K and $\Theta_{D,\text{VER3}}= 497 \pm 4$ K for the VER1,3 crystals and $ \alpha_{\text{EFG1}}=8.18(6) \times 10^{-5}$ J$\,$K$^{-4}\,$mol$^{-1}$ corresponding to a Debye temperature  $\Theta_{D,\text{EFG1}}= 457 \pm 3$ K for the EFG1 crystal. We note that the identical analysis, which explicitly isolates the phononic contribution to the specific heat $\alpha T$ from the electronic contribution $\gamma T$, leads to two different Debye temperatures $\Theta_{D}$ which differ by $40 \pm 5$ K. We contend that this difference cannot be explained by a systematic experimental error.   

The fitted Sommerfeld coefficients $\gamma_{\text{VER1}}=-1.3(5) \times 10^{-5}$ J$\,$K$^{-2}\,$mol$^{-1}$  and $\gamma_{\text{VER3}}=0.2(1.9) \times 10^{-5}$ J$\,$K$^{-2}\,$mol$^{-1}$ are within error compatible with zero as expected for the nominally undoped crystals. For the EFG1 crystal we obtain a small, but finite $\gamma_{\text{EFG1}}=8(2) \times 10^{-5}$ J$\,$K$^{-2}\,$mol$^{-1}$ indicating a finite charge carrier concentration. Assuming a density-of-states effective mass $m_{{\text{DOS}}}= 6m_{e}$ \cite{Ahrens2007} we obtain an upper limit of the charge carrier concentration for the Verneuil-grown crystals $n_{\text{VER1,3}} < 2\times 10^{13}$ cm$^{-3}$ and $n_{\text{EFG1}}=1.25(5)\times 10^{16}$ cm$^{-3}$. Assuming that the charge carrier concentration exclusively results from doubly ionized oxygen vacancies \ch{V_{O}^{..}} and neglecting electron localization as well as charge compensation by ionic defects, upper limits for the oxygen vacancy concentrations $c_{\ch{V_{O}^{..}},\text{EFG1}} < 2 \times 10^{-4}$ at.ppm and $c_{\ch{V_{O}^{..}},\text{VER1,3}} < 1.2 \times 10^{-2}$ at.ppm can be derived.  We conclude that the concentration of non-charge compensated oxygen vacancies is negligible in both samples investigated by IXS, EFG1 and VER1.

\subsubsection{Discussion of sample-dependent shifts in $T_{C}$}

Despite some variation in the absolute phase-transition temperature $T_{C}$, $T_{C}$ is 3.2 K larger for the EFG1 sample than for the VER1 sample. Several experimental \cite{Huennefeld2002}, \cite{Bauerle1978} and theoretical \cite{RiccaAschauer2020}, \cite{Buban2004} studies have shown that $T_{C}$ decreases with increasing oxygen-vacancy concentration. Following B{\"{a}}uerle and Rehwald \cite{Bauerle1978}, an oxygen vacancy locally reduces the hybridization between oxygen $p$-orbitals and titanium $d$-orbitals. For lower hybridization the R-point soft mode frequency is locally \emph{increased}, and the cubic phase is stabilized accompanied by an overall \emph{decrease} in the transition temperature $T_{C}$. Thus, the smaller $T_{C}$ for the VER1 sample would require its oxygen-vacancy concentration to be larger. Given that the charge carrier concentration is negligible, as is demonstrated by the Sommerfeld coefficient extracted from our $c_{p}$ data, this scenario appears unlikely. 

On the other hand dopants like e.g. niobium, are known to shift the phase-transition temperature $T_{C}$ to larger temperatures with increasing dopant concentration. Experimental work has demonstrated that in the case of Nb doping, Nb$^{5+}$ ions predominantly substitute Ti$^{4+}$ on the B-sublattice of the ABO\textsubscript {3} perovskite structure \cite{Karczewski2010}. The, roughly by factor 2, larger mass of the Nb ions reduces the eigenfrequencies of phonon modes which involve the motion of Ti$^{4+}$ ions. As a consequence, the transition temperature of the structural phase transition driven by the R-point soft phonons, which do involve Ti-ion vibrations, is increased. In addition, Nb doping is reported to lead to the creation of Sr$^{2+}$ vacancies in order to fulfill the electroneutrality condition or else excessive oxygen ions are incorporated at interstitial positions in the structure \cite{Karczewski2010}. These effects obviously do not counterbalance the effect of an increased $T_{C}$. However, impurities in our EFG sample which act as efficient electron donors must also be ruled out by our electrical-conductivity and Sommerfeld-coefficient data.  

ICP-OES identifies Fe impurities to be significantly present in our samples (see section \ref{sectionICP_OES}). Fe-doped SrTiO\textsubscript {3} is a known photocatalyst and has been recently investigated by x-ray photoelectron spectroscopies (XPS) \cite{Kubacki2018}. In particular, it has been experimentally demonstrated that Fe ions with valence state Fe$^{2+}$ or Fe$^{3+}$  substitute Ti$^{4+}$ ions on the Ti sublattice \cite{Kubacki2018}. The absence of a significant charge-carrier concentration as concluded from the Sommerfeld coefficient suggests localized electronic states rather than itinerant states. It is thus interesting to relate the observed shift in $T_{C}$ to the presence of Fe impurities.  

Although experimental investigations of $T_{C}$ as a function of Fe-doping concentration to our knowledge do not exist, McCalla's work based on ionic valence mismatch and a bond valence sum approach \cite{McCalla2016} provides a route to obtain an estimate for the Fe-concentration. Once the differential rate of change of the transition temperature with substitution $dT_{C}/dx$ is known, the substitution concentration $x$ is also known for our experimentally observed $\Delta T_{C}=3.2$ K. The former quantity is related by the empirically established formula
\begin{equation}
\frac{dT_{C}}{dx}=\gamma \varepsilon ^{g}+\eta 
\label{eq:EmpiricaldTCdxFormulaMcCalla}
\end{equation}
to the relative ionic valence mismatch
\begin{equation}
\varepsilon = \Delta V_{\text{max}} - \Delta V_{\text{X}}
\label{eq:RelativeIonicValenceMismatchMcCalla}
\end{equation}
with empirical parameters $\gamma=0.138$ K/at.\%, $\eta=-30.2$ K/at.\%, $g=3.97$ \cite{McCalla2016}. For B-site substitution, the valence mismatch of cation X is $\Delta V_{\text{X}}=V_{\text{ideal}}-V_{\text{X}}$. Here, $V_{\text{ideal}}$ is the ideal formal valence of the cation, i.e. for Fe$^{2+}$, $V_{\text{ideal}}=2.0$  and for Fe$^{3+}$, $V_{\text{ideal}}=3.0$. The valence of cation X, $V_{\text{X}}$ is obtained by summing the bond valence
\begin{equation}
S_{\text{XO}}=\exp \left( \frac{R_{c}-R_{\text{XO}}}{b_{c}}\right) 
\label{eq:BondValenceMcCalla}
\end{equation}
over all nearest neighbors of cation X, i.e. for the six nearest neighbors of the B-site in the tetragonal phase
\begin{equation}
V_{\text{X}}=\sum\limits_{i=1}^{6}S_{\text{XO}}=
2\exp \left( \frac{R_{c}-R_{\text{XO,1}}}{b_{c}}\right) +4\exp \left( \frac{R_{c}-R_{\text{XO,2}}}{b_{c}}\right), 
\label{eq:BondValenceSumMcCalla}
\end{equation}
where we assume $R_{\text{XO,1}}=1.958$ $\text{\AA}$  and $R_{\text{XO,2}}=1.962$ $\text{\AA}$  (see Tab. S2 in the Supporting Information to reference \cite{McCalla2016}). The maximal possible ionic valence mismatch $\Delta V_{\text{max}}$ is given by setting $R_{\text{XO,1,2}}=2R_{c}$, i.e. evaluating 
\begin{equation}
\Delta V_{\text{max}}=V_{\text{ideal}}-V_{\text{max}}=V_{\text{ideal}}-6\exp \left( -\frac{R_{c}}{b_{c}}\right) 
\label{eq:MaximumBondValenceSumMcCalla}
\end{equation}
The bond-valence parameters $R_{c}$ and $b_{c}$ are taken from Gagné and Hawthorne \cite{Gagne2015}, who have calculated bond-valence parameters for ion pairs involving oxygen based on their generalized reduced gradient method. We use their tabulated values for Fe$^{2+}$: $R_{c}=1.658$ $\text{\AA}$ and $b_{c}=0.447$ $\text{\AA}$ and for Fe$^{3+}$: $R_{c}=1.766$ $\text{\AA}$ and $b_{c}=0.360$ $\text{\AA}$. For B-site substitution we obtain $dT_{C}/dx = 45.7$ K/at.\% and  $dT_{C}/dx = 21.5$ K/at.\%  for Fe$^{2+}$ and Fe$^{3+}$, respectively. We note that Fe$^{2+}$ would effectively be in a Fe$^{3.05+}$ valence state and Fe$^{3+}$ would effectively be in a Fe$^{3.49+}$ valence state. Alternatively, if we assume A-site substitution we obtain valence states in favor of a significant charge carrier concentration at odds with the experimental results. Both, effective Fe$^{3.05+}$ and Fe$^{3.49+}$ valence states are possible scenarios, in line with both, absence of charge carriers and a positive  $dT_{C}/dx$. To account for the $3.2$ K shift in $T_{C}$ would require an amount of Fe$^{2+}$ (Fe$^{3+}$) ions at a substitution level of $x=0.07$ at.\% ($x=0.15$ at.\%) resulting in defect concentrations $n_{D}=1.2 \times 10^{19}$ cm$^{3}$ to $n_{D}=2.5 \times 10^{19}$ cm$^{3}$, these numbers supporting the scenario that Fe is responsible for the shift in phase transition temperature $T_{C}$.  

\subsection{Electrical conductivity and charge carrier concentration}

The charge carrier concentrations as obtained from the Sommerfeld analysis of the specific heat data identifies electronically insulating behavior for the VER1,3 samples ($n_{\text{VER1,3}} < 2 \times 10^{13}$ cm$^{-3}$) and electronically conducting behavior for the EFG1 sample with $n_{\text{EFG1}} = 1.25(5) \times 10^{16}$ cm$^{-3}$, a carrier concentration identified in reference \cite{Spinelli2010} 
to be very close to the crossover from impurity-band to conduction-band transport for $n$-doped STO. This result is supporting the conclusion drawn from the thermal conductivity data, namely that the point defect level is larger for the EFG1 sample.  
In order to investigate whether the central peak intensity is sensitive to a finite concentration of free charge carriers, thus pointing towards a mechanism involving electronic degrees of freedom, an independent assessment of the charge carrier concentration is mandatory.

\begin{figure}
\centering
\includegraphics[width=1.0\linewidth]{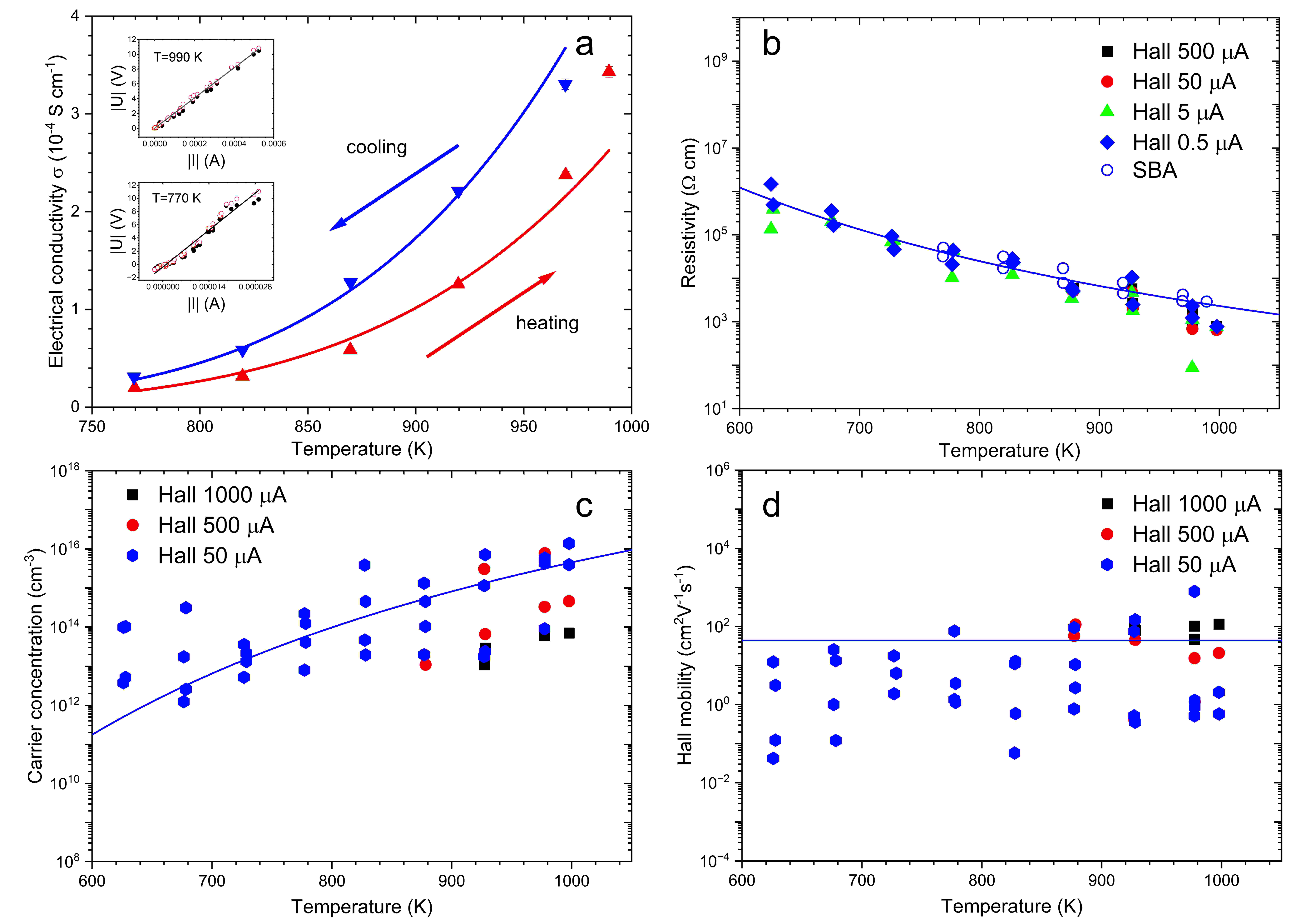}
\caption{Panel a: Temperature-dependent electrical conductivity of sample EFG1 as obtained from measurements with the Seebeck analyzer (SBA). Lines are fits to a model based on electrons excited into the conduction band by thermal activation. Error bars as extracted from linear fits to the $U-I$ curves are smaller than the symbol size. The insets show exemplary $U-I$ curves for $T=\SI{990}{\kelvin}$ (top inset) and $T=\SI{770}{\kelvin}$ (bottom inset) each for positive excitation current (black filled circles) and negative excitation current (red open circles). Only magnitudes of $I$ and $U$ are shown. Panel b: Temperature-dependent electrical resistivity of sample EFG1 as obtained from Hall measurements for various excitation currents. SBA data is shown for comparison. The line is a fit to an exponential. Panel c: carrier concentration from Hall measurements. Panel d: Hall mobility data.}
\label{fig:figS08X_SBA_Hall_EFG}
\end{figure}

Measurements of the electrical conductivity $\sigma$ of a \SI{1.25}{\milli \meter} thick, disk-shaped (\SI{15.3}{\milli \meter} diameter) EFG sample were performed with a commercial Seebeck analyzer device (SBA458 Nemesis, NETZSCH-Ger\"atebau GmbH) in the temperature range \SIrange{320}{990}{\kelvin} where a heating cycle, increasing the temperature from room temperature to the maximum temperature, was followed by a cooling cycle (Fig.~\ref{fig:figS08X_SBA_Hall_EFG}, panel a). Below \SI{770}{\kelvin} these attempts failed with data lacking a reliable linear relationship between applied current $I$ and probed voltage signal $U$. This is attributed to the low charge carrier concentration leading to electrical conductivities far below the lower sensitivity limit of the Seebeck analyzer device specified by the manufacturer as $\sigma \sim$ \SI{5}{S \meter^{-1}}. Above \SI{770}{\kelvin}, a linear voltage response $U(I)$ was observed when applying a positive current $I$ to the sample, and reliable electrical conductivity data could be extracted from linear fits to the $U-I$ curves (see Fig.~\ref{fig:figS08X_SBA_Hall_EFG}, panel a, insets), albeit still below the specified sensitivity limit.

Subsequently, the charge-carrier concentration $n$ and the Hall mobility $\mu$ were probed by a commercial Hall effect measurement system (Lake Shore 8400 Series) on the same \SI{1.25}{\milli \meter} thick EFG sample in van der Pauw geometry (Fig.~\ref{fig:figS08X_SBA_Hall_EFG}, panels c, d). The resistivity as determined with the Hall setup (Fig.~\ref{fig:figS08X_SBA_Hall_EFG}, panel b) used excitation currents in the range \SIrange{0.5}{1000}{\micro \ampere}. DC field Hall measurements were performed at magnetic fields $B=\SI{0.9}{\tesla}$. Additional AC field Hall measurements were performed using an excitation frequency of \SI{100}{\milli \hertz}. The Hall factor $h_{f}$ was assumed to be 1 for all analyses. Given the negligible charge carrier concentration for the VER1,3 samples as derived from the specific heat measurements, no attempts were made to collect Hall data of the Verneuil-grown samples.

We note that any assessment of the charge carrier concentration identifies only those defects which act as electron donors (e.g. O vacancies) or acceptors (e.g. Fe$^{3+}$ substituting Ti$^{4+}$ on the B-site) while being insensitive to electrically inactive, i.e. charge neutral defects (e.g. Ca$^{2+}$ substituting Sr$^{2+}$ on the A-site). In contrast, the thermal transport measurements are sensitive to mass differences irrespective of the defect valence state. However, in the low charge carrier concentration regime in which the EFG1 sample falls, compensation effects prevent the defect concentration from being determined quantitatively even from a known charge carrier concentration.

Owing to the rather low electrical conductivity and the limited sensitivity of the measurement equipment used, reliable resistivity and Hall effect data at room temperature are lacking. However, reliable conductivity and Hall data of the EFG1 sample were collected at elevated temperatures above 625 K (see Fig.~\ref{fig:figS08X_SBA_Hall_EFG}, panel b). Conductivity and resistivity data show thermally activated behavior corresponding to an activation energy of \SI{1.5(3)}{eV} in the temperature range $\SI{625}{\kelvin} < T < \SI{1000}{\kelvin}$. While the Hall mobility $\mu$ appears to be rather constant in the temperature range investigated, the carrier concentration $n$ is responsible for the observed thermally activated behavior of the conductivity. Despite significant scatter in the carrier concentration data, it is clear that the maximum $n$ does not exceed $\sim 1 \times 10^{16}$ cm$^{-3}$ even at temperatures as large as \SI{1000}{\kelvin}. Using a simple hydrogenic model of shallow donors which implies a drastic decrease of the activation energy by several orders of magnitude with decreasing temperature because of the increase in dielectric constant, justifies the assumption that this charge carrier density establishes also an upper limit to the charge carrier concentration at low temperatures, i.e. in the temperature regime where the specific heat measurements were performed. We note that this approach ignores oxygen-vacancy generation at elevated temperatures which would generate two electrons per oxygen vacancy. This is likely the mechanism responsible for the observed increase in electrical conductivity in the cooling cycle, i.e. after the sample has been exposed to high temperatures for an extended period of time (Fig.~\ref{fig:figS08X_SBA_Hall_EFG}, panel a). Experimentally, the heating and cooling cycles for the SBA measurements took both 6 hours from $T = \SI{770}{\kelvin}$ to $T = \SI{990}{\kelvin}$ and 6 hours from $T = \SI{990}{\kelvin}$ to $T = \SI{770}{\kelvin}$ and were performed with nitrogen as purge gas. Keeping the samples at elevated temperatures in inert gas atmosphere for hours provides favorable conditions for oxygen-vacancy generation. However, given the rather small increase in electrical conductivity (up to factors of 2) we conclude that oxygen-vacancy generation is not the dominant mechanism underlying the exponential increase of the conductivity, rather it is thermal activation of localized charge carriers from substitutional defects or, most likely, valence-band-to-conduction-band transitions in the intrinsic semiconductor regime. The latter scenario is supported by the fact that the activation energy coincides with half the band gap of pristine STO ($\sim \SI{3.25}{eV}$). Using these arguments, we conclude that the charge carrier density as obtained from high temperature resistivity and Hall measurements is compatible with the carrier concentration extracted from specific heat data. Hence, the Hall measurements underpin the statement that the larger central peak intensity is observed for a sample with lower charge carrier concentration.

However, we argue that our EFG1 sample is not a conductor as the bare low-temperature charge carrier density would suggest with reference to the results of Spinelli {\it{et al.}}~\cite{Spinelli2010}. These authors have investigated Nb-doped and oxygen-deficient STO samples all of which exhibit $n$-type conductivity. Our Hall measurements demonstrate that the EFG1 sample has a rather high resistivity of $\sim 7 \times 10^{5}$ \SI{}{\ohm cm} at \SI{625}{\kelvin}, which is three orders of magnitude larger than the room temperature resistivity $\sim 2 \times 10^{2}$ \SI{}{\ohm cm} of a reduced, i.e. oxygen deficient \ce{SrTiO_{3-x}} sample with a comparable charge carrier concentration as stated in reference \cite{Spinelli2010}. We attribute this apparent discrepancy to the fact that the EGF1 sample does not exhibit pure $n$-type conductivity. In fact, this is in line with the presence of Fe-defects as identified by the ICP-OES measurements, since Fe$^{3+}$ substitution on the Ti$^{4+}$ site implies $p$-type conductivity. Even if charge compensation effects prevent a straightforward interpretation of Hall effect measurements in terms of separating charge carrier populations in the conduction and the valence band and the observed activation energy in the high temperature regime is compatible with intrinsic semiconductor behavior, we contend that also the EFG1 sample should be assigned to the impurity-band transport regime rather than to a conduction-band transport regime. Despite measurable differences in the electronic transport behavior  
between samples EFG1 (finite charge carrier concentration) and VER1 (negligible charge carrier concentration), no clear-cut metal-insulator difference can be identified.

\subsection{Elemental analysis of point defects}

\subsubsection{Micro X-ray fluorescence spectrometry ($\mu$XRF)}

In order to obtain information on the major impurities the elemental composition of specimens EFG1 and VER3 was studied by $\mu$XRF using a commercial M4 TORNADO spectrometer from Bruker Corporation. Measurements were performed with an x-ray tube with a Rh-anode operated at 50 kV voltage and 150 $\mu$A current. No further filters were applied. Parameters of the polycapillary x-ray optics were chosen to provide a spot size of 170 $\mu$m. The samples were placed on a \SI{0.025}{mm} thick W sheet. The emission spectra were recorded by a silicon drift detector which was cooled to \SI{-45}{\celsius}. The detector is located behind an 8 $\mu$m thick Be window.  Acquisition time for the spectra was typically \SI{240}{\second}. 

\begin{figure}
\centering
\includegraphics[width=0.53\linewidth]{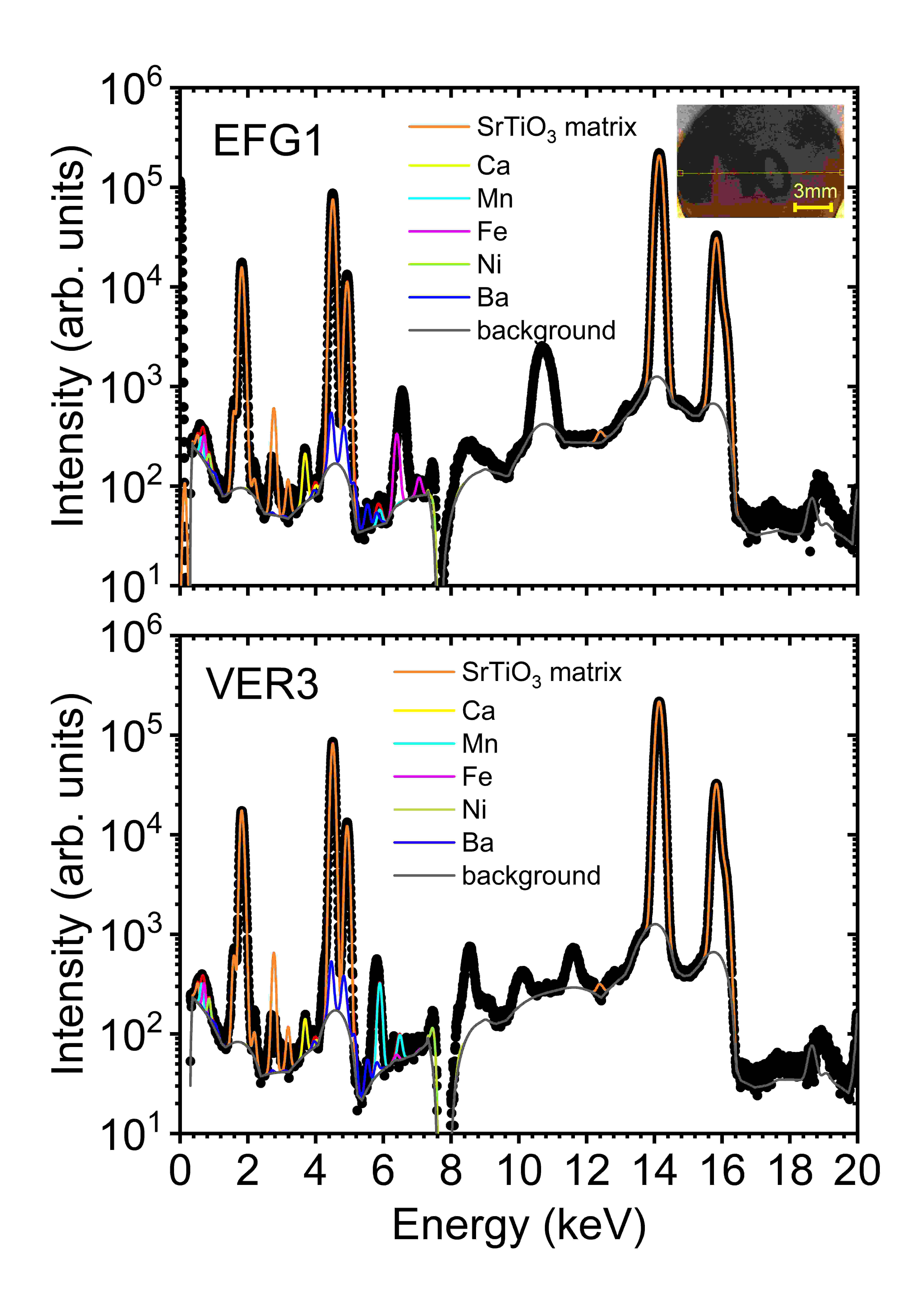}
\caption{Experimental $\mu$XRF spectrum for sample EFG1 (black data points). Lines correspond to fit results for K- and L-edge emission lines of the \ce{SrTiO_{3}} matrix and Ca, Mn, Fe and Ni impurities and background as indicated in the legend. Inset: Sample positions (red crosses) probed across the diameter of the disk-shaped EFG1 sample.}
\label{fig:fig_S08_muXRF_spectrum_EFG1_AND_VER3}
\end{figure}

The data were analyzed using the PyMca software package from ESRF (version 5.9.2) \cite{Sole2007}. The spectra for EFG1 and VER3 specimens are shown in Fig.~\ref{fig:fig_S08_muXRF_spectrum_EFG1_AND_VER3}, along with fits of the most intense peaks corresponding to K- and L-transitions of the constituents of the matrix, Sr and Ti. These peaks, located at \SI{1.8038}{keV} (Sr, L3-M5), \SI{4.5105}{keV} (Ti, K-L3), \SI{4.9369}{keV} (Ti, K-M3b), \SI{14.1573}{keV} (Sr, K-L3), and \SI{15.8412}{keV} (Sr, K-M3b), were also used for the energy calibration of the spectra. The background was fitted using the non-analytical background algorithm 'strip' provided with the PyMca software. The grey curves in Fig.~\ref{fig:fig_S08_muXRF_spectrum_EFG1_AND_VER3}, top and bottom, labelled 'background' include both, continuum and pileup signal. The XRF signal between \SI{7.5}{keV} and \SI{13.5}{keV} is contaminated with Bragg scattering, and intensities in this energy range strongly depend on the relative orientation of x-ray source, detector and the crystallographic axes of the sample. 

Impurities could be identified using the XRF signal above background between 1.2 keV and 7.5 keV. Ca can be unambiguously identified to be present in both, EFG1 and VER3 specimens by the fluorescence line at \SI{3.692}{keV} which is attributed to the K-L3 transition. Further candidate impurities are Fe, Mn, Ni and Ba. 

\begin{table*}[hb]

\centering
\begin{ruledtabular}
\begin{tabular} {C{1.5cm} |   |  C{1.2cm}   |    C{1.2cm}    |   C{1.2cm}    |  C{1.2cm}    |    C{1.2cm}    |   C{1.2cm}    |   C{1.2cm}    |  C{1.2cm} }
               & c(Sr)                 & c(Ti)                 & c(O)                  & c(Ca)                 & c(Ba)                 & c(Fe)                 & c(Mn)                 & c(Ni)                  \\
sample         & wt.\% & wt.\% & wt.\% & wt.\% & wt.\% & wt.\% & wt.\% & wt.\%  \\ 
\hline
&                 &               &                &                  &                 &                &               &                  \\[-2mm]
reference & 47.75                 & 26.09                 & 26.16                 & 0                     & 0                     & 0                     & 0                     & 0                      \\
EFG1           & 47.75                 & 25.61                 & 35.24                 & 0.081                 & 0.37                  & 0.061                 & 0.005                 & 0.006                  \\
VER3           & 47.75                 & 25.78                 & 34.33                 & 0.041                 & 0.33                  & 0.002                 & 0.070                 & 0.011                  \\

\hline\hline
&                 &               &                &                  &                 &                &               &                  \\[-2mm]
               & c(Sr)                 & c(Ti)                 & c(O)                  & c(Ca)                 & c(Ba)                 & c(Fe)                 & c(Mn)                 & c(Ni)                  \\
      & at.\% & at.\% & at.\% & at.\% & at.\% & at.\% & at.\% & at.\%  \\ 
\hline
&                 &               &                &                  &                 &                &               &                  \\[-2mm]
reference    & 100                   & 100                   & 100                   & 0                       & 0                       & 0                        & 0                        & 0                      \\
EFG1           & 100                   & 97.5                  & 154                   & 0.373                 & 1.70                  & 0.201                 & 0.016                 & 0.017                  \\
VER3           & 100                   & 98.4                  & 148                   & 0.190                 & 1.51                  & 0.005                 & 0.235                 & 0.036                  \\

\end{tabular}
\end{ruledtabular}
\caption{Fit results from $\mu$XRF measurements.}
\label{Tab_muXRF_concentrations}
\end{table*}

Fit results for the mass fraction concentrations in units of wt.\%, for convenience also converted to at.\% are given in Tab.\ref{Tab_muXRF_concentrations}. However, these values, though indicating the major impurities, are only rough estimates. Intensities are normalized to match the strongest peak from the \ce{SrTiO_{3}} matrix observed for the Sr, K-L3 transition at \SI{14.1573}{keV}. Hence, the mass fraction for Sr derived from this transition matches the nominal mass fraction for all samples. The fitted mass fraction for Ti for EFG1 and VER3 specimens is close to the nominal value. The O mass fraction is heavily overestimated since the weak signal of the O, K-L3 transition at \SI{0.525}{keV} suffers from a small signal-to-background ratio.     

The peaks at \SI{6.56}{keV} in the EFG1 data and \SI{5.8}{keV} in the VER3 data are close to the energies of the Fe, K-L3 transition at \SI{6.404}{keV} and the Mn, K-L3 transition at \SI{5.899}{keV}, respectively. However, the peak positions do not match exactly, likely due to the modification of the electronic states in the \ce{SrTiO_{3}} matrix, so that the fit does not match the peaks, implying a larger error in the estimated concentrations. The precise determination of the concentration of Ni, which displays only a very weak signal, and of that of Ba, whose weak signatures overlap with the strong Sr fluorescence lines, is also challenging. This is calling for an alternative approach to impurity determination, which we have chosen to be ICP-OES (see section \ref{sectionICP_OES}).

Additional measurements were performed with Rh-anode and 20 $\mu$m spot size, the samples being placed on paper. Spectra recorded for different positions of the EFG1 sample (see inset to Fig.~\ref{fig:fig_S08_muXRF_spectrum_EFG1_AND_VER3}) did not reveal a significant variation of the Ca and Fe lines across the diameter of the disk-shaped specimen. 

\subsubsection{Inductively coupled plasma optical emission spectrometry (ICP-OES)}\label{sectionICP_OES}
ICP-OES has been performed on specimens EFG1, VER1 and VER2. The single-crystalline samples were ball-milled with a WC grinding medium, preparing two \SI{10}{\milli \gram} batches of each specimen for double determination. 
After microwave-assisted acid digestion of the two \SI{10}{\milli \gram} samples in \ce{HCl} for \SI{20}{\minute} at $T=\SI{220}{\celsius}$ using a commercial MLS-MWS microwave system, spectroscopic measurements were performed probing the emission lines of selected elements. Spectroscopy was performed using an iCAP 7400 Duo ICP-OES analyzer from Thermo Fisher Scientific GmbH. Particular care was taken with the calibration to quantify of the limit of detection (LOD) and the limit of quantification (LOQ). The LOD values were determined by an experimental procedure utilizing the blank value method, which entailed a ten-fold measurement of the blank value after an initial calibration with synthetic solution standards. To obtain the LOD (LOQ), the standard deviation was multiplied by a factor of three (nine). The wavelength with the lowest LOD / LOQ was used for quantification. 
For each impurity element, three wavelengths were measured by threefold determination. Accordingly, the concentrations and standard deviations given in Tab.~\ref{Tab_ICP_OES_concentrations} are obtained by averaging the six data values. 

The selection of emission lines was guided by those candidate impurities identified by $\mu$XRF, namely Ca, Ba, Fe, Mn, Ni. In addition, the spectroscopy was probing the emission lines of Co, Cu, Ir, Pt and Rh. These latter elements are either known contaminants in the starting materials, or else the melt has been in contact with them during synthesis, e.g.~platinum crucibles have been used for calcination and an iridium die was used in the EFG synthesis process \cite{Guguschev2015}. Rh has been chosen, since the  $\mu$XRF measurements have been performed with a Rh anode rendering Rh detection at low impurity levels by $\mu$XRF challenging.

\squeezetable

\begin{table*}[]

\centering
\begin{ruledtabular}

\begin{tabular}{c| | c | c | c | c | c | c | c | c | c | c }  
                  &   &     &   &   &         &   &   &   &   &    \\ [-2mm]
 $\lambda$ (nm) & 230.4  & 396.8    & 239.5  & 257.6  & 231.6        & 238.8  & 224.7  & 212.6  & 203.6  & 343.4   \\ 
\hline
                  &   &     &   &   &         &   &   &   &   &    \\ [-2mm]
                                & c(Ca)  & c(Ba)    & c(Fe)  & c(Mn)  & c(Ni)        & c(Co)  & c(Cu)  & c(Ir)  & c(Pt)  & c(Rh)   \\ 

sample                          & wt.ppm & wt.ppm   & wt.ppm & wt.ppm & wt.ppm       & wt.ppm & wt.ppm & wt.ppm & wt.ppm & wt.ppm  \\ 
\hline
                  &   &     &   &   &         &   &   &   &   &    \\ 
EFG1                            & 59(1)  & $<$3        & 147(1) & 1.0(1) 	& 14.5(4)      & $<$7.8    &$<$ 57     & $<$21    &$<$81     & $<$18     \\
VER1                            & 60(1)  & $<$3        & 107(2) & 3.0(8) 	& $<$9 		 & $<$7.8    & $<$57     &$<$21     & $<$81     & $<$18      \\
VER2                            & 81(1)  & 149.0(4) & 12(2)  & $<$0.9    & $<$9            & $<$7.8    & $<$57     & $<$21     & $<$81     & $<$18     \\

\hline
\hline
                  &   &     &   &   &         &   &   &   &   &    \\ [-1mm]
                                & c(Ca)  & c(Ba)    & c(Fe)  & c(Mn)  & c(Ni)        & c(Co)  & c(Cu)  & c(Ir)  & c(Pt)  & c(Rh)   \\ 

                          & at.ppm & at.ppm   & at.ppm & at.ppm & at.ppm        & at.ppm & at.ppm & at.ppm & at.ppm & at.ppm  \\ 
\hline
                  &   &     &   &   &         &   &   &   &   &    \\ 
EFG1                            & 270(5) & $<$4        & 484(3) & 3.3(4) 	& 45(1)         	& $<$24     & $<$165    & $<$20     & $<$76     & $<$32      \\
VER1                            & 274(6) & $<$4        & 351(7) & 10(3) 	& $<$28 		& $<$24     & $<$165    & $<$20     & $<$76     & $<$32      \\
VER2                            & 371(5) & 199.1(5) & 39(8)  & $<$3     	& $<$28 	          & $<$24     & $<$165    & $<$20     & $<$76     & $<$32     

\end{tabular}
\end{ruledtabular}
\caption{Impurity concentrations obtained by ICP-OES measurements. Standard deviations are given in parenthesis. Whenever no signal was detected, the limit of quantification is given quantifying an upper limit for the impurity concentration. Top row: wavelengths of the emission lines. }
\label{Tab_ICP_OES_concentrations}
\end{table*}


Results for the mass fraction concentrations in units of wt.ppm, for convenience also converted to at.ppm are given in Tab.~\ref{Tab_ICP_OES_concentrations}. Co, Cu, Ir, Pt and Rh could not be detected by ICP-OES in any of the specimens above the specified LOD. In addition, glow discharge mass spectrometry (GDMS) has been performed by Eurofins EAG Toulouse, France, confirming that Co, Cu and Pt are absent with a sensitivity limit at $\sim\SI{3}{at.ppm}$. GDMS is less sensitive to Rh than ICP-OES. A very small 5(1) at.ppm contamination with Ir is identified by GDMS for the EFG1 sample while VER1,2 samples have Ir impurities below the \SI{0.1}{at.ppm} level. The Mn contamination, albeit detectable, is also rather small. For the further discussion we neglect Co,Cu, Ir, Pt, Rh and Mn focusing on the most prominent impurities. 

While Ba is absent in the VER1 specimen, the VER2 specimen does show a significant concentration of Ba. The situation is reversed for Fe which is absent in VER1 and present in VER2. This demonstrates that at impurity levels around 100 at.ppm a charge-to-charge variation of trace contaminations for our samples exists. Turning to the specimens EFG1 and VER1 which are taken from the very same charge as the IXS samples, it is evident that Ba is absent in both. These specimens have also the same level of Ca impurity concentration, while Fe and Ni impurities have a larger concentration in the EFG1 specimen. Thus ICP-OES confirms that the EFG1 sample has a larger concentration of point defects. 

Fe predominantly exits in the Fe$^{3+}$ valence state and substitutes Ti$^{4+}$ on the B-sublattice of the ABO\textsubscript {3} perovskite structure \cite{Li2019, Kharton2001}, requiring one oxygen vacancy to satisfy charge balance. Ni also substitutes on the B-site of the perovskite unit cell, but may take valence states Ni$^{2+}$, Ni$^{3+}$, Ni$^{4+}$ corresponding to double, single or zero oxygen vacancies. The charge-neutral substitution of Sr$^{2+}$ ions by Ca$^{2+}$ on the A-site does not introduce additional oxygen vacancies. We conclude that in total the EFG1 sample has 529(1) at.ppm B-site impurities compared to only 351 at.ppm B-site impurities for the VER1 sample, corresponding to a factor 1.5 increased impurity level on the B-site. Using the charge-balance rule, we conclude that approximately the same concentration of oxygen vacancies is present. 

\subsection{Assessment of mosaicity by Larmor diffraction}\label{larmor}

The strong suppression of the thermal conductivity peak at $T\simeq\SI{25}{\kelvin}$ observed for the EFG1 sample clearly indicates enhanced phonon scattering. Nevertheless, from thermal conductivity data it is not straightforward to differentiate between point and surface defects that both act to scatter phonons. Below, we provide evidence that the density of surface defects is lower for crystals grown by the EFG method.   

EFG-grown crystals show a much higher structural perfection as compared to Verneuil-, optical floating zone and Czochralski-grown crystals. This has been confirmed by earlier measurements of the etch-pit density (EPD), probing the dislocation density $\rho_{D}$, and x-ray diffraction, probing the mosaicity $\eta_{200}$ at (200) Bragg peaks \cite{Guguschev2015}. Smaller specimens cut from EFG-grown crystals and afterwards chemo-mechanically polished have been used for these experiments. Typically, EPD values ranged between $8.2 \times 10^{4}$ cm$^{-2}$ and $2.8 \times 10^{5}$ cm$^{-2}$ \cite{Guguschev2015} superior to the EPD of Verneuil-grown bulk crystals which typically exceeds $10^{6}$ cm$^{-2}$. The x-ray diffraction experiments used Cu K$\alpha_{1}$ radiation with a collimation corresponding to an illuminated crystal area of approximately $2 \times 10$ mm$^{2}$. Mosaicity values between \SI{1.1}{\arcminute} and  \SI{2.6}{\arcminute} were determined. Although not representative of the mosaic of the entire crystal, since only a small volume is probed in the x-ray experiments, these experimental results indicate a much narrower crystal mosaic distribution for EFG-grown samples superior to the mosaicity of  Verneuil-grown samples. 

We have investigated specimens EFG2 and VER4 with elastic neutron scattering using Larmor diffraction (LD) \cite{Rekveldt2001}. These specimens are large single crystals with a high sample mass. Sample EFG2 is a cylindrically shaped single boule with a length of 11 mm and a diameter of about 15 mm which has a total sample mass of 9.9 g. Sample VER4 is a single boule shaped as truncated ellipsoid with a length of 38 mm and a diameter varying from 13 mm to 32 mm. The total sample mass amounts to 115.2 g. 

\begin{figure}
\centering
\includegraphics[width=0.75\linewidth]{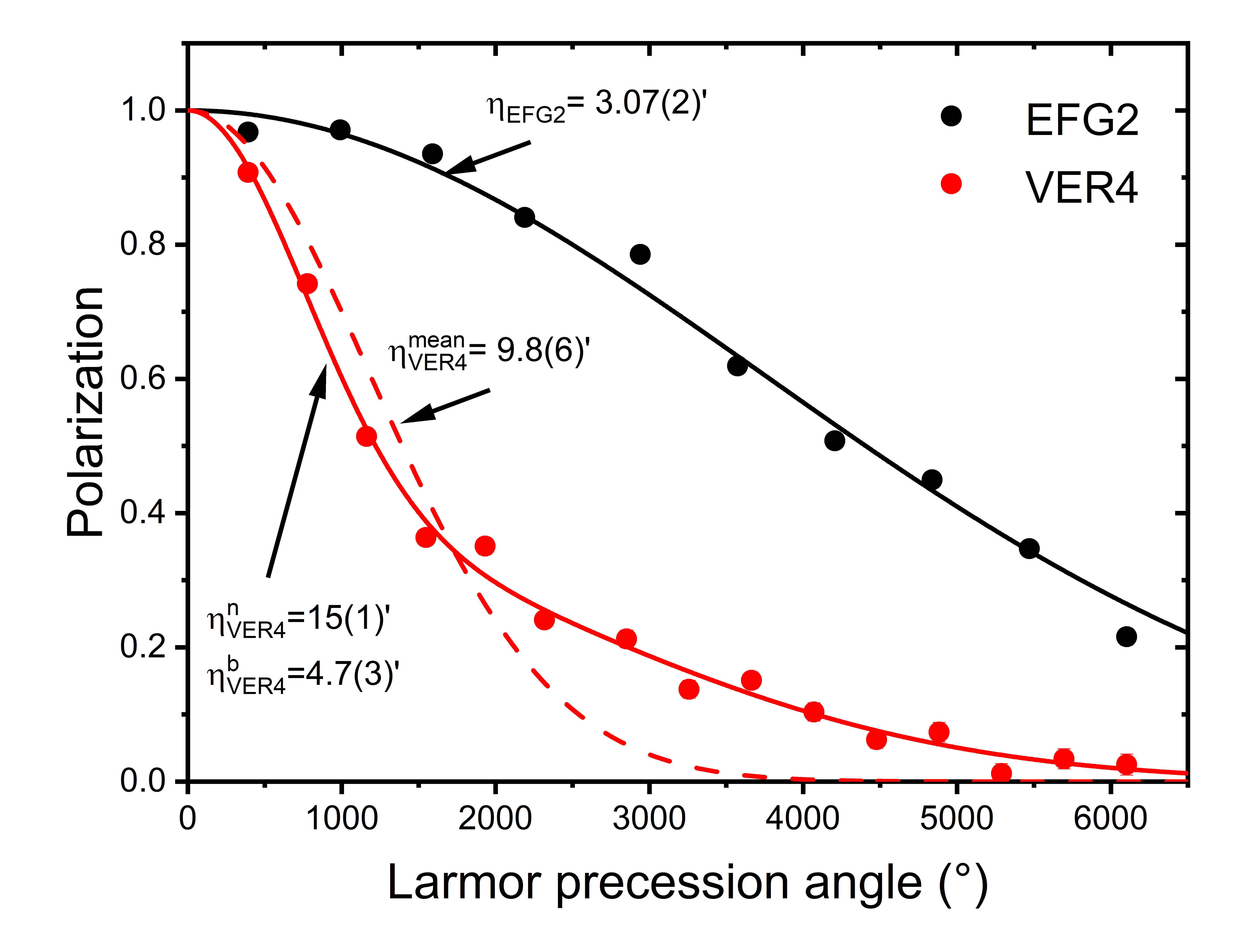}
\caption{Polarization as a function of Larmor precession angle $\phi$ probing the sample mosaic distribution.}
\label{fig:fig_S09_Larmor_Diffraction_Mosaicity_EFG2_VER4}
\end{figure}

The neutron scattering experiments were performed at the TRISP triple axis spectrometer with neutron resonance spin echo (NRSE) located at the Heinz Maier-Leibnitz Zentrum (MLZ), Garching, Germany \cite{Keller2007, TRISP2015}. The NRSE coils, with tilt angles $\theta_{1,2}$ and coil distances $L_{1,2}$ in spectrometer arms 1 and 2, were operated in a configuration which provides sensitivity to the sample mosaicity $\eta$ \cite{Habicht2016}. In this magnetic field geometry, the total Larmor precession angle $\phi$ for Bragg scattering of neutrons with velocity $v_{I}$ from a specific mosaic block depends on the block's angular orientation $\Delta \eta $
\begin{equation}
\phi =\pm 2\frac{\omega _{1}L_{1}}{v_{I}}\Delta \eta \tan \theta _{1},
\end{equation}
where $\Delta \eta $ is the angular difference between the reciprocal lattice vector $\mathbf{G}$ of the mosaic block and the mean reciprocal lattice vector $\mathbf{G}_{0}$ of the entire crystal. Hence, the ensemble-averaged neutron beam polarization directly encodes the distribution in mosaic angles $\Delta \eta $. The Larmor frequency 
\begin{equation}
\omega _{1}=\frac{2\mu B_{1}}{\hbar }
\end{equation}%
depends on the chosen magnetic field $B_{1}$. Here, $\mu $ is the magnetic moment of the neutron $\mu =-1.9130427$ $\mu _{N}$,  $\mu _{N}$ being the nuclear magneton. For a given sample mosaic, increasing the field strength, and thus increasing the Larmor precession angle, leads to an increased depolarization of the neutron beam. At fixed Larmor precession angle $\phi$, the degree of depolarization of the neutron beam reflects the mosaicity.

Assuming a Gaussian mosaicity distribution with FWHM $\eta$, the mosaicity is obtained by fitting the polarization to  
\begin{equation}
P= e^{-\phi_{0}^{2}\eta^{2}},
\label{eq:PolarizationMosaicity}
\end{equation}
with 
\begin{equation}
\phi_{0}=\frac{2\omega _{1}L_{1}}{v_{I}}\tan \theta _{1}.
\label{eq:Phi0Definition}
\end{equation}

Fig.~\ref{fig:fig_S09_Larmor_Diffraction_Mosaicity_EFG2_VER4}  shows the polarization data collected at the [002] Bragg peak along with fits to Eq.~(\ref{eq:PolarizationMosaicity}). The data have been correcting for instrument-related depolarization. A single Gaussian distribution ($\eta_{\text{EFG2}}=\SI{3.07(2)}{\arcminute}$) is sufficient to model the data for the EFG2 crystal, while two Gaussians, a narrow  ($\eta^{n}_{\text{VER4}}=\SI{4.7(3)}{\arcminute}$) and a broad  ($\eta^{b}_{\text{VER4}}=\SI{15(1)}{\arcminute}$) mosaic distribution better fit the VER4 data. Forcing a single Gaussian fit with fixed $P(\phi=0)=1$ gives ($\eta_{\text{VER4}}=\SI{9.8(6)}{\arcminute}$).  Clearly, the data demonstrates the superior sample quality of the EFG2 sample compared to the VER4 sample. 
This is in line with rocking scans across Bragg peaks performed during the course of the IXS experiments, where $\eta_{\text{EFG1}}=\SI{2.22(2)}{\arcminute}$ and $\eta_{\text{VER1}}=\SI{3.4(1)}{\arcminute}$ has been experimentally determined. The much smaller $\eta$ values observed in the IXS experiment are attributed to the much smaller sample volume probed as compared to the neutron experiment.

\newpage
\section{Specific heat model}\label{SI_B}

We model the total specific heat at constant pressure $c_{p}$ by summing over weighted contributions from several distinct phonon branches: two low-energy transverse acoustic branches, a single longitudinal acoustic branch, 8 dispersionless branches, modeled as Einstein modes, and an additional soft-mode branch, mapping the phonon dispersion along the $\Gamma$-R direction. The soft-mode branch hosts the R-point phonon which softens at the cubic-to-tetragonal structural phase transition and is responsible for the specific heat anomaly observed at $T_C$. 

\subsection{Low-energy acoustic phonon contribution}

The specific heat at constant volume of the transverse and longitudinal acoustic branches is given by
\begin{eqnarray}
c_{V}^{A}(T) &=& 2\int\limits_{0}^{\omega_{0T}}D_{T}^{\prime }(\omega )W_{ph}(\omega ,T)d\omega +
\int\limits_{0}^{\omega_{0L}}D_{L}^{\prime }(\omega )W_{ph}(\omega ,T)d\omega,
\label{eq:cpacousticModelC4}
\end{eqnarray}
where $D_{i}^{\prime }(\omega)$ is the phonon density of states (PDOS) of the corresponding branch $i=T,L$, and $W_{ph}$ is a weighting function related to the temperature derivative of the Bose-Einstein distribution function 
 \begin{equation} 
W_{ph}(\omega ,T)=\frac{\partial }{\partial T}\frac{\hbar \omega }{e^{\frac{
\hbar \omega }{k_{B}T}}-1}.
\label{eq:WindowfunctionModelC2}
\end{equation} 
The phonon dispersion for each of these acoustic branches is assumed to be isotropic and temperature-independent with a sinusoidal dependence of the phonon frequencies on wavevector $q$ 
\begin{equation}
\omega_{i} =\frac{2}{\pi}\omega _{Di}\sin \left( \frac{\pi }{2}\frac{q}{q_{D}}\right).
\label{eq:sinedispersionModelC2}
\end{equation} 
Here, $\omega_{Di}= v_{Si} q_{D}/ \hbar$ is the Debye frequency of the $i$-th branch which is obtained from the velocity of sound $v_{Si}$. Following the conventional Debye approach of mapping a 3D dispersion to 1D wavevector space, the dispersion is truncated at the Debye wavevector 
\begin{equation}
q_{D}=\left( 6\pi ^{2}\frac{rN_{A}}{V_{\text{mol}}}\right) ^{1/3},
\label{eq:Debyewavevector}
\end{equation}
where $N_{A}$ is the Avogadro constant, $r$ is the number of ions in the unit cell, here, $r=5$ for SrTiO$_{3}$, and $V_{\text{mol}}$ is the molar volume.
For $\omega \leq \omega _{Di}$ the PDOS {\emph{per branch}} is given by 
\begin{equation} 
D^{\prime }_{i}(\omega )=\frac{1}{2\pi ^{2}}V_{\text{mol}}\frac{\left[ q(\omega )\right]^{2}}{v_{gi}(\omega )}
\label{eq:PDOSModelC2}
\end{equation} 
with phonon group velocity    
\begin{equation} 
v_{gi}(\omega )=v_{Si}\cos \left( \frac{\pi }{2}\frac{q(\omega )}{q_{D}}\right).
\label{eq:groupvelocityModelC2}
\end{equation}

\subsection{High-energy optical phonon contribution}

Several high-energy optical phonon branches with moderate dispersion are approximated by Einstein modes at discrete frequencies $\omega_{{E}i}$ 
, $i=1$ ... $n$. The corresponding PDOS is given by a set of $\delta$-functions. 

These modes contribute to the specific heat at constant volume with \cite{GrossMarx}
\begin{equation} 
c_{V}^{{E}i}(T)=3rN_{A}k_{B}\left( \frac{\hbar \omega_{{E}i}}{k_{B}T}\right) ^{2}\frac{e^{\frac{\hbar \omega_{{E}i}}{k_{B}T}}}{\left( e^{\frac{\hbar \omega_{{E}i}}{k_{B}T}}-1\right) ^{2}}.
\label{eq:cpEinsteinModelC1}
\end{equation}

\subsection{Soft-mode phonon contribution}\label{SoftModePhononContribution}

Most important for the extraction of the transition temperature $T_{C}$ are those phonons which are responsible for the specific heat anomaly. These soft phonons are located at the R-point of the Brillouin zone and are considered here by taking into account an additional acoustic phonon branch representing the dispersion along the $\Gamma$-R direction. This branch, degenerate in the cubic phase, splits into two separate branches in the tetragonal phase. 

\subsubsection{Specific heat} 

The contribution of the soft-mode dispersion branch to the specific heat at constant volume is given by 
\begin{equation}
c_{V}^{S}(T)=\left\{ 
\begin{array}{cc}
2c_{V}^{E_{g}}(T)+c_{V}^{A_{1g}}(T)\text{,} & T<T_{C} \\ 
3c_{V}^{\Gamma _{25}}(T)\text{,} & T\geq T_{C}%
\end{array},
\right. 
\label{eq:CVSMonlyModelC4}
\end{equation}
where the
\begin{equation}
c_{V}^{M}(T)=\sum_{q=\frac{1}{2}\Delta q}^{\left( N+\frac{1}{2}\right)\Delta q }
\frac{V_{\text{mol}}}{2\pi ^{2}}q^{2}\frac{\partial }{\partial T}\left( 
\frac{\hbar \omega ^{M}\left( q,T\right) }{e^{\frac{\hbar \omega ^{M}(q.T)}{
k_{B}T}}-1}\right) \Delta q
\label{eq:CVMonlyBelowTCAndAboveTCSModelC4}
\end{equation}
are evaluated for all phonon frequencies $\omega ^{M}$ along the $\Gamma$-R dispersion branches associated with modes $M=E_{g}\text{, }A_{1g}\text{, }\Gamma _{25}$. The prefactors in Eq.~(\ref{eq:CVSMonlyModelC4}) take into account the correct mode degeneracy. 
In contrast to Eq.~(\ref{eq:cpacousticModelC4}), which is based on the PDOS, in Eq.~(\ref{eq:CVMonlyBelowTCAndAboveTCSModelC4}) we have chosen to evaluate the discrete summation in $q$-space. This allows us to minimize the computational time needed to achieve adequate convergence given the van Hove singularities in $D^{\prime}(\omega)$. We take a 1D $q$-'mesh' with $N=10000$ points, which turns out sufficient to ensure proper convergence. 

\subsubsection{Phonon dispersion}

We approximate the phonon dispersion along the $\Gamma$-R direction with a Born-von Kármán parameterization \cite{Brockhouse1962}, which allows us to include the soft phonons in the R-corner of the Brillouin zone in our fit model for the specific heat. Explicitly, the wavevector-dependence of the phonon frequency is parameterized by
\begin{equation}
\omega ^{2}=\frac{1}{\hbar ^{2}}\sum_{n=1}^{4}\phi _{n}\left( 1-\cos \left(2\pi \,n\,\frac{q}{q_{D}}\right) \right). 
\label{eq:softphonondispersionModelC4}
\end{equation}
This dispersion is defined in the wavevector interval $ 0 \leq q \leq q_{D}$, and the R-point is mapped to the Debye wavevector $q_D$. The coefficients $\phi _{n}$ are obtained by solving a system of four linear equations utilizing Eq.~(\ref{eq:softphonondispersionModelC4}) for four energies $E_{0.125}$, $E_{0.25}$, $E_{0.33}$ and $E_{0.5}$ chosen at wavevectors $q=0.125\,q_{D}, 0.25\,q_{D}, 0.33 \,q_{D}$ and $0.5 \,q_{D}$, respectively. 
The solution is given by 
\begin{equation}
\left( 
\begin{array}{c}
\phi _{1} \\ 
\phi _{2} \\ 
\phi _{3} \\ 
\phi _{4}%
\end{array}%
\right) =\left( 
\begin{array}{cccc}
-\frac{1}{2}\allowbreak a & -\frac{1}{4}a & \frac{2}{3}a & \frac{1}{8}c \\ 
0 & \frac{1}{2} & 0 & -\frac{1}{4} \\ 
\frac{1}{2}a & \frac{1}{4}a & -\frac{2}{3}a & \frac{1}{8}\sqrt{2} \\ 
\frac{1}{2}\allowbreak a & \frac{1}{4}\sqrt{2} & -\frac{2}{3}b & 
-\allowbreak \frac{1}{2}b%
\end{array}%
\right) \left( 
\begin{array}{c}
E_{0.125}^{2} \\ 
E_{0.25}^{2} \\ 
E_{0.33}^{2} \\ 
E_{0.5}^{2}
\end{array}
\right) ,
\label{eq:PhiMatrixEquationSoftModeModelC4}
\end{equation}
where $a =2-\sqrt{2}$, $b =1- \sqrt{2}$ and $c =4-\sqrt{2}$. The energies $E_{0.125}$, $E_{0.25}$ and $E_{0.33}$ are kept temperature-independent and are obtained from fits to inelastic neutron scattering (INS) data measured at $T=300$ K and $T=114$ K as
\begin{equation}
E_{0.125}=8.90\text{ meV, }E_{0.25}=13.97\text{ meV, } E_{0.33}=15.60\text{ meV}. 
\label{eq:FixedEnergiesSoftModeModelC4}
\end{equation}

\subsubsection{Temperature dependence of the R-point phonon}

The energy $E_{0.5}$ is the only temperature-dependent parameter in Eq.~(\ref{eq:PhiMatrixEquationSoftModeModelC4}) and is chosen to follow the temperature dependence of the R-point soft phonon mode at $\mathbf{q}=\left( 0.5,0.5,0.5\right)$ in both, the high-temperature cubic phase and the low-temperature tetragonal phase. Following Cowley {\it{et al.}}~\cite{Cowley1969} and Shirane {\it{et al.}}~\cite{Shirane1969} the soft mode temperature dependence is described by a generalized Curie-Weiss law
 \begin{equation}
E_{0.5}(T)=\left( A+\frac{C}{\left\vert T_{C}-T\right\vert }\right) ^{\eta },
\label{eq:SoftModeEnergyParameterizationSoftModeModelC4}
\end{equation}
where $T_{C}$ is the transition temperature of the structural phase transition. The parameters $A$, $C$ and the critical exponent $\eta$ are obtained from fits to INS and Raman spectroscopy data. 

There is debate in literature \cite{OtnesShirane1971} whether to include the parameter $A$ in the modified Curie-Weiss law Eq.~(\ref{eq:SoftModeEnergyParameterizationSoftModeModelC4}) or not. From our fits to the temperature dependence of the soft mode energy we find $A$ to be zero within error or very close to zero. Consequently, we set $A=0$ for our fits. The critical exponent for the $\Gamma_{25}$ mode is so close to the mean-field value $\left\vert \eta \right\vert =1/2$ that we have fixed this parameter to $1/2$. Thus the temperature behavior of the $\Gamma_{25}$ mode is described by a bare Curie-Weiss law without modifications. In contrast, this is not possible for  modes $E_{g}$ and $A_{1g}$ where the data fit critical exponents $\eta=-0.26$ and $\eta=-0.30$, respectively. 

A priori it is not clear whether or not the R-point soft mode energy is exactly zero at $T_{C}$ as were expected for the order parameter from mean-field theory. Likewise, it is not evident whether there is a discontinuity in the first-order derivative $dE_{0.5}/dT$ at $T_{C}$ as expected for a first-order phase transition, or if $dE_{0.5}/dT$ is a continuous function of $T$. Due to the finite instrumental resolution in earlier investigations by INS and Raman spectroscopy, data is lacking or is unreliable with large systematic errors in a roughly $\pm 5$K wide temperature interval close to $T_{C}$, so that these experiments appear to be not sensitive enough to clarify the exact temperature behavior of $E_{0.5}$ in the transition region.

We have therefore included an additional, phenomenological parameterization of the temperature dependence of the soft mode energy $E_{0.5}$ in the interval $98.6$ K $\leq T \leq 112$ K by a polynomial of 8th-order in $T$
 \begin{equation}
E_{0.5}\left( T\right) =\sum_{n=1}^{9}a_{n}\left( T-T_{C}\right) ^{n-1}.
\label{eq:8thOrderPolynomialSoftModeModelC4}
\end{equation}
Below  $98.6$ K and above $112$ K the temperature evolution remains unchanged and is described by the generalized Curie-Weiss law of Eq.~(\ref{eq:SoftModeEnergyParameterizationSoftModeModelC4}). This choice of temperature interval has been made on the basis of the availability of sufficiently precise neutron scattering data from reference \cite{Shirane1969}. 
The coefficients $a_{n}$ in Eq.~(\ref{eq:8thOrderPolynomialSoftModeModelC4}) are determined by six continuity conditions for the soft mode energy and its first and second derivatives each at $T=98.6$ K and $T=112$ K. At $T=98.6$ K, i.e.~in the tetragonal phase, this procedure is applied for modes $E_{g}$ and $A_{1g}$. 
The remaining three conditions to uniquely define the $a_{n}$ are set by  
\begin{eqnarray}
\left. E_{0.5}\right\vert _{T_{C}} &=&E_{\min }\text{, } \label{eq:BoundaryConditions7to9PolynomialSoftModeModelC4_1}
\\
\left. \frac{dE_{0.5}}{dT}\right\vert _{T_{C}} &=&0\text{, } 
\label{eq:BoundaryConditions7to9PolynomialSoftModeModelC4_2} \\
\left. \frac{d^{2}E_{0.5}}{dT^{2}}\right\vert _{T_{C}} &=&\kappa _{C}\text{. 
}
\label{eq:BoundaryConditions7to9PolynomialSoftModeModelC4_3}
\end{eqnarray}

With the condition given by Eq.~(\ref{eq:BoundaryConditions7to9PolynomialSoftModeModelC4_1}) we allow for the soft mode energy at $T_{C}$ to take a minimum value $E_{\text{min}}$, here chosen to be $E_{\text{min}}=0.5$ meV (rationale see below in section \ref{CouplingBarePhononFrequency}). The slope of $E_{0.5}(T)$ at $T_{C}$ is required to be zero, forcing the minimum of the polynomial to coincide with the critical temperature $T_{C}$. Finally, we allow for a curvature parameter $\kappa_{C}$ at $T_{C}$. Its numerical value is found by iteration and selected to ensure that the two polynomials for $E_{g}$ and $A_{1g}$ which join the curves defined by the modified Curie-Weiss law for $E_{g}$ and $A_{1g}$ lead to reasonably degenerate curves in the interval  $T_{C} \leq T \leq 112$ K. Solving the linear set of equations defined by the imposed boundary conditions leads to a unique set of coefficients $a_{n}$. 

\subsubsection{Coupling and bare phonon frequency}\label{CouplingBarePhononFrequency}

The physical motivation for our polynomial approximation roots in the experimental work by Shapiro {\it{et al.}} \cite{ShapiroAxeShiraneRiste1972}, \cite{ShiraneCowleyMatsudaShapiro1993} who have studied critical scattering in SrTiO\textsubscript{3} by thermal-neutron triple-axis spectroscopy. Based on anharmonic perturbation theory, the observed frequency spectrum of the fluctuations was interpreted to consist of a central-peak part and a soft-mode phonon part. Coupling of the soft mode to some - in Shapiro's work unspecified - degree of freedom with coupling parameter $\delta$ leads to a renormalization of the bare phonon frequency $\omega_{\infty} \left( \mathbf{q}, T \right)$  
\begin{equation}
\omega _{0}^{2}\left( \mathbf{q},T\right) =\omega _{\infty }^{2}\left( 
\mathbf{q},T\right) -\delta^{2} \left( T\right) .
\label{eq:ShapiroRenormalizationModelC4}
\end{equation}
Interestingly, the structural phase transition occurs when the renormalized frequency $\omega_{0}$ goes to zero at $T_{C}$, but the bare phonon frequency $\omega_{\infty}(T_{C})=\delta$ remains at finite energy for finite coupling constant $\delta$. Mean-field theory predicts that the square of the phonon frequency depends linearly on $\left( T-T_{C} \right)^{-1}$. In contrast, a deviation from mean-field behavior was experimentally observed by Shapiro {\it{et al.}} for $\left( T-T_{C} \right) < 10$ K. The experimental data shown in Fig. 9 of reference \cite{ShapiroAxeShiraneRiste1972} clearly demonstrates that $\partial \omega_{\infty} /\partial T = 0$ at $T_{C}$, while the soft-mode phonon energy levels off to finite energy $\omega_{\infty}(T_{C})=\delta$ upon cooling towards $T_{C}$. For the particular sample studied, the INS data allowed to extract a coupling constant $\delta=0.55(16)$ meV. This is in line with our model for the specific heat and the specific choice made for $E_{\text{min}}=0.5$ meV.

The coupling $\delta$ can be thought of as a decay channel for the soft phonons, phonon-phonon interaction being an obvious candidate. Silberglitt \cite{Silberglitt1972} has shown that the central peak behavior is consistent with a coupling, through quartic anharmonicity, of the R-corner soft phonon to a collective mode composed of $(-\mathbf{q},\mathbf{q})$-pairs of acoustic phonons. Treating the soft phonon self-consistently, he could show that the soft phonon mode remains at finite energy. The aforementioned INS studies have been later complemented by x-ray scattering experiments by Hirota {\it{et al.}} \cite{HirotaShapiroShirane1995}. Their experiments demonstrated that the coupling constant $\delta$ is sample-dependent, and they concluded that defects have an influence on the strength of the coupling parameter. Consistent with investigations which showed that an \emph{increasing} defect concentration causes a systematic \emph{increase} of central-peak intensity \cite{HastingsShapiro1978}, \cite{Currat1978}, a \emph{larger} defect concentration was thus suspected to lead to a \emph{larger} coupling parameter $\delta$ and consequently a \emph{larger} bare soft-mode energy $\hbar\omega_{\infty}(T_{C})$. Thus the coupling parameter $\delta$ can be identified with our parameter $E_{\text{min}}(T_{C})$ defined in Eq.~ (\ref{eq:BoundaryConditions7to9PolynomialSoftModeModelC4_1}). We note that the central peak appears to be elastic with an experimentally (by neutron backscattering experiments) demonstrated linewidth below $0.08$ $\mu$eV \cite{Topler1977}. Hence, any spectral component related to the central peak will not contribute to the specific heat. 

\subsubsection{Weighting function}

Since the phonon frequencies are temperature-dependent, the weighting function [defined in Eq.~(\ref{eq:WindowfunctionModelC2})]
now includes an explicit dependence on the temperature derivative of the phonon frequency $\partial \omega /\partial T$. 
\begin{eqnarray}
{W}_{ph}\left( \omega ,T\right) =k_{B}\frac{x^{2}e^{x}}{\left( e^{x}-1\right) ^{2}}+ 
\hbar \frac{\partial \omega }{\partial T}\frac{xe^{x}+e^{x}-1}{\left( e^{x}-1\right) ^{2}}
= k_{B} w_{ph} +\hbar \frac{
\partial \omega }{\partial T} w_{ph}^{\prime }
\label{eq:cVSoftModeBranchWindowFunctionModelC4}
\end{eqnarray}
where 
\begin{equation} 
x=\frac{\hbar \omega (q)}{k_{B}T}.
\label{eq:xDefinitionModelC4}
\end{equation} 
and
\begin{eqnarray}
w_{ph} &=&\frac{x^{2}e^{x}}{\left( e^{x}-1\right) ^{2}}, \label{eq:littlewsDefinitionModelC4_A} \\
w_{ph}^{\prime } &=&\frac{xe^{x}+e^{x}-1}{\left( e^{x}-1\right) ^{2}}.
\label{eq:littlewsDefinitionModelC4_B}
\end{eqnarray}

The $\left(\partial \omega /\partial T\right)$-dependent term in Eq.~(\ref{eq:cVSoftModeBranchWindowFunctionModelC4}) is significant for the evaluation of $c_{V}$ and, in case of the soft zone boundary modes close to $T_{C}$, is strongly dominating. 

\subsection{Total specific heat}

Adding the different contributions from all phonon modes, we obtain for the total specific heat at constant pressure $c_{p}$
\begin{eqnarray}
c_{p} &=& a_{{A}} c_{V}^{A}
+\sum\limits_{i=1}^{8}a_{Ei}\,c_{V}^{Ei}
+ a_{S} c_{V}^{S}
+V_{\text{mol}}TB_{T}\alpha _{V}^{2},
\label{eq:cptotalModelC4}
\end{eqnarray}
where the coefficients $a_{A,S,Ei}$ ensure appropriate weighting of the individual contributions.

The last term 
\begin{equation}  
V_{\text{mol}} T  B_{T} \alpha_{V}^{2} = c_{p}(T)-c_{V}(T)
\label{eq:cpminuscVTerm}
\end{equation}
accounts for the fact that the experiment provides a measure of specific heat at constant pressure $c_{p}$ and work is performed when the sample expands at constant pressure. In Eq.~(\ref{eq:cpminuscVTerm}) $B_{T}$ is the isothermal bulk modulus, i.e.~the inverse of the isothermal compressibility 
\begin{equation}  
\frac{1}{B_{T}}=-\frac{1}{V}\left. \frac{\partial V}{\partial p}\right\vert
_{T},
\label{eq:BulkModulus}
\end{equation}
and $\alpha_{V}$ is the volume coefficient of expansion. We assume an isotropic medium. Hence, $\alpha_{V}$ is simply a multiple of the linear coefficient of expansion $\alpha_{L}$, $L$ being the length of the sample, i.e. 
\begin{equation}  
\alpha _{V}=-\frac{1}{V}\left. \frac{\partial V}{\partial T}\right\vert
_{p}=-\frac{3}{L}\left. \frac{\partial L}{\partial T}\right\vert _{p}.
\label{eq:CoeffThermalExpansion}
\end{equation}
In order to determine the contribution to the specific heat according to Eq.~(\ref{eq:cpminuscVTerm}) we use computational data of Lu {\textit{et al.}}  \cite{YanliLu2015} who have computed $B_{T}$ and $\alpha_{V}$ over a wide temperature range.

\subsection{Fitting strategy to extract $T_{C}$}

In order to to extract the transition temperature $T_{C}$ for Verneuil-grown and EFG SrTiO\textsubscript {3} samples, we fit the $c_{p}$ data in a limited temperature range in the vicinity of the phase-transition temperature $T_{C}$. 

We introduce a baseline function 
\begin{equation}
c_{p,0}^{\text{tot}}\equiv \left. c_{p}\right\vert _{w_{ph}^{\prime }=0}.
\label{eq:C0AllPhononsModelC4}
\end{equation}
which evaluates $ c_{p}$ as defined in Eq.~(\ref{eq:cptotalModelC4}) albeit without the $\frac{\partial \omega }{\partial T}w_{ph}^{\prime }$ term in ${W}_{ph}$ [Eq.~ (\ref{eq:cVSoftModeBranchWindowFunctionModelC4})].
Omitting this term renders $c_{p,0}^{\text{tot}}$ to be a smooth baseline function devoid of any anomaly.

In order to obtain the parameters for  $c^{\text{tot}}_{p,0}$, we first fit the total specific heat function $c_{p}$ [Eq.~ (\ref{eq:cptotalModelC4})] divided by $T$, i.e. $c_{p}/T$ including both, $w_{ph}$ and $\frac{\partial \omega }{\partial T}w_{ph}^{\prime }$ terms to the experimental $c_{p}/T$ data for an arbitrarily chosen reference $T^{\prime}_{C}=107$ K. Only a subset of specific heat data is used for those fits, namely $c_{p}$ data for temperatures within 60 K $\leq T \leq $ 90 K and 115 K $\leq T \leq $ 145 K. Above 145 K significant amounts of excess specific heat are present, tentatively attributed to oxygen-octahedron rotation introduced by oxygen vacancies deep in the cubic phase. Since those processes are not included in our model, we exclude the $T$-region above 145 K from our fits. We find a good match between data and fit even below 60 K. However, we have chosen to exclude data below 60 K from the fits to avoid bias of the baseline function towards better fitting the low temperature $c_{p}/T$ data rather than the $c_{p}/T$ data in the $T$-region close to $T_{C}$.  The interval 90 K $ < T < 115$ K is excluded since the fit is obviously sensitive to $T^{\prime}_{C}$ here. 

Indeed, we find the best match between data and fit model if we fit $c_{p}/T$ instead of $c_{p}$. By normalizing with $T$, the experimental data points are implicitly weighted differently, so that more weight is attributed to the data points close to $T_{C}$ resulting in a better least-squares minimization in this $T$ region than would be obtained by fitting $c_{p}$ directly. The fitted values for the velocity of sound for transverse and longitudinal modes agree very well with the velocity of sound extracted from INS data. The low energy Einstein-mode frequencies match very well with those regions of the PDOS where an enhanced spectral weight is present. 

Subsequently to this fits, the baseline function is obtained by evaluating $c_{p,0}^{\text{tot}}$ using the parameters obtained from the fits, but excluding the $\frac{\partial \omega }{\partial T}w_{ph}^{\prime }$ term. Results of the fits (solid lines) and the total phonon baseline functions (dashed lines) are shown for Verneuil-grown and EFG samples in Fig.~1(a,b) in the main text by a $c_{p}/T$ representation of the data.

Next, we subtract the so-obtained baseline function $c_{p,0}^{\text{tot}}$ from the experimental data providing us with the difference specific heat 
\begin{equation}
\Delta c_{V}\equiv c_{p}^{\text{exp}}-c_{p,0}^{\text{tot}}.
\label{eq:DefinitionDifferenceSpecificHeatModelC4}
\end{equation}
The baseline-subtracted experimental data is shown in Fig.~1(c,d) in the main text for Verneuil-grown and EFG crystals. For temperatures below the phase transition $\Delta c_{V}$ is positive and slowly increasing towards $T_{C}$, a maximum is reached followed by a steep decrease of $\Delta c_{V}$ until a shallow minimum is reached with negative $\Delta c_{V}$. This is because of the sign change of $\frac{\partial \omega }{\partial T}$ at $T_{C}$. For temperatures above the phase transition $\Delta c_{V}$ is negative and remains almost constant up to $T \sim 130$ K. Above 130 K $\Delta c_{V}$ increases again due to the availability of other states which allow to accommodate thermal energy. 

Obviously the experimental data does not exhibit a true singularity in contrast to the model based on the generalized Curie-Weiss parameterization. This calls for a further extension of the model to reproduce the smooth temperature evolution of the anomaly. The temperature resolution in our short-pulse relaxation calorimetry experiments amounts to $\Delta T / T =1$ \%. At $T_{C}$ we thus expect a broadening due to experimental resolution by at most 1 K which by itself cannot explain the experimentally observed broadening by $\sim 10$ K. With confidence, we can thus expect a sample-related physical origin underlying the smooth evolution of the anomaly which we attribute to a spatial variation of the transition temperature.  

We therefore allow for a distribution of transition temperatures over a finite temperature interval, included in the model by convoluting the \emph{soft-mode anomaly function} $c_{V,1}^{SM}$ with a temperature-dependent Gaussian distribution $g\left( T-T_{0} \right)$ 
\begin{equation}
\left( c_{V,1}^{SM}\ast g\right) \left( T\right) =\int_{T_{\min }}^{T_{\max
}}c_{V,1}^{SM}\left( T_{0}\right) g(T-T_{0})dT_{0},
\label{eq:ConvolutionDeltacVModelC4}
\end{equation}
where the Gaussian distribution 
\begin{equation}
g(T-T_{0})=\frac{1}{\sqrt{2\pi }\sigma }e^{-\frac{1}{2\sigma ^{2}}\left(
T-T_{0}\right) ^{2}}
\label{eq:ConvolutionGaussianModelC4}
\end{equation}
is assumed to have a temperature-independent variance $\sigma$ corresponding to a full width at half maximum (FWHM) $\sigma _{\text{FWHM}}=\sqrt{8\ln 2}\sigma$. 

The \emph{soft-mode anomaly function} $c_{V,1}^{SM}$, defined as 
\begin{equation}
c_{V,1}^{SM}\equiv \left. c_{V}^{SM}\right\vert _{w_{ph}=0},
\label{eq:C1SMonlyModelC4}
\end{equation}
has no $w_{ph}$ term but includes the $\frac{\partial \omega }{\partial T}w_{ph}^{\prime }$ term in ${W}_{ph}$.
$c_{V,1}^{SM}$ is the only specific heat term which enters into the difference specific heat $\Delta c_{V}$. It is this term which is responsible for the anomaly observed in the specific heat.

In our fits of $\Delta c_{V}$ we allow for three free fit parameters which are: (1) the mean phase-transition temperature $T_{C}$, (2) the width $\sigma _{\text{FWHM}}$ of the $T_{C}$ distribution, and (3) an amplitude factor $f_{SM}$. The latter quantity is the fraction of phonon modes located on the $\Gamma$-R branch which contributes to the specific heat $c^{2S+SM}_{V}$ provided by the low-energy acoustic phonon branches (two transverse branches, one longitudinal branch and the soft-mode branch).
In order to minimize the sensitivity to variations of the baseline, our fits are limited to the temperature interval $90$ K $\leq T \leq 120$ K.

\clearpage

\section{Background subtraction from elastic scans}\label{SI_B1}

$(\textbf{Q},T)-$dependent background subtraction was performed to remove the phonon contribution from the $\textbf{Q}-$integrated intensities of the elastic scans at \textbf{\textit{Q$_R$}}$=(3/2,3/2,5/2)$. As an example, here below we describe the background subtraction procedure for the EFG sample for a set of elastic scans along [111]-direction. We followed exactly the same steps for the other set of elastic scans for both EFG and VER samples.

\begin{figure}[htb]
\includegraphics[width=0.85\textwidth]{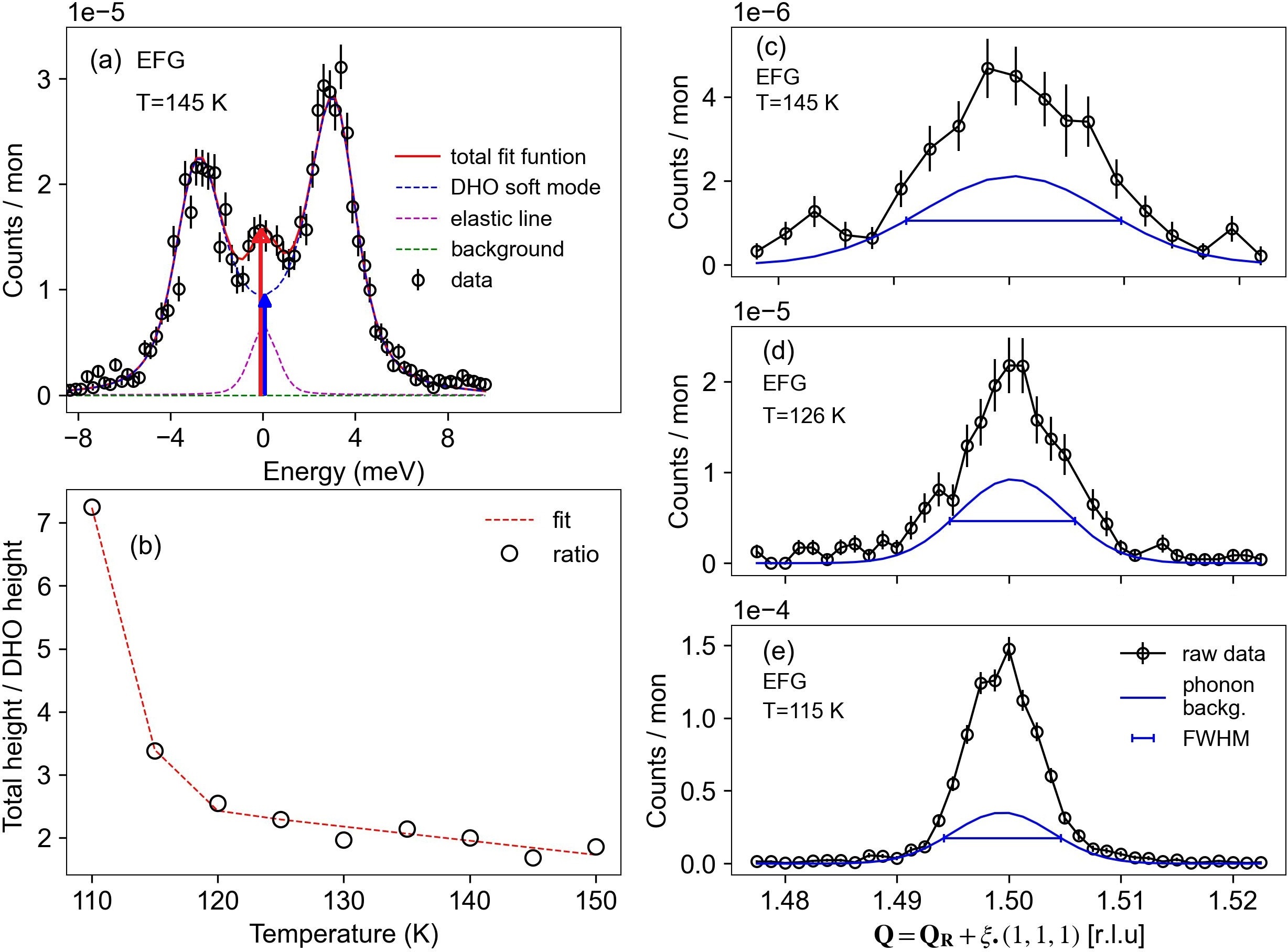}
\caption{\label{fig_S5}(a) Constant$-$\textbf{\textit{Q}} scan performed at \textbf{\textit{Q$_R$}}$=(3/2,3/2,5/2)$ at $T=145$ K for the EFG grown sample. Solid (red) line represents a total fit consisting of a DHO function convoluted with the experimental resolution (blue dashed line), estimated background (straight green dashed line) and a resolution limited pseudo-Voigt function for the elastic line (purple dashed line). (b) Height ratios of the total fit function and the DHO soft mode at the $E=0$ for different temperatures, red dashed line presents a Gaussian fit to the ratios. (c,d,e) Estimated phonon background presented in blue solid line corresponding to the elastic scans along $[111]-$direction at three different temperatures $T=145$ K, $T=126$ K and $T=115$ K, respectively.}
\end{figure}

In order to estimate the soft mode contribution to the elastic peak we first extracted the heights of the total fit function (red arrow) and DHO soft mode (blue arrow) from each energy scans at $E=0$ using the fit model described in main text (see Fig. \ref{fig_S5}(a)). The ratio of the heights (total/DHO) is presented as a function of temperature in the Fig. \ref{fig_S5}(b). It is evident that the DHO soft mode contribution to the elastic intensity is significant especially at higher temperatures. With lowering temperature the contribution starts to diminish as the central peak intensity dominates the energy scans. The $T-$dependence of the ratio was extracted by fitting with a Gaussian function with a peak fixed at 108 K.

The $Q-$ dependence of the soft mode background has been estimated using a Gaussian function with relative height compared to the elastic peak following the ratio in Fig. \ref{fig_S5}(b), and with maximum possible FWHM avoiding the negative intensity of the subtracted scans. Fig. \ref{fig_S5}(c,d,e) present the estimated phonon background for elastic scans along the [111]-direction at different temperatures. The intensity ratio of the raw data and the phonon background at $Q_R = [3/2,3/2,5/2]$ is similar to that in Fig. \ref{fig_S5}(b) at different temperatures. As can be seen the phonon background contribution is significant at high temperature and becomes negligible near to the transition temperature $T_c$ below 115 K.

\section{Comparison of soft mode intensity}\label{SI_C}

\begin{figure}[htb]
\includegraphics[width=0.45\textwidth]{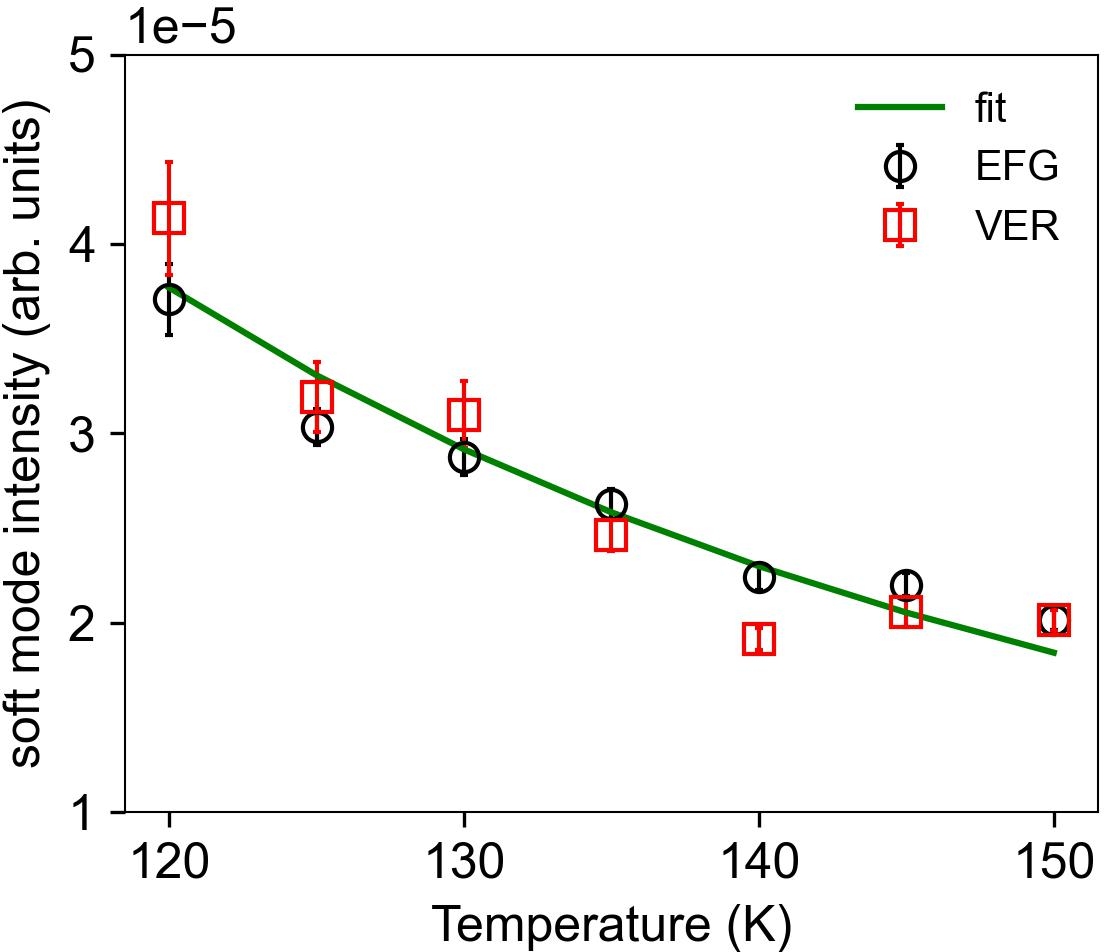}
\caption{\label{fig_S6}Integrated intensities of the DHO soft mode for the EFG and VER samples as a function of temperature.}
\end{figure}

We compared the integrated intensities of the DHO soft mode for both the EFG and VER samples in Fig. \ref{fig_S6}. The plot shows that the phonon intensities are the same for both samples although the central peak intensity in the VER sample containing lower defects is 4 times higher compared to EFG sample containing lower defects. This indicates that the collective lattice excitations are insensitive to the local defects of the EFG and VER samples.

\section{Assessment of mosaicity from IXS}\label{SI_D}
\begin{figure}[htb]
\includegraphics[width=0.5\textwidth]{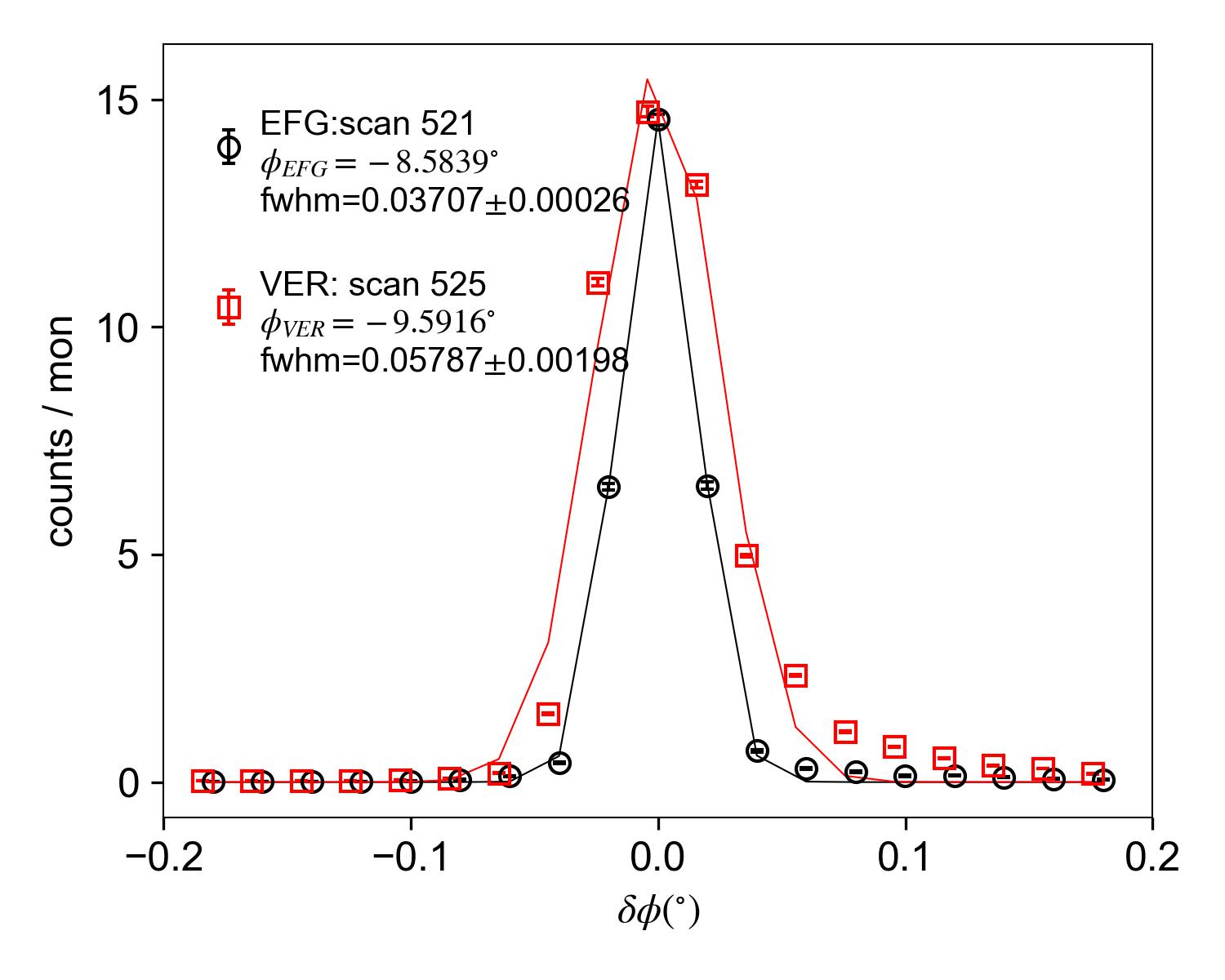}
\caption{\label{fig_S7} Rocking scans across the (222) Bragg peak from the EFG and VER samples.}
\end{figure}

Fig.  \ref{fig_S7} presents the rocking scans performed across (222) Bragg peak for both samples in the high momentum resolution setting of our setup (\textit{i.e.} with a 15 mm analyzer opening).  From the Gaussian fit to the data, we get the respective mosaicity $\delta\phi_{EFG} = 0.0371(3)^{\circ}$ and $\delta\phi_{VER} = 0.058(2)^{\circ}$, indicating the EFG sample is crystallographically more perfect compared to the VER sample. Our findings are in line with the findings of neutron Larmor diffraction results (see supplementary information section: \ref{larmor}).

\section{Assessment of Q-resolution from IXS}\label{SI_e}

\begin{figure}[htb]
\includegraphics[width=0.85\textwidth]{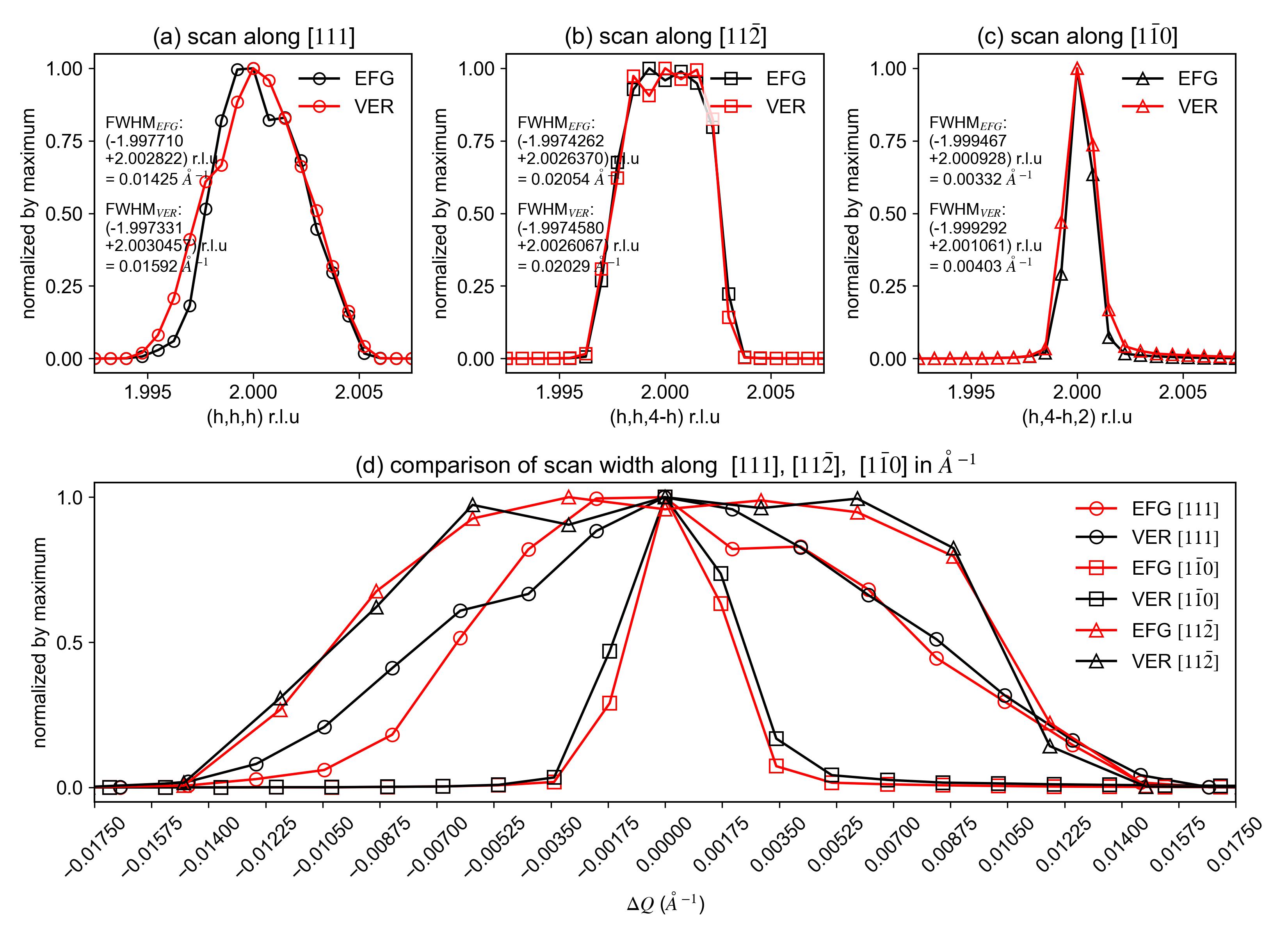}
\caption{\label{fig_S8} Scans through (222) Bragg reflection along (a) [$111$] (b) [$11\bar{2}$] and (c) [$1\bar{1}0$] directions. (d) Scans presented in absolute units (\AA$^{-1}$) relative to the peak centers.}
\end{figure}

In order to assess the instrumental $Q-$resolution in the high resolution setting, \textit{i.e.} with analyzer opening $\Phi=15$ mm,  we have performed $Q-$scans across the (222) Bragg peak. Fig. \ref{fig_S8}(a,b,c) show the scans along the three orthogonal directions [$111$], [$11\bar{2}$] and [$1\bar{1}0$]. Fig. \ref{fig_S8}(d) presents the scans in absolute units (\AA$^{-1}$) relative to the peak centers. It is evident that the instrumental $Q-$resolution along the [$1\bar{1}0$] direction is better by a factor of 4 compared to the other two directions. We used the extracted resolutions for EFG and VER samples (value indicated in the figure panels a,b,c) to determine the correlation lengths of the superlattice peak at $\boldsymbol{Q}_{\mathrm{R}} = (3/2, 3/2, 5/2)$ in the three orthogonal directions as described in the main text.

\section{$T-$dependent energy scans}\label{SI_f}
\begin{figure}[h]
\includegraphics[width=0.65\textwidth]{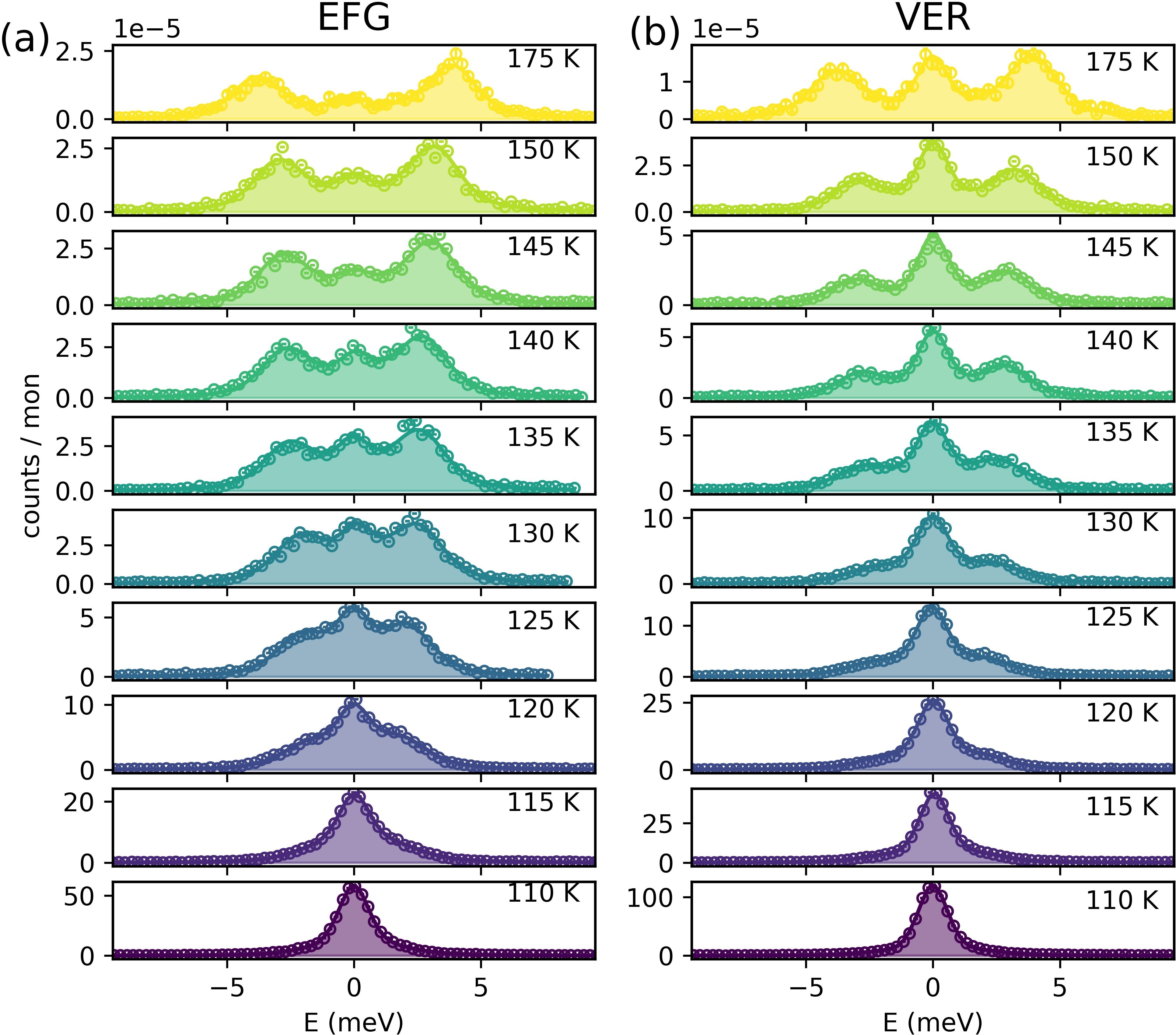}
\caption{\label{fig_S9} $T-$dependent energy scans at $\boldsymbol{Q}_{\mathrm{R}} = (3/2, 3/2, 5/2)$ in the temperature range from 175 K to 110 K for (a) EFG and (b) VER samples.}
\end{figure}

Fig. \ref{fig_S9} presents the energy scans performed for the EFG and VER samples in the temperature range from 175 K to 110 K. The scans are normalized with respect to the monitor. Comparing the intensity at $E=0$ for the two samples, it is evident that the central peak is stronger in the VER sample containing less defects with respect to the EFG sample. We have extracted the $T-$dependence of the phonon energies $(E_{ph})$ and linewidths $(\Gamma_{ph})$ of the soft phonon mode (as presented in Fig. 2(c) in the main text) from the fits of the energy scans in Fig. \ref{fig_S9}.  Solid lines [color-coded] represent the fits to the energy scans consisting of a DHO function convoluted with the experimental resolution, an estimated background and a resolution-limited pseudo-Voigt function for the elastic line ($E=0$). All data corresponding to these energy scans are available at the open data repository KITopen \citep{SKITopen_STO_2025}.

\section{$T-$dependent momentum scans along [111]-direction}\label{SI_g}

\begin{figure}[h]
\includegraphics[width=0.85\textwidth]{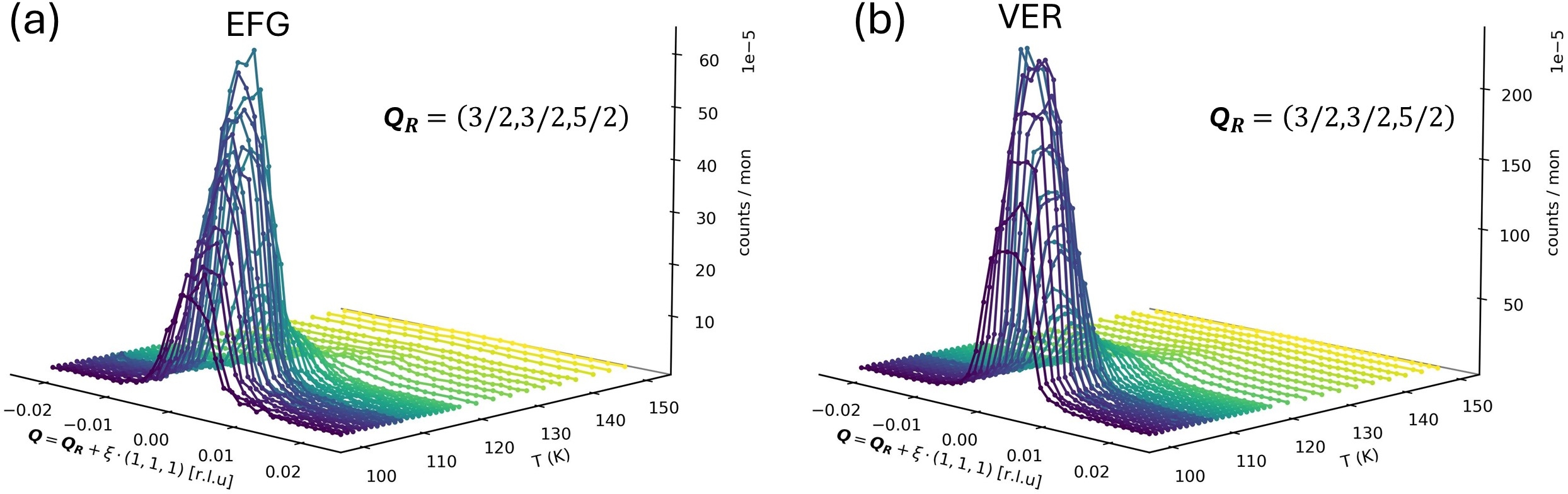}
\caption{\label{fig_S10} $T-$dependent momentum scans at $\boldsymbol{Q}_{\mathrm{R}} = (3/2,3/2,5/2)$ along [$111$]-direction in the temperature range from 175 K to 110 K for (a) EFG and (b) VER samples.}
\end{figure}

Fig. \ref{fig_S10} presents the momentum scans performed along [111]-direction for the EFG and VER samples in the temperature range of 150 K to 100 K. The data sets presented here are normalized with respect to the monitor. Comparing the peak intensity from the waterfall plots for the two samples close to their respective AFD transitions [T$_{c,EFG} = 109.5(1)$ K and T$_{c,VER} = 106.3(2)$ K], it is evident that the intensity at $\boldsymbol{Q}_{\mathrm{R}} = (3/2, 3/2, 5/2)$ is almost four times stronger in the VER sample containing less defects with respect to the EFG sample. We have extracted the $T-$dependence of the integrated intensities and the full width at half maxima (as presented in Fig. 3 (c,d) in the main text) from the Gaussian fits of the momentum scans in Fig. \ref{fig_S10}. A phonon background was subtracted from each scan, as described in section \ref{SI_B1} of the supplementary information, before the Gaussian fitting was performed. All data corresponding to these momentum scans along [111] are available at the open data repository KITopen \citep{SKITopen_STO_2025}.

\section{High-resolution momentum scans}\label{SI_h}
\begin{figure}[h]
\includegraphics[width=0.95\textwidth]{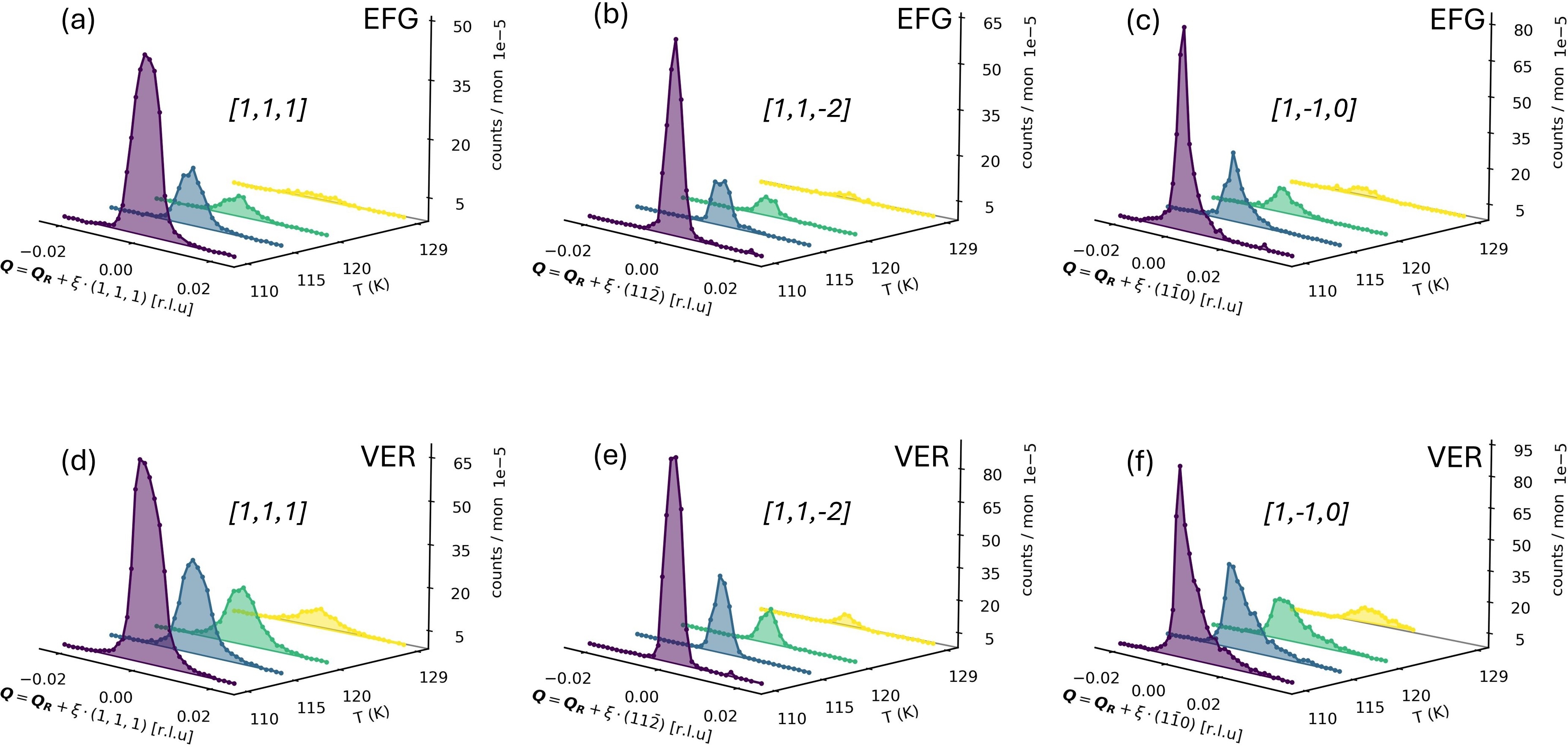}
\caption{\label{fig_S11} High-resolution momentum scans at $\boldsymbol{Q}_{\mathrm{R}} = (3/2, 3/2, 5/2)$ in the temperature range 130 K to 110 K along [$111$], [$11\bar{2}$] and [$1\bar{1}0$] directions are presented respectively in (a, b, c) for EFG sample and in (c, d, e) for VER sample.}
\end{figure}

Fig. \ref{fig_S11} presents the high-resolution momentum scans at $\boldsymbol{Q}_{\mathrm{R}} = (3/2,3/2,5/2)$  along three orthogonal directions [$111$], [$11\bar{2}$] and [$1\bar{1}0$] for the EFG and VER samples in the temperature range from 130 K to 110 K. The data sets presented here are normalized with respect to the monitor. Comparing the peak shapes from the waterfall plots of the two samples at 100 K, close to the AFD transitions ($T_c$s), it is evident that the peak shapes (representing the anti-ferrodistortive correlations) present anisotropy in the three orthogonal directions. We have extracted the Lorentzian width [inverse of correlation length](as presented in Fig. 4 (c-f) in the main text) by careful analysis of the scans using a Voigt fit function with a fixed Gaussian width for instrument resolution, which reveals that the anisotropic correlation is comparatively stronger in the VER sample containing the less defect with respect to the EFG sample. All data corresponding to these high resolution momentum scans along [$111$], [$11\bar{2}$], and [$1\bar{1}0$] directions are available at the open data repository KITopen \citep{SKITopen_STO_2025}.

\newpage

\end{document}